\documentclass[twocolumn]{aastex63}
\usepackage{longtable}
\usepackage{supertabular}

\received{June 1, 2019}
\revised{January 10, 2019}
\accepted{\today}

\shorttitle{Electron temperatures of planetary nebulae derived with the LOFAR}
\shortauthors{Hajduk et al.}

\graphicspath{{./}{figures/}}

\begin{document}

\title{Evidence for cold plasma in planetary nebulae from radio observations with the LOw Frequency ARray (LOFAR)}

\correspondingauthor{Marcin Hajduk}
\email{marcin.hajduk@uwm.edu.pl}

\author[0000-0001-6028-9932]{Marcin Hajduk}
\affiliation{Department of Astrophysics/IMAPP, Radboud University, P.O. Box 9010, 6500 GL Nijmegen, The Netherlands}
\affiliation{Space Radio-Diagnostics Research Centre, University of Warmia and Mazury, ul.Oczapowskiego 2, 10-719 Olsztyn, Poland}

\author{Marijke Haverkorn}
\affiliation{Department of Astrophysics/IMAPP, Radboud University, P.O. Box 9010, 6500 GL Nijmegen, The Netherlands}

\author{Timothy Shimwell}
\affiliation{ASTRON, Netherlands Institute for Radio Astronomy, Oude Hoogeveensedijk 4, Dwingeloo, 7991 PD, The Netherlands}
\affiliation{Leiden Observatory, Leiden University, PO Box 9513, 2300 RA, Leiden, The Netherlands}

\author{Mateusz Olech}
\affiliation{Space Radio-Diagnostics Research Centre, University of Warmia and Mazury, ul.Oczapowskiego 2, 10-719 Olsztyn, Poland}

\author{Joseph R. Callingham}
\affiliation{ASTRON, Netherlands Institute for Radio Astronomy, Oude Hoogeveensedijk 4, Dwingeloo, 7991 PD, The Netherlands}
\affiliation{Leiden Observatory, Leiden University, PO Box 9513, 2300 RA, Leiden, The Netherlands}

\author{Harish K. Vedantham}
\affiliation{ASTRON, Netherlands Institute for Radio Astronomy, Oude Hoogeveensedijk 4, Dwingeloo, 7991 PD, The Netherlands}
\affiliation{Kapteyn Astronomical Institute, University of Groningen, PO Box 72, 97200 AB, Groningen, The Netherlands}

\author{Glenn J. White}
\affiliation{Department of Physics and Astronomy, The Open University, Walton Hall, Milton Keynes, MK7 6AA, UK}
\affiliation{RAL Space, STFC Rutherford Appleton Laboratory, Chilton, Didcot, Oxfordshire, OX11 0QX, UK}


\author{Marco Iacobelli}
\affiliation{ASTRON, Netherlands Institute for Radio Astronomy, Oude Hoogeveensedijk 4, Dwingeloo, 7991 PD, The Netherlands}

\author{Alexander Drabent}
\affiliation{Th\"{u}ringer Landessternwarte, Sternwarte 5, D-07778 Tautenburg, Germany} 



\begin{abstract}
We present observations of planetary nebulae with the LOw Frequency ARray (LOFAR) between 120 and 168\,MHz. The images show thermal free-free emission from the nebular shells. We have determined the electron temperatures for spatially resolved, optically thick nebulae. These temperatures are 20 to 60\% lower than those estimated from collisionally excited optical emission lines. This strongly supports the existence of a cold plasma component, which co-exists with hot plasma in planetary nebulae. This cold plasma does not contribute to the collisionally excited lines, but does contribute to recombination lines and radio flux. Neither of the plasma components are spatially resolved in our images, although we infer that the cold plasma extends to the outer radii of planetary nebulae. However, more cold plasma appears to exist at smaller radii. The presence of cold plasma should be taken into account in modeling of radio emission of planetary nebulae. Modelling of radio emission usually uses electron temperatures calculated from collisionally excited optical and/or infrared lines.  This may lead to an underestimate of the ionized mass and an overestimate of the extinction correction from planetary nebulae when derived from the radio flux alone. The correction improves the consistency of extinction derived from the radio fluxes when compared to estimates from the Balmer decrement flux ratios.
\end{abstract}



\section{Introduction} \label{sec:intro}

Planetary nebulae (PNe) are detectable at a broad range of wavelengths, from X-rays up to radio frequencies. Continuum radio emission originates from thermal free-free emission of ionized elements. It traces all of the ionized ejecta in PNe. Radio emission is not affected by interstellar or circumstellar extinction caused by dust.

Radio observations constrain the physical parameters of  astrophysical plasma. In particular, optically thick free-free emission allows the electron temperature to be determined from the Rayleigh-Jeans law. Optically thick free-free emission decreases very quickly as frequency squared making it difficult to detect. However, the LOw Frequency ARray \citep[LOFAR][]{2013A&A...556A...2V} provides enough sensitivity and spatial resolution to image optically thick radio emission of PNe.

The electron temperature is one of the most important parameters in studying PNe. It governs the energy balance and is a very important parameter when assessing the chemical composition of PNe \citep{2002RMxAC..12...62S}. Electron temperatures can be measured from the flux ratios of collisionally excited lines (CEL) \citep{1986ApJ...308..322K}. These depend on the electron density and rely on the accuracy of the determination of transition probabilities and collision strengths. CELs are suppressed by collisional de-excitation when the critical density is exceeded in plasma. They are also weak at low electron temperatures. Thus, electron temperatures measured from CELs are weighted towards hot regions that do not exceed the critical density, and are less sensitive to dense and cool plasma.

Electron temperatures can be alternatively derived from recombination lines (RLs) or from free-bound emission (e.g. Balmer jump). RLs are in general much fainter than CELs, and therefore more difficult to measure. However, electron temperatures derived from RLs are systematically lower than from CELs. Moreover, the abundances derived from RLs are higher than the abundances obtained from CELs \citep{1971BOTT....6...29P,2001A&A...379.1024S,2004MNRAS.351..935Z,2005MNRAS.362..424W}. The difference between these two determinations is referred to as an abundance discrepancy factor. \citet{1971BOTT....6...29P} attributed this discrepancy to temperature fluctuations in PNe. If large temperature fluctuations occur, then the temperature measured from CELs is overestimated and RLs appear stronger. However, photoionization models failed to reproduce temperature fluctuations sufficiently large enough to account for the temperature and abundance discrepancy \citep{1995ApJ...450..691K}. 

\citet{2000MNRAS.312..585L} showed that the inclusion of high-density hydrogen-deficient plasma can explain the RL and CEL temperature and abundance discrepancies. This has been subsequently confirmed by \citet{2004MNRAS.353..953T}, \citet{2005MNRAS.358..457Z}, and
\citet{2005MNRAS.362..424W}. A further insight came from \citet{2015ApJ...803...99C}, who linked the abundance discrepancy with binarity of their central stars. \citet{2019arXiv190406763G} present the most recent review of the topic. 

Optically thick free-free radio emission provides an alternative method to assess electron temperatures in PNe. Brightness temperature is simply equal to electron temperature for optically thick thermal radiation. Unlike other methods, this determination does not depend on electron density. The only assumption is Maxwellian distribution of electrons in nebular plasma, which is most likely fulfilled \citep{2018ApJ...862...30D}. In this paper we report electron temperatures for a sample of PNe using LOFAR observations of optically thick free-free emission.

\section{Radio emission from planetary nebulae}

\begin{table*}
\centering
\caption{The flux densities, deconvolved diameters $\Theta_\mathrm{d}$, corrected diameters $\Theta$, and optical diameters taken from \citet{2016MNRAS.455.1459F} of PNe detected in the LoTSS survey. The deconvolved and corrected diameters are not given for unresolved PNe. Large PNe were not fitted with Gaussian and their diameters $\Theta_\mathrm{d}$ refer to the size which exceeded 3$\sigma$ (see text).} \label{tab:observations}
\begin{tabular}{ccccc}
\hline
\hline
\\[-1.5ex]
Name & F(144\,MHz) [mJy] & $\Theta_\mathrm{d}$ [$\mathrm{arcsec} \times \mathrm{arcsec}$] & $\Theta$ [arcsec] &  $\Theta_\mathrm{opt}$ [arcsec]  \\
\\[-1.5ex]
\hline
\\[-1.5ex] 
BV\,5-1    & $6.1 \pm 1.0$   & $12.4 \times 5.1$   & $14.4 \pm 7.8$  & $42 \times 10$  \\
BV\,5-2    & $4.0 \pm 1.1$   & $19.5 \times 14.3$   & $24 \pm 11$  &  \\
H\,3-29     & $21.6 \pm 2.8 $   &   $18.1 \times 15.0$    & $22.8 \pm 1.6$ & $23.8 \times 23$ \\
H\,4-1    & $0.95 \pm 0.20$  &   & & $2.7 \pm 2.7$   \\
IC\,2149    & $20.4 \pm 4.3$ &  $13.7 \times 8.0$ & $14.7 \pm 3.4$ & $12.5 \times 8.0$  \\
IC\,3568    & $20.1 \pm 2.2$ &  $12.5  \times	11.9$ & $17.24 \pm 0.57$  & $17.8 \times 17.8$  \\
IC\,4593    & $16.7 \pm 2.0$ &  $13.2  \times	9.4$ & $15.6 \pm 1.1$ & $15.3 \times 14.7$  \\
J\,320 & $10.1 \pm 1.7$ & $7.0  \times	3.9$  & $8.9 \pm 1.9$ & $9.4 \times 6.3$ \\
K\,3-17    & $52 \pm 12$ &  $31.9  \times 14.5$ & $29.6 \pm 3.8$   & $18.6 \times 11.9$ \\
K\,3-80    & $6.7 \pm 1.2$ &  $5.7  \times 5.2$ & $8.0 \pm 1.5$  & $6 \times 6$ \\
M\,1-1    & $7.6 \pm 1.4$ &  $8.1  \times 3.6$ & $8.2 \pm 1.9$   & $7 \times 6$ \\
M\,2-51  & $24.4 \pm 3.8$    & $36.9	\times	27.4$ & $45.8 \pm 5.6$ & $64 \times 48$ \\
M\,2-52  & $13.8 \pm 2.2$    & $12.2	\times	10.1$ & $15.9 \pm 1.9$ & $16 \times 13$ \\
NGC\,1514  &  $191 \pm 69$ & $108.0 \times 69.2$  & & $188 \times 182$ \\
NGC\,2242   &  $8.3 \pm 1.4$ & $15.8 \times 14.2$   & $20.1 \pm 2.1$ & $20 \times 20$ \\
NGC\,2371   & $33.6 \pm 8.1$  &  $56.3 \times 36.6$  & & $48.9 \times 30.6$ \\
NGC\,3587   &  $105 \pm 21$ &    $185.9 \times 182.8$  & & $208 \times 202$ \\
NGC\,40   &  $115 \pm 30$ &    $47.2 \times 42.4$  & & $56 \times 34$   \\
NGC\,6058   & $5.8 \pm 1.0$  & $16.1 \times 14.1$   & $24.4 \pm 4.5$ & $36 \times 28$ \\
NGC\,6210   & $26.7 \pm 3.6$  & $19.2 \times 13.9$   & $24.5 \pm 3.4$ & $14 \times 14$ \\
NGC\,650/651   & $86 \pm 15$  &  $126 \pm	68$ & & $168 \times 111$ \\
NGC\,6543   & $56.1 \pm 6.1$  & $23.0 \times 18.2$   & $28.2 \pm 1.1$ & $26.5 \times 23.5$ \\
NGC\,6572   & $11.5 \pm 2.1$  &    &  & $15 \times 13$ \\
NGC\,6720   &  $155 \pm 51$ &    $84.6 \times	59.1$  & & $89 \times 66$ \\
NGC\,6826   & $77.6 \pm 8.4$  & $22.3 \times 18.7$   & $28.1 \pm 1.0$ & $27 \times 24$ \\
NGC\,7027   & $16.6 \pm 2.2$  & $12.7 \times 5.6$   & $12.2 \pm 1.5$ & $15.6 \times 12.0$  \\
PM\,1-305   & $7.7 \pm 2.5$  &    & &    \\
We\,1-1   & $6.4 \pm 1.7$  &    &  &   \\
RA\,24     & $9.1 \pm 1.7$  & $21.7 \times 5.0$  &  $17.7 \pm 6.6$ &  \\
Vy\,1-2     & $1.76 \pm 0.64$  &   &  & $6 \times 4$  \\
\\[-1.5ex]
\hline
\end{tabular}
\end{table*}

In the case of an ionized nebula with constant electron density and temperature $T_\mathrm{e}$ (hereinafter referred to as an homogeneous nebula or a homogeneous model) which covers a solid angle of $\Omega$, the free-free flux density is given by

\begin{equation} \label{eq:snu} S_\nu = \frac{2 \nu^2 k T_\mathrm{e}}{c^2}(1-e^{-\tau_\nu}) \Omega \end{equation}
in the Rayleigh-Jeans approximation. The optical depth is given as 
$\tau_{\nu} = 0.0544 \times T_\mathrm{e}^{-1.5} \, \nu^{-2} g_{\mathrm{ff}} (\nu, T_\mathrm{e}) \, EM$ \citep{1975A&A....39..217O}. $EM$ denotes emission measure, $k$ - Boltzmann constant, $\nu$ - frequency, and c - the speed of light. \citet{2014MNRAS.444..420V} computed non-relativistic Gauntt factors $g_{\mathrm{ff}}$ for a wide range of frequencies.

The radio spectra of PNe appear nearly flat ($S_\nu \propto \nu^{-0.1}$) in the optically thin part, at $\tau_\nu \rightarrow 0$. The spectrum steeply declines with decreasing frequency squared ($S_\nu \propto \nu^2$) when $\tau_\nu \gg 1$. The emission peaks at the turnover frequency close to $\tau_\nu \approx 1$. 

The brightness temperature depends on the surface brightness of the object
\begin{equation} \label{eq:tb}
T_\mathrm{B} (\nu) = \frac{\mathrm{c}^2}{2 \mathrm{k} \nu^2} \frac{S_{\nu}}{\Omega}.
\end{equation}
The brightness temperature approaches electron temperature in an optically thick case. It is often assumed, that PNe with $T_\mathrm{B}$ determined from Equation \ref{eq:tb} lower than some arbitrary value (e.g. 1\,kK in \citet{2004MNRAS.353..796R}, 3\,kK in \citet{1992A&A...266..486S}) are optically thin, which is well justified in the case of homogeneous nebula for a typical electron temperature of 10\,kK. However, the study of the radio spectral indices ($SI$) $F(5\,{\mathrm{ GHz}})/F(1.4\,{\mathrm{GHz}})$ and brightness temperatures by \citet{2007MNRAS.378..231P} revealed that the majority of PNe show an excess of the $F(5\,{\mathrm{ GHz}})/F(1.4\,{\mathrm{GHz}})$ ratios over the value predicted by homogeneous model. \citet{2007MNRAS.378..231P} has attributed this excess to the existence of strong radial density gradients in the nebulae. In such cases, nebulae become partially optically thick over a wide range of frequencies \citep{1975MNRAS.170...41W}. 

\citet{2001A&A...373.1032S} attempted to explain the  $F(5\,{\mathrm{ GHz}})/F(1.4\,{\mathrm{GHz}})$ index excess using an alternative approach. They used two components instead of one in Equation\,\ref{eq:snu}. One of the components covers only a fraction of the solid angle $\xi \Omega$ and has an optical thickness of $\tau_\nu$. The other component covers the rest of the solid angle $(1 - \xi \Omega)$ and has an optical thickness of $\eta \tau_\nu$: \begin{equation} \label{eq:snu2} S_\nu = \frac{2 \nu^2 k T_\mathrm{e}}{c^2}((1-e^{-\tau_\nu}) \xi \Omega + (1-e^{-\eta \tau_\nu}) (1 - \xi \Omega)). \end{equation}
\citet{2001A&A...373.1032S} achieved the best fit for Equation\,\ref{eq:snu2} using $\xi = 0.27$ and $\eta = 0.19$.

\citet{2018MNRAS.479.5657H} studied radio spectra of PNe and excluded the presence of strong density gradients. We showed that a prolate ellipsoidal shell model \citep{1990ApJ...348..580M,1996ApJ...462..813A} has a higher $F(5\,{\mathrm{ GHz}})/F(1.4\,{\mathrm{GHz}})$ index compared to the homogeneous model. Other studies have shown that the prolate ellipsoidal shell model provides a better fit to the observed surface brightness distribution of PNe than the homogeneous model. However, ellipsoidal shells would have to be enormously elongated to account for the high excesses observed in some PNe.

Equation\,\ref{eq:snu} gives a satisfactory fit to most of the PNe using electron temperatures derived from CELs given that $\Omega$ is smaller than the observed size of the nebula, i.e. when the bulk of the emission comes from a fraction of the solid angle. This is equivalent to Equation\,\ref{eq:snu2} for $\eta = 0$ and $0 < \xi < 1$. It is impossible to find a single value of $EM$ which would allow fitting the optically thin and optically thick parts of the spectrum simultaneously for $\xi = 1$ in most cases. With higher $EM$ \citet{2018MNRAS.479.5657H} could reproduce the turnover frequency, but overestimated the optically thin flux. Lower values of $EM$ allowed us to fit the optically thin part of the spectrum, but shifted the turnover to lower frequencies than observed.

\section{Observations and data analysis}

\begin{table*}
\centering
\caption{Comparison of mean electron temperatures derived from radio observations optically thick PNe at 144\,MHz, Balmer jump, helium lines, [O\,{\sc iii}] and [N\,{\sc ii}] CELs, and RL of O\,{\sc ii} in Kelvin.} \label{tab:temperatures}
\begin{tabular}{cccccccc}
\hline
\hline
\\[-1.5ex]
Name & $\mathrm{144\,MHz}^a$ & BJ$^b$ & He\,{\sc i} $\lambda 7281 / \lambda 6678^c$ & He\,{\sc i} $\lambda 7281 / \lambda 5876^c$ & [O\,{\sc iii}]$^{d,e}$ & [N\,{\sc ii}]$^{d,e}$ & O\,{\sc ii}$^f$ \\
\\[-1.5ex]
\hline
\\[-1.5ex]
IC\,2149    & $7700 \pm 3000$ & & & & 10300  & 8700 & \\
IC\,3568    & $5760 \pm 680$   & $9500 \pm 900$ & $8100 \pm 1000$ & $7800 \pm 1450$ & 10400 &  & 400 \\
IC\,4593    & $5680 \pm 850$ & & & & 8900 & 11400 & 630 \\
K\,3-17   & $5100 \pm 1500$ & & & (11900) & 13300 & & \\
NGC\,40   & $4700 \pm 1300$ & $7000 \pm 700$ & $10240 \pm 1900$ & $10580 \pm 4200$ & 11000 & 7900 & 400 \\
NGC\,6543   & $6010 \pm 730$ & $6800 \pm 1400$ & $6010 \pm 1400$ & $5450 \pm 1400$ & 8100 & 9000 & 500 \\
NGC\,6826   & $8360 \pm 980$ & $8700 \pm 700$  & $8290 \pm 1500$ & $8520 \pm 2000$ & 11200 & 12200 & 800 \\
NGC\,7027   & $9600 \pm 2000$ & $12000 \pm 400$ & $10360 \pm 1100$ & $9030 \pm 2200$ & 12400 & (13700) & 450 \\
\\[-1.5ex]
\hline
\end{tabular}
\\
$^a$ This work 
$^b$ \citet{2004MNRAS.351..935Z}
$^c$ \citet{2005MNRAS.358..457Z}
$^d$ \citet{1986ApJ...308..322K} 
$^e$ \citet{1996PASP..108..980K} 
$^f$ \citet{2013MNRAS.428.3443M}
\end{table*}

LOFAR is a radio interferometer which consists of 52 stations distributed in Europe. The Netherlands host 24 core and 14 remote stations operating at the shortest baselines. The remaining 14 stations are located in other countries and provide the longest baselines. Each single station consists of a set of low-band (LBA) and high-band (HBA) antennas observing in 30-80 and 110-240 frequency ranges, respectively \citep{2013A&A...556A...2V}.

We used the radio continuum 120-168\,MHz images (central frequency of 144\,MHz) of PNe collected by the LOFAR Two-Metre Survey (LoTSS) \citep{2019A&A...622A...1S}. The survey uses only the data from core and remote stations. The collected visibilities are processed with direction-dependent calibration \citep{2016ApJS..223....2V}. The clean algorithm is replaced with a spectral-dependent deconvolution algorithm, which improves dynamic range of the obtained images \citep{2018A&A...611A..87T}. \citet{2019A&A...622A...5D} presents the calibration strategy and examples. 

The survey provides low- and high-resolution images with the full width at half maximum of the restoring beam being 20 and 6\,arcsec, respectively. We assumed the absolute flux density scale accuracy of 10\% \citep{2019A&A...622A...1S}. The median positional accuracy of the high-resolution images is 0.2\,arcsec, though it may range from 0.1\,arcsec to 4.8\,arcsec for individual fields. LoTSS fields reach a flux accuracy of $100-500\,\mu \mathrm{Jy}\,\mathrm{beam}^{-1}$. 

Good sampling of the $uv$ plane by short baselines provides LOFAR with an excellent sensitivity to extended emission. Some examples are presented in \citet{2019A&A...622A...1S}. An upgraded pipeline improved the reduction of extended emission and removed artifacts which were present in the preliminary LoTSS release \citep{2017A&A...598A.104S}. With the shortest baseline of about 80\,m the largest angular scale of LoTSS reaches 40 arc mins \citep{2018MNRAS.474.5023S}.

The LoTSS observations and data processing are still ongoing. We included observations which were processed before April 2021\footnote{The present coverage of LoTSS is shown in \url{ https://lofar-surveys.org/lotss-tier1.html}}. This largely overlapped with the upcoming LoTSS-DR2 (T. Shimwell et al., 2021, in prep.). LoTSS-DR2 includes overlapping fields which are mosaiced to produce the final survey images. We also included additional pointings which have not been yet mosaiced. Their quality will improve in the future after LoTSS completes observations and produces final mosaics.

We selected 165 PNe in the observed part of sky using the SIMBAD database \citep{2000A&AS..143....9W} and the catalogue by \citet{2016JPhCS.728c2008P}. Out of them, 30 were detected. Table\,\ref{tab:observations} presents the nebular sizes and flux densities of these PNe at 144\,MHz. The fluxes and diameters of compact PNe $\Theta_\mathrm{d}$ were measured with Gaussian deconvolution using {\sc CASA} \citep{2007ASPC..376..127M}. We multiplied the deconvolved diameters by correction factors to account for more realistic surface brightness distribution than a simple Gaussian \citep{2000MNRAS.314...99V}. We applied the correction factors computed for disk geometry for optically thick PNe. This choice is justified by a flat surface brightness profile of a spherically symmetric nebular model at optically thick 20\,cm (equivalent to frequency of about 1400\,MHz) computed by \citet{2000MNRAS.314...99V}. Flat surface brightness profile represents a circular, constant surface brightness disk. 

The correction factors were not applied for the large PNe and for unresolved PNe. For well-resolved PNe we fitted an ellipse to the emission which exceeded background by $3 \sigma$, which is marked with a thick line in Figures\,\ref{fig:spectra1} - \ref{fig:spectra4}. To measure the flux density, we integrated the emission within this area.

\section{Results}

\subsection{Spectral fitting}

\begin{figure*}
\caption{Images and spectra of PNe - continued.} \includegraphics[height=5cm]{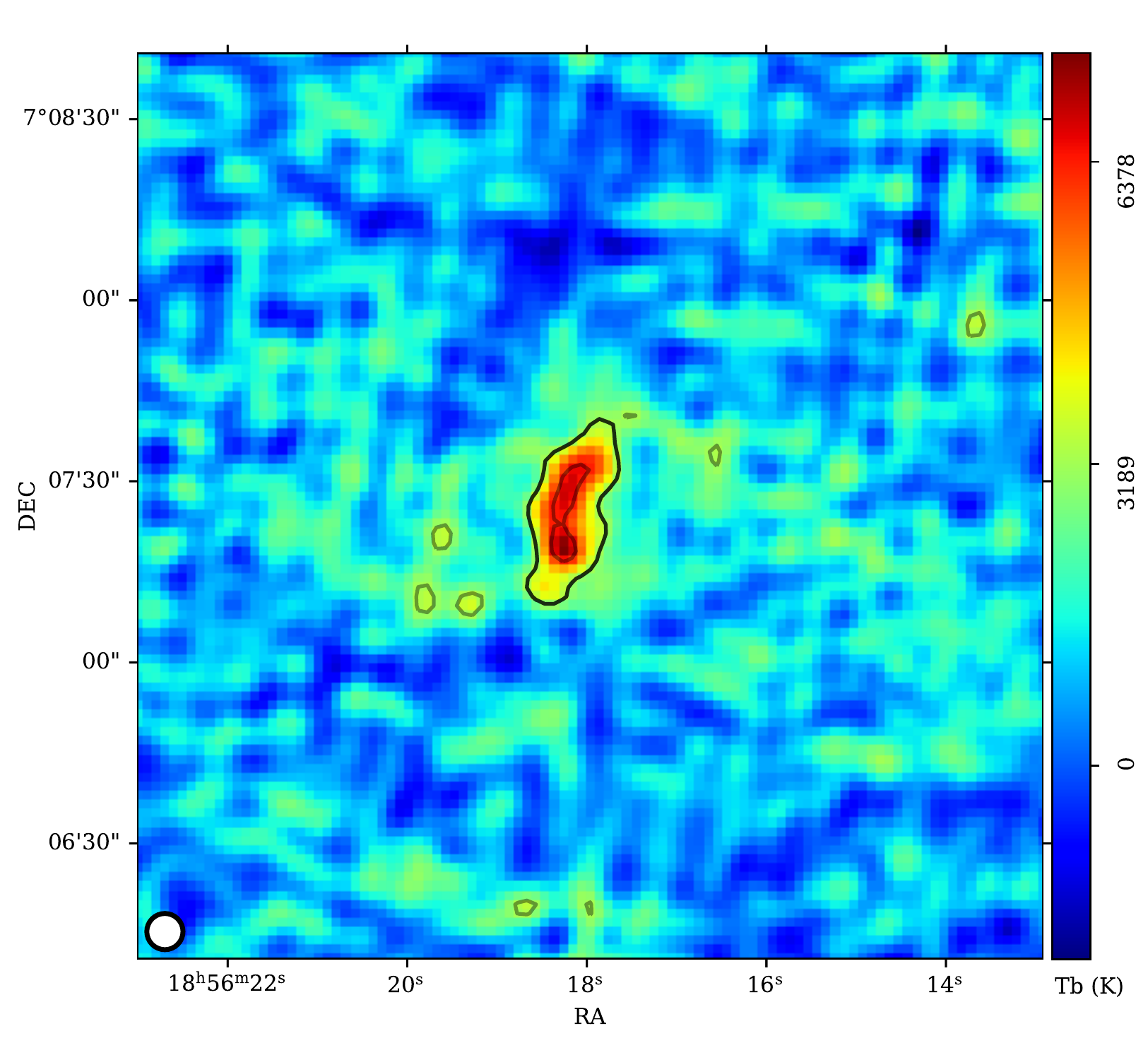}\includegraphics[height=5cm]{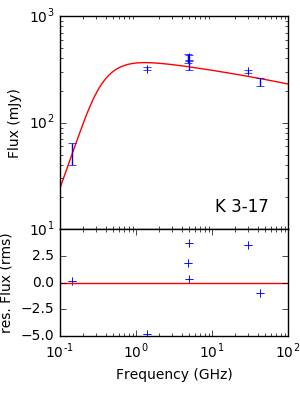}\includegraphics[height=5cm]{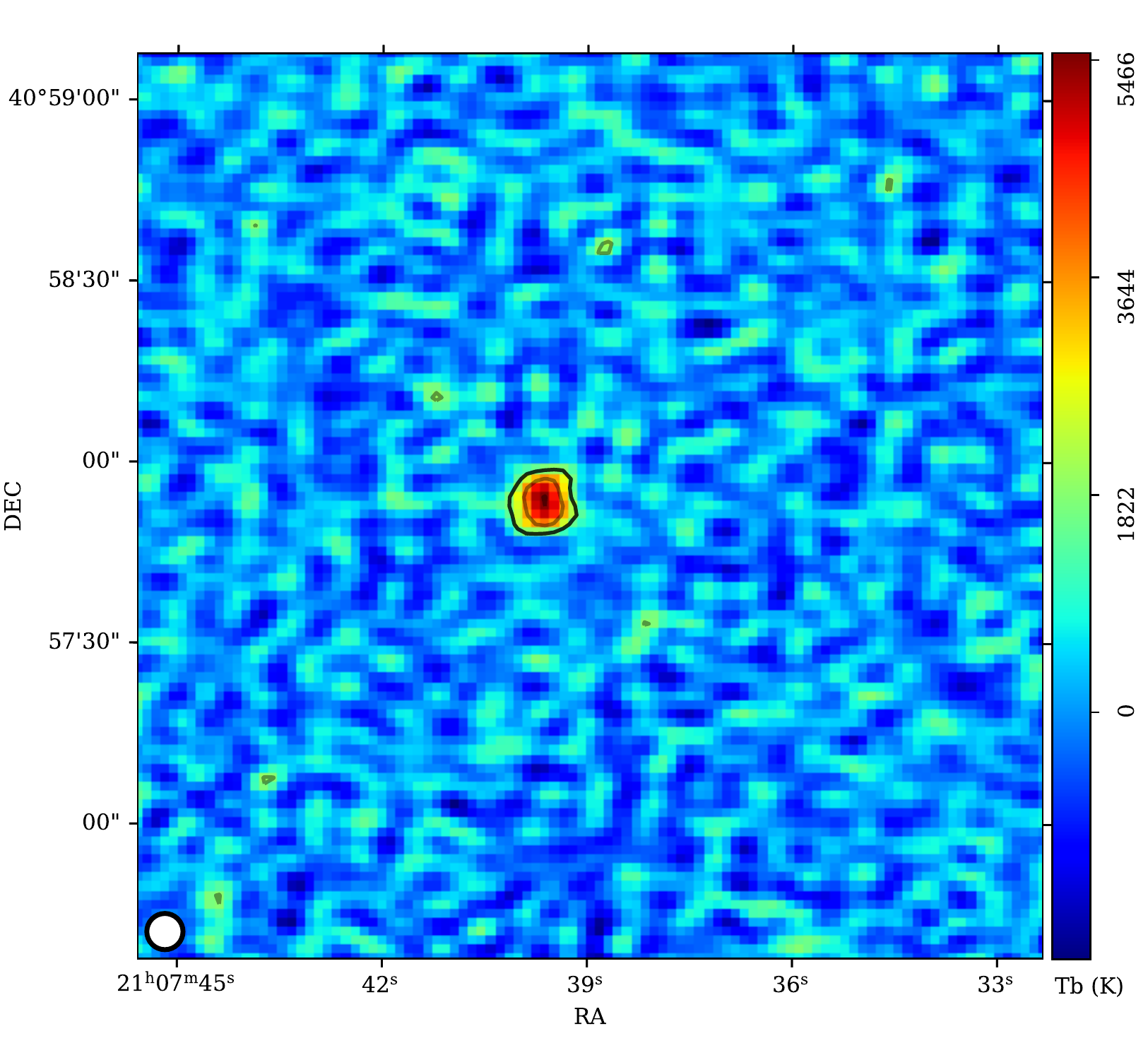}\includegraphics[height=5cm]{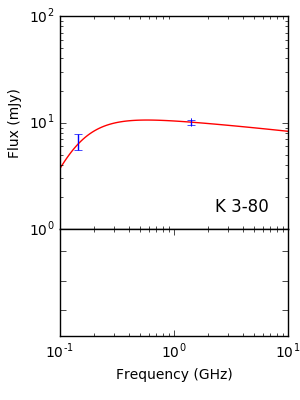}

\includegraphics[height=5cm]{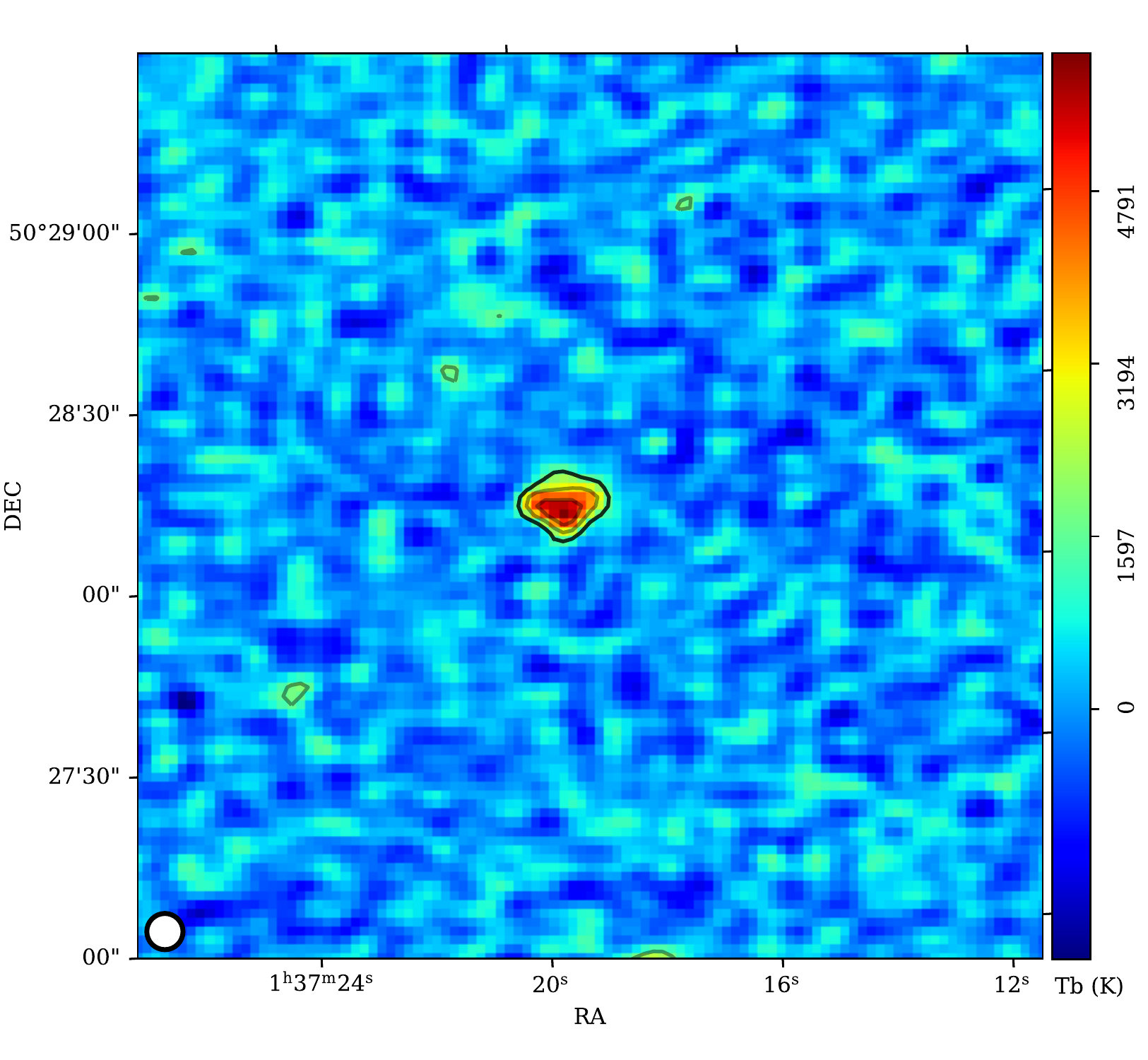}\includegraphics[height=5cm]{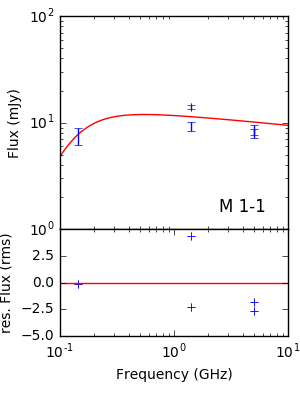}\includegraphics[height=5cm]{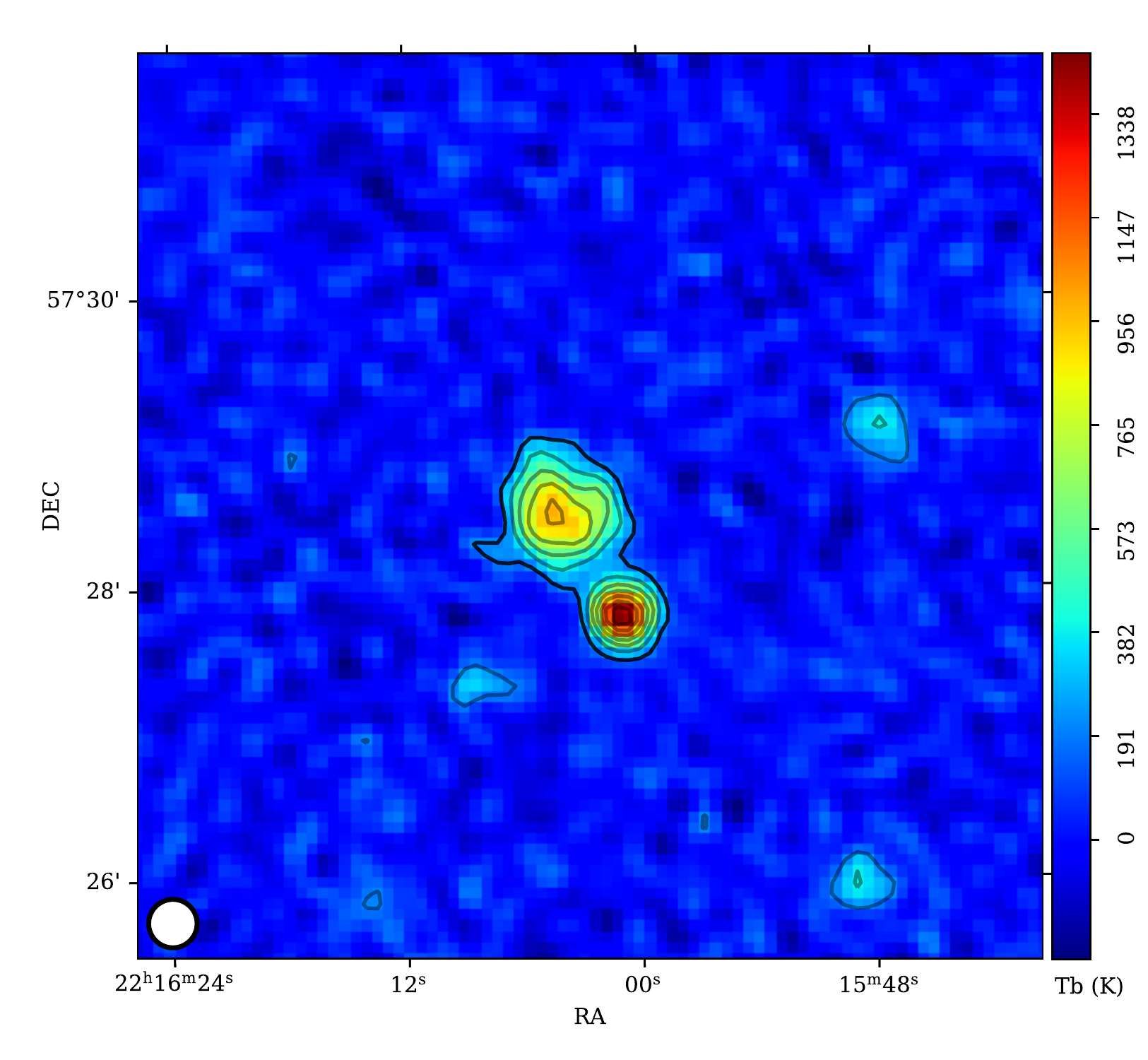}\includegraphics[height=5cm]{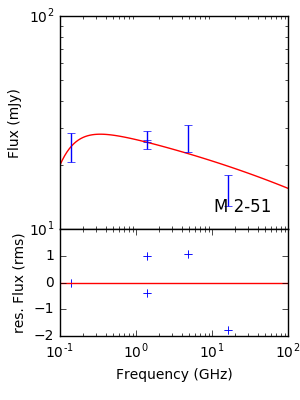}

\includegraphics[height=5cm]{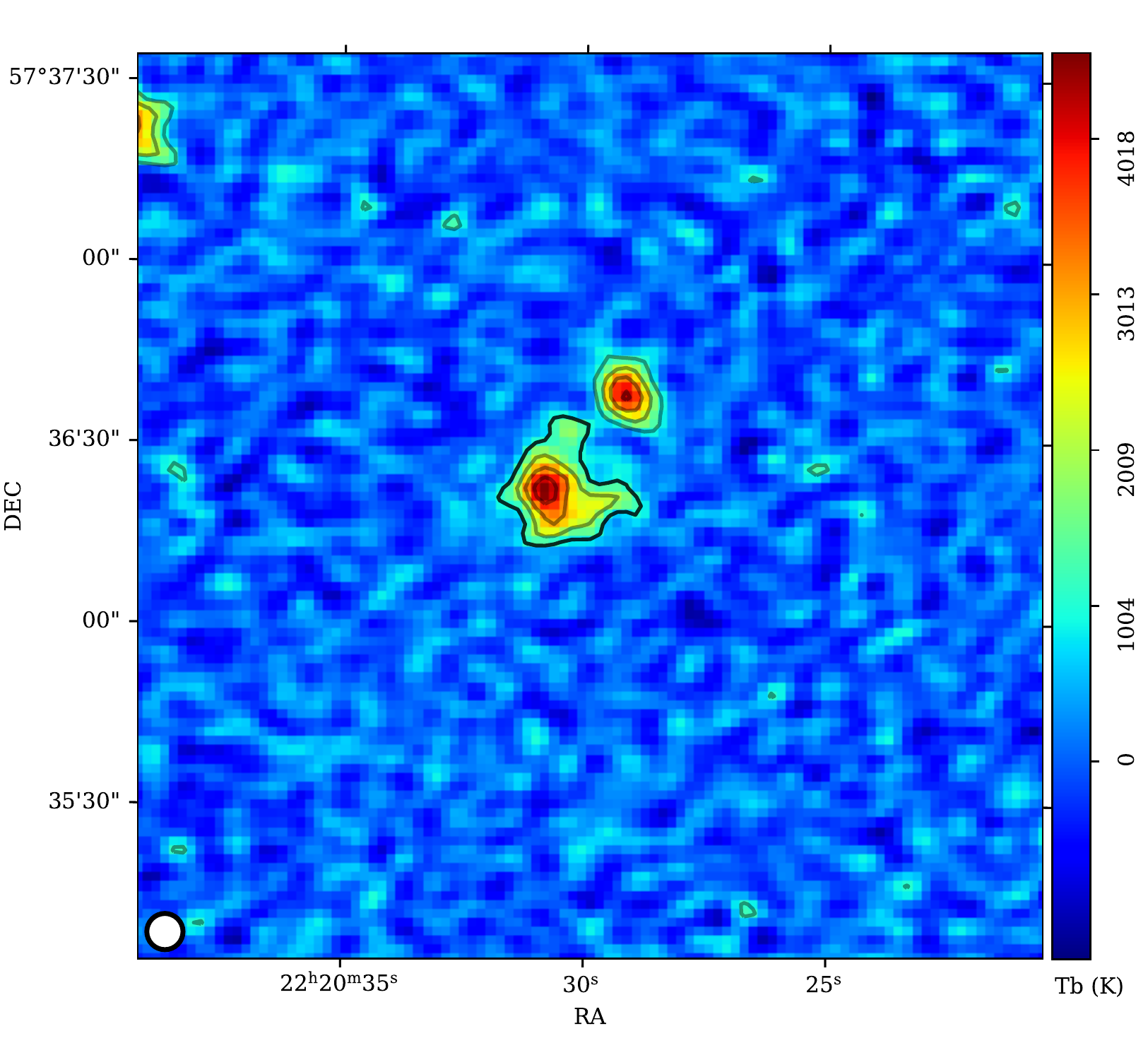}\includegraphics[height=5cm]{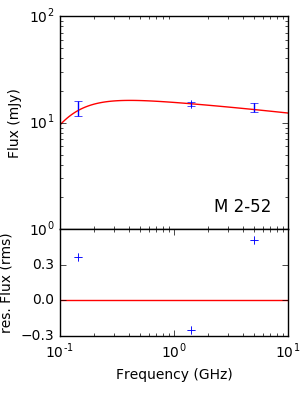}\includegraphics[height=5cm]{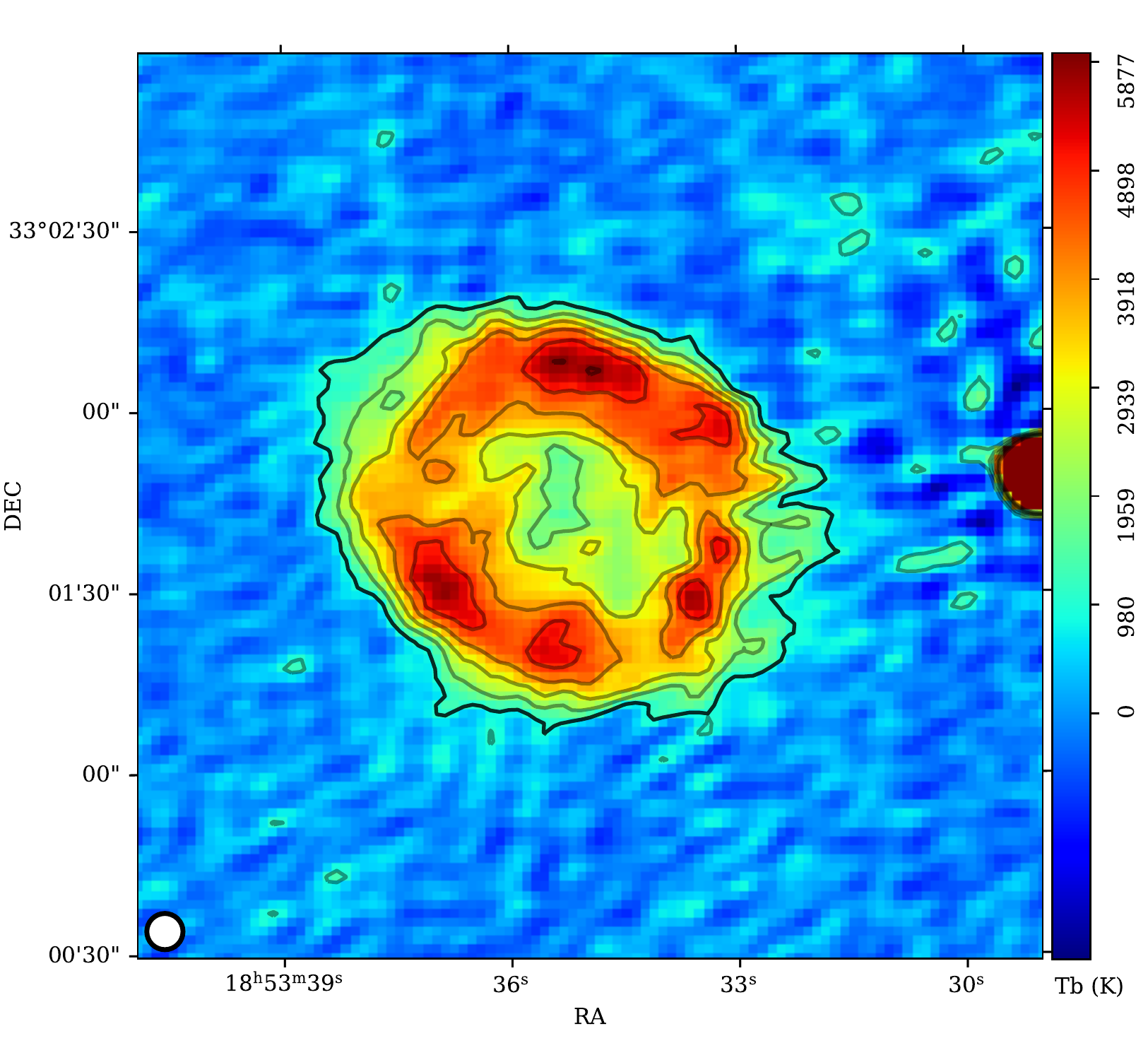}\includegraphics[height=5cm]{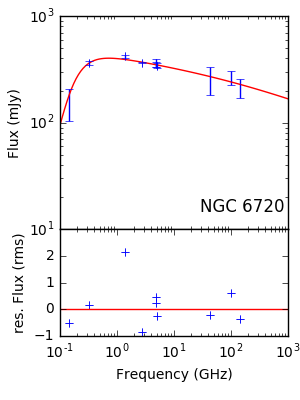}

\includegraphics[height=5cm]{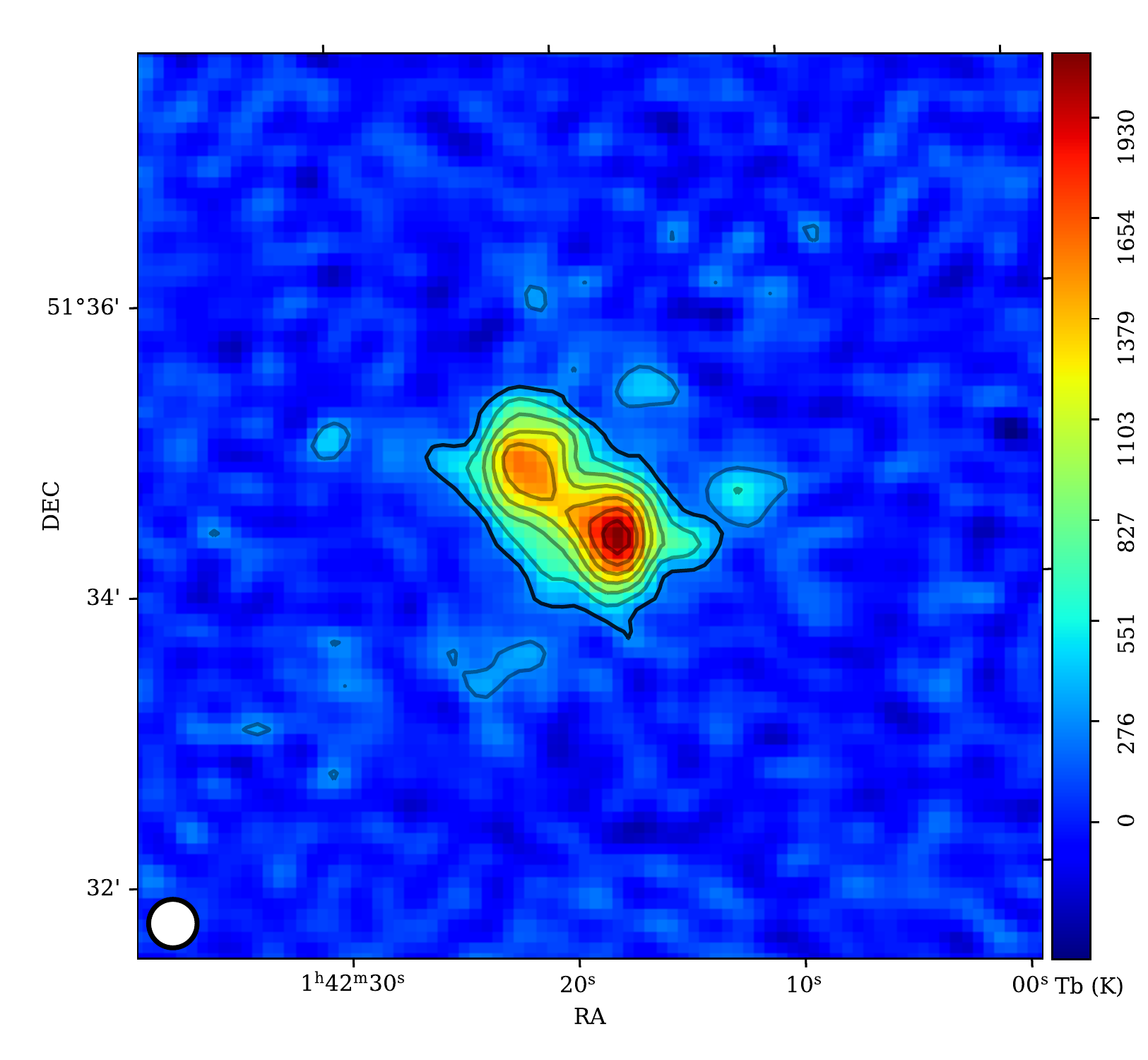}\includegraphics[height=5cm]{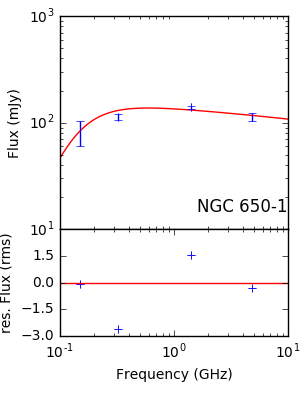}\includegraphics[height=5cm]{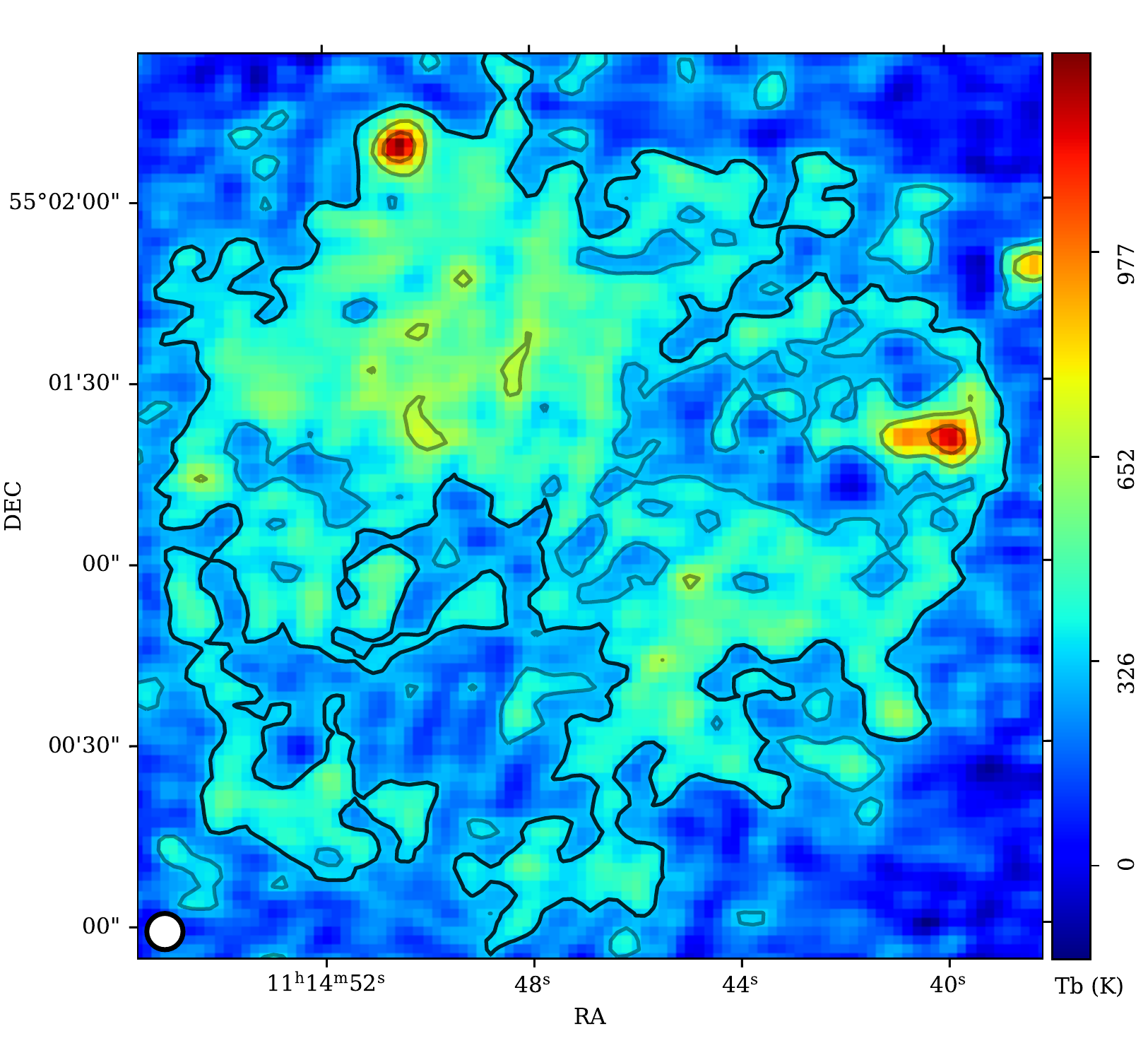}\includegraphics[height=5cm]{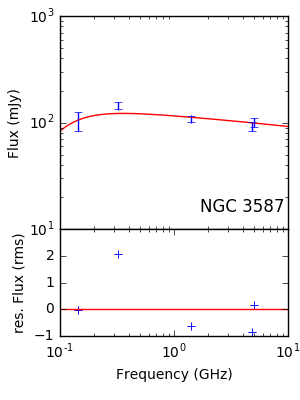}

\label{fig:spectra2}
\end{figure*}

\begin{figure*}
\caption{Images and spectra of PNe - continued.} \includegraphics[height=5cm]{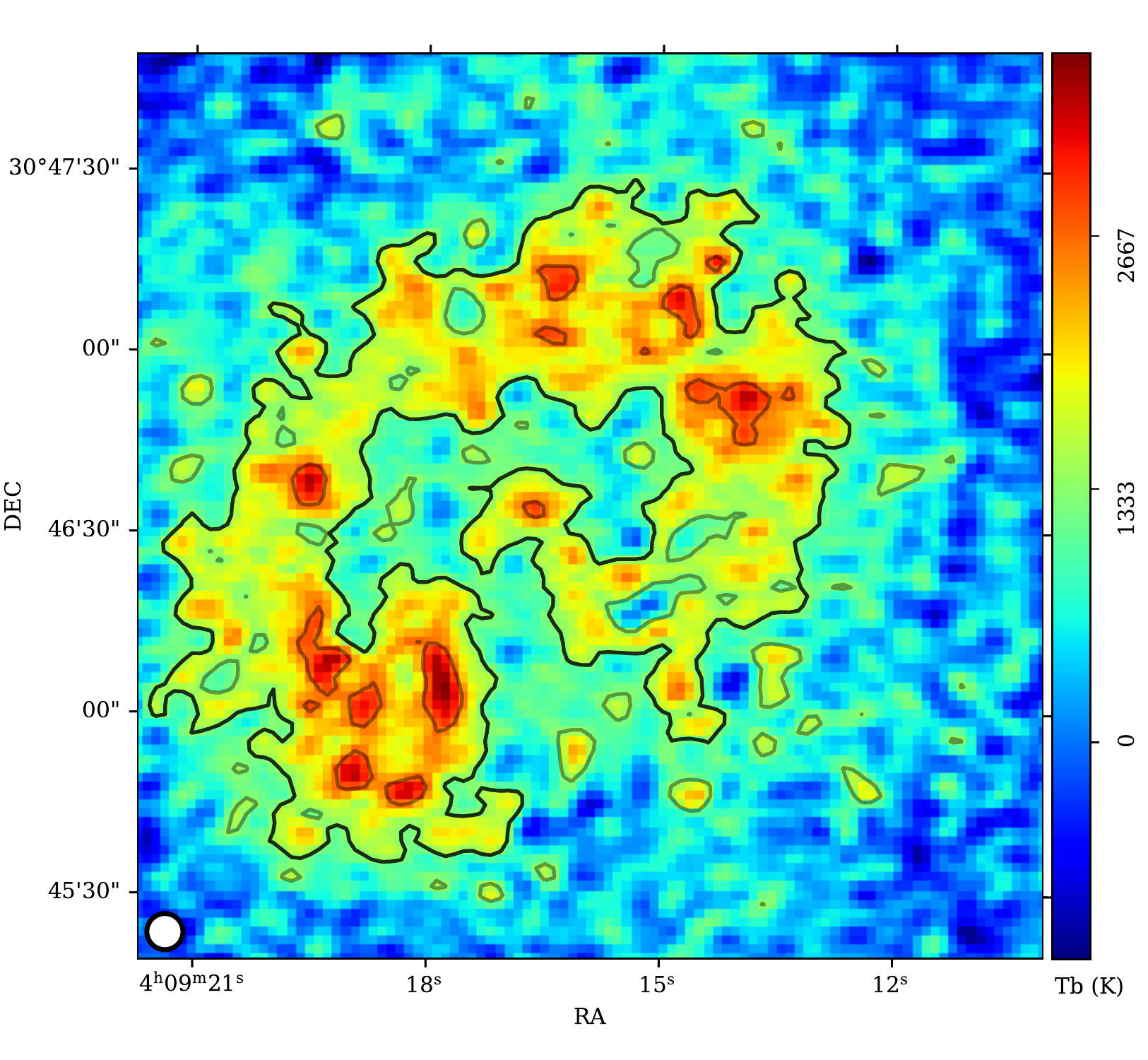}\includegraphics[height=5cm]{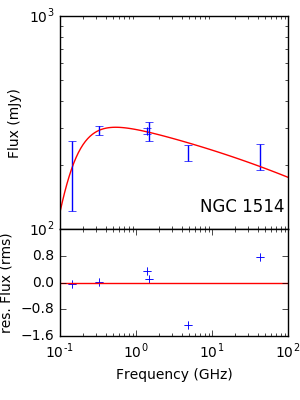}\includegraphics[height=5cm]{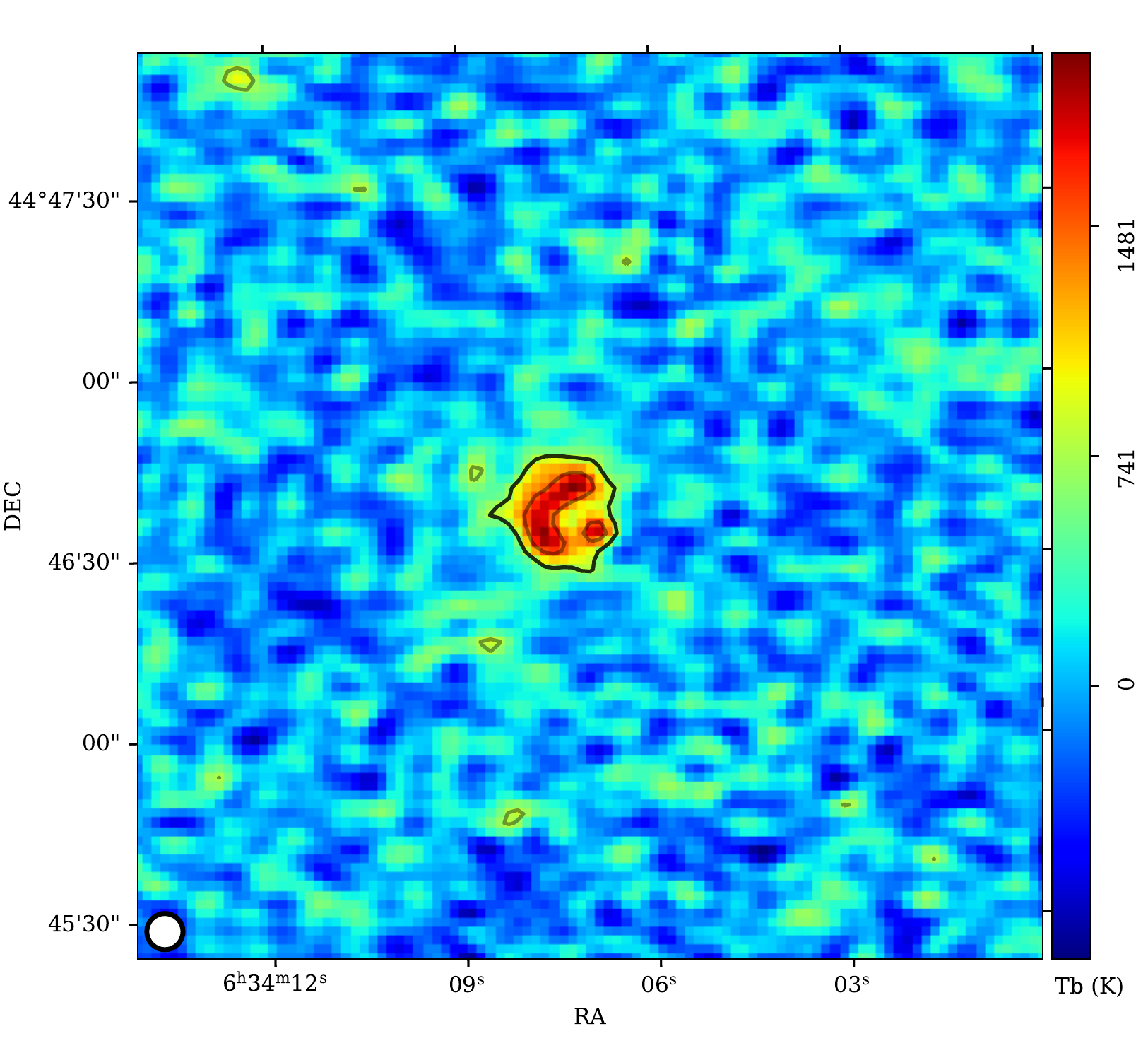}\includegraphics[height=5cm]{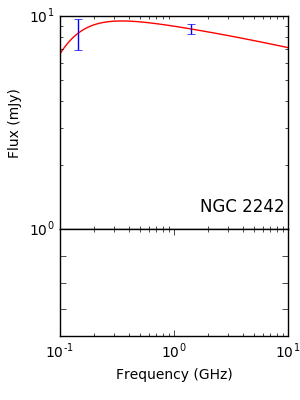}

\includegraphics[height=5cm]{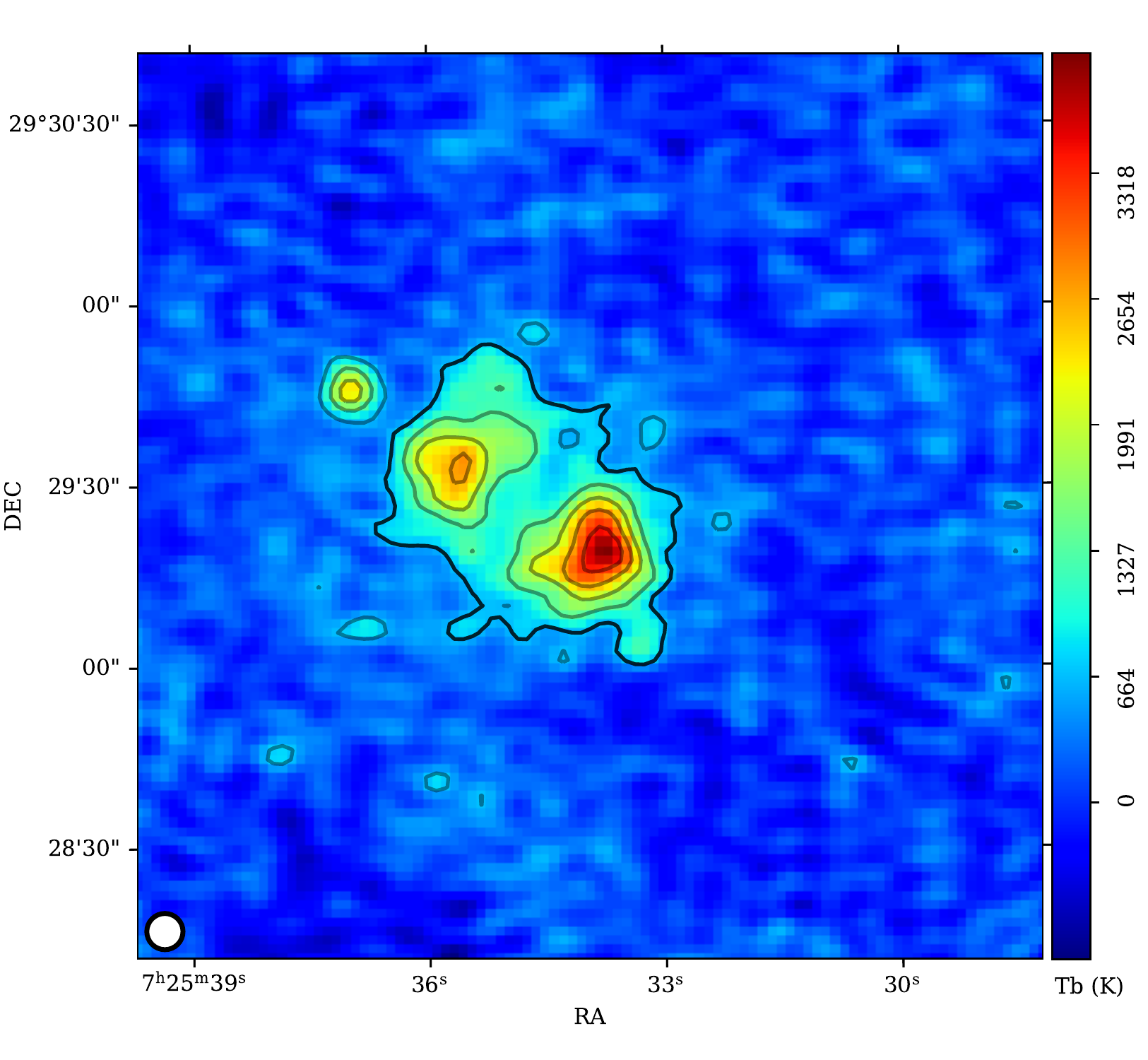}\includegraphics[height=5cm]{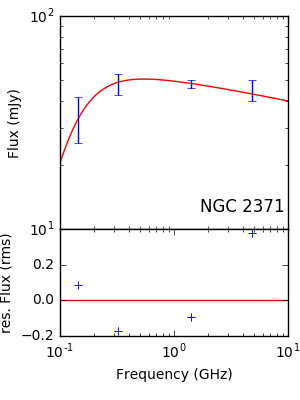}\includegraphics[height=5cm]{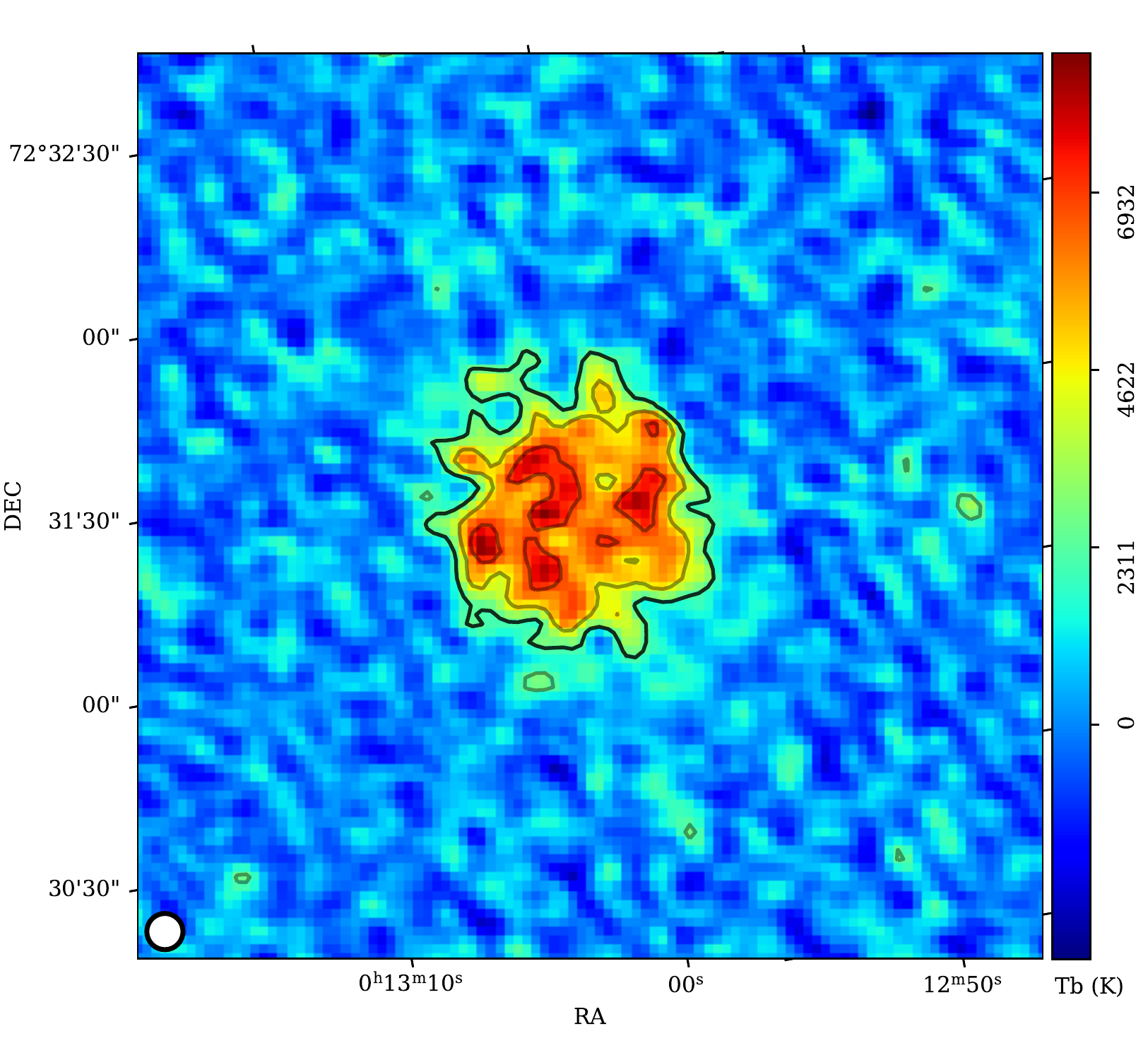}\includegraphics[height=5cm]{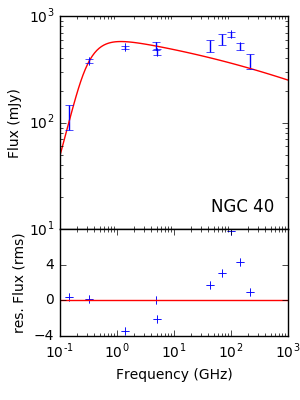}

\includegraphics[height=5cm]{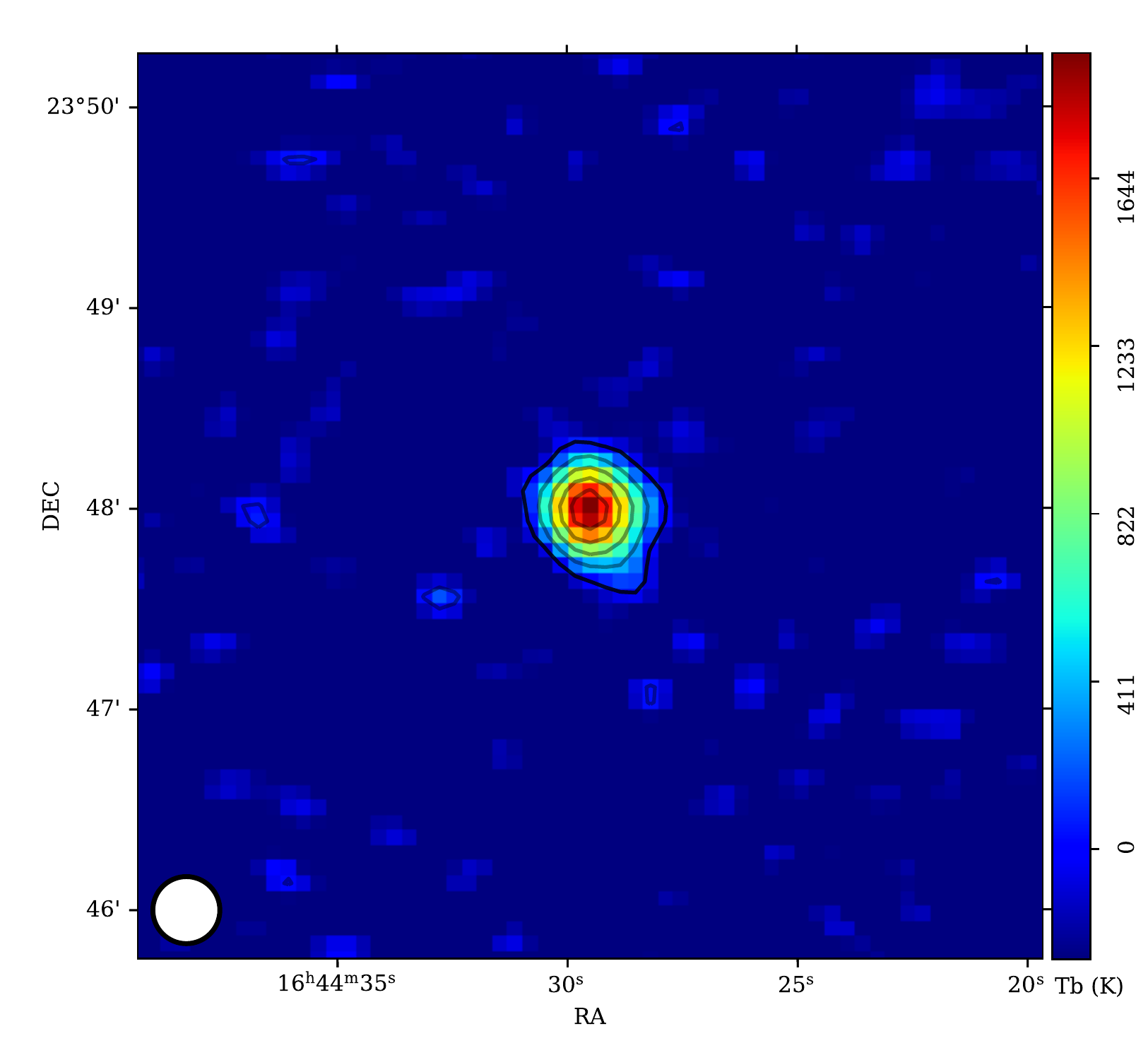}\includegraphics[height=5cm]{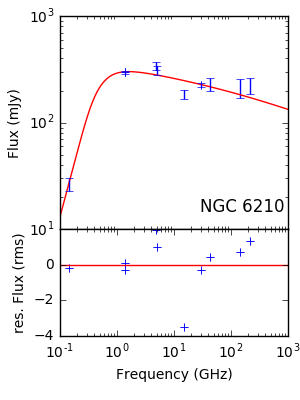}\includegraphics[height=5cm]{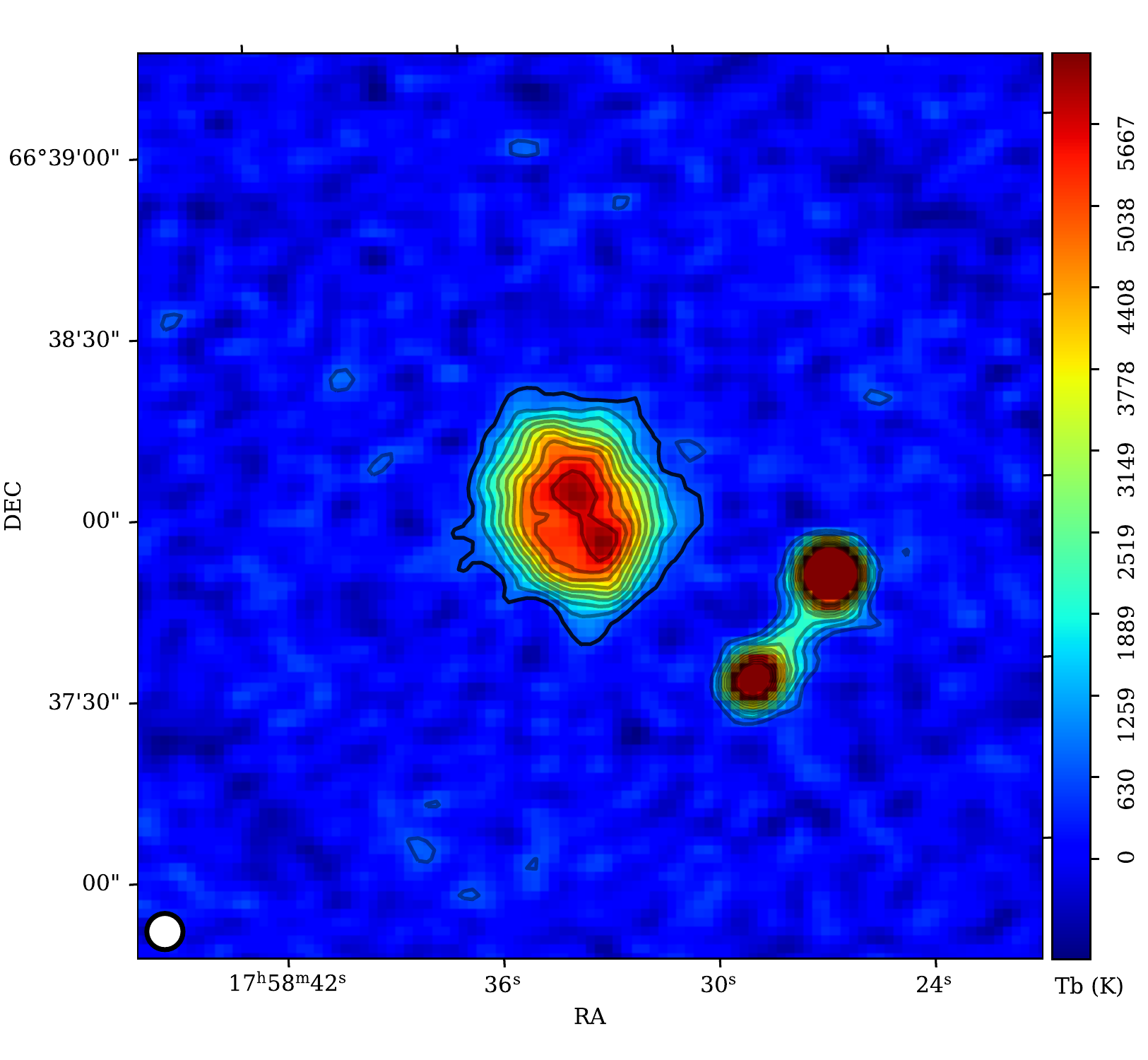}\includegraphics[height=5cm]{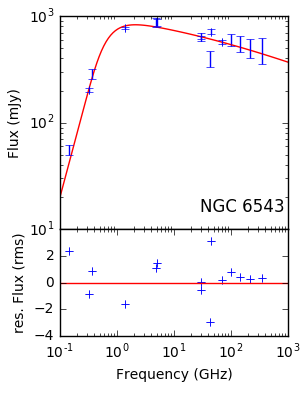}

\includegraphics[height=5cm]{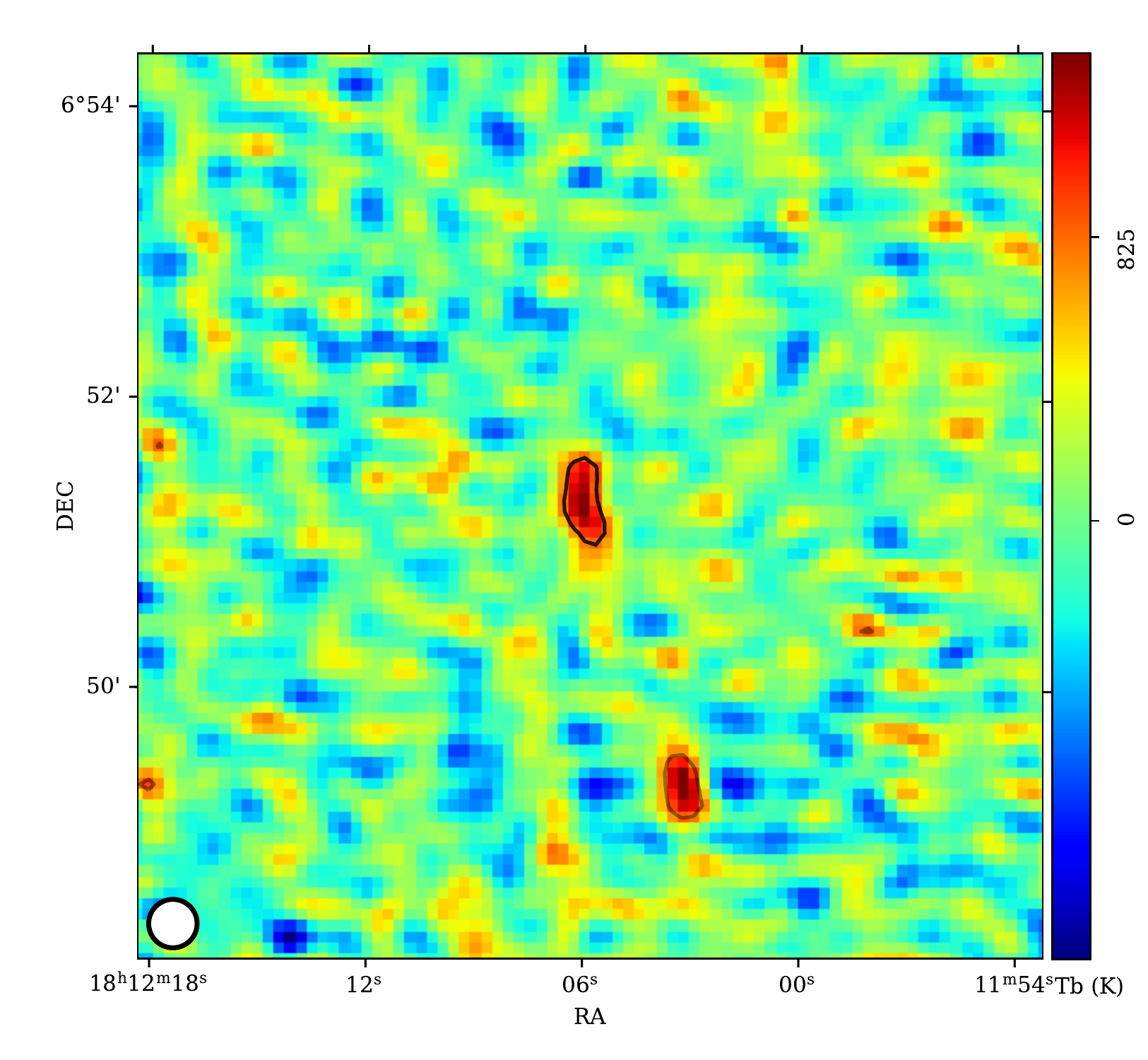}\includegraphics[height=5cm]{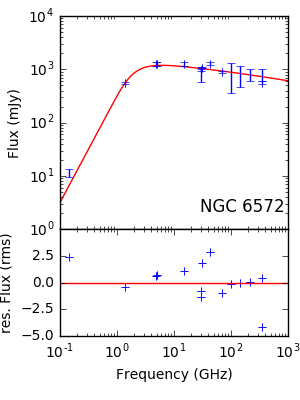}\includegraphics[height=5cm]{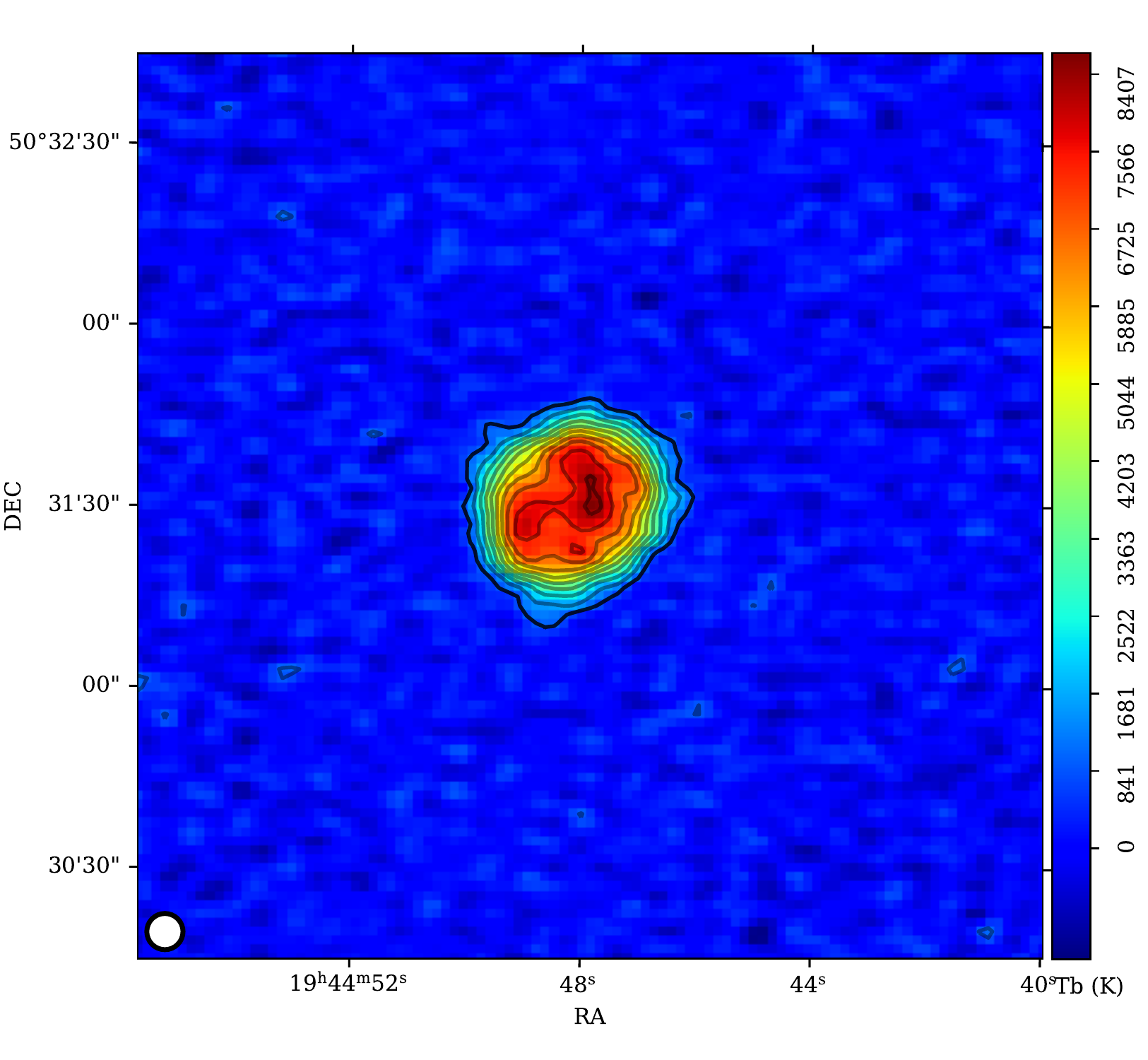}\includegraphics[height=5cm]{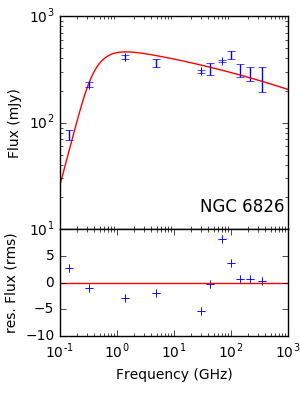}

\label{fig:spectra3}
\end{figure*}

We combined our new 144\,MHz flux densities with flux densities collected at different frequencies with other surveys, which are listed in \citet{2018MNRAS.479.5657H}. We fitted the spectra with Equation\,\ref{eq:snu2} for $\eta=0$ using the derived sizes (Figures\,\ref{fig:spectra1} - \ref{fig:spectra4}). Only two of the three unknown parameters: $\xi$, $EM$, and $T_\mathrm{e}$ could be fitted independently. $EM$ is parametrized in the optical depth term. We used electron temperatures derived from CELs by \citet{1986ApJ...308..322K}, leaving $\xi$ and $EM$ as free parameters. Using electron temperature derived from CELs or fixing it to an arbitrary value (e.g. $10^4$K) is a common practice in fitting radio spectral energy distribution (SED) of PNe \citep{2009A&A...498..463P,2018MNRAS.479.5657H,2021MNRAS.503.2887B}. This reduces the number of unknown variables in the fit to two. However - as we will show later - cool plasma component may also contribute to radio emission and bias the results. Optical depth does not strongly depend on the assumed temperature and can be robustly derived from the fit. Lower temperature would lead to higher $\xi$ and lower $EM$.

We fitted the spectra to check if PNe are optically thick at 144\,MHz. In such a case electron temperature could be determined from the brightness temperature. Our fits also confirm that PN radio SEDs are consistent with free-free emission. The LoTSS source density is 770 per square degree \citep{2019A&A...622A...1S}. The probability of finding one confusing source closer than 6 arcsec in the sample of 165 objects is about 30\%. Some PNe show background sources nearby, but far enough to be separated.

Table\,\ref{tab:upper} lists upper flux limits measured from the maps for 135 undetected PNe. The median root-mean square (rms) is about 700\,$\mu$Jy/beam (Table\,\ref{tab:upper}). A 3\,rms upper limit of 2.1\,mJy/beam at 144\,MHz corresponds to the brightness temperature of about 4800\,K for a 6 arc sec beam. Out of 135 undetected PNe 32 objects have been detected at 1.4\,GHz with the NRAO VLA Sky Survey (NVSS) by \citet{1998ApJS..117..361C}. The sensitivity of NVSS is about 450\,$\mu$Jy/beam. We measured upper limits for spectral indices between 1.4\,GHz and 144\,MHz (Table\,\ref{tab:si}), assuming the PN sizes of $\leq 6$ arcsec. All but two were between optically thin ($SI = -0.1$) and optically thick ($SI = 2$) limit. This indicates that most of the PNe detected in NVSS show optically thick effects at 144\,MHz. However, PNe with higher optical thickness are brighter in radio and more likely to be detected.

\subsection{Electron temperatures}\label{te}

\begin{figure*}
\caption{Left: Images of PNe at 144\,MHz. We converted the flux intensity scale to brightness temperature, represented by the colorbar. The contours levels are spaced by $3\,\sigma$. White circle marks the size of the beam. Right: observed and fitted radio spectra of PNe. Lower panel shows the difference between the fit and the observed fluxes.} \includegraphics[height=5cm]{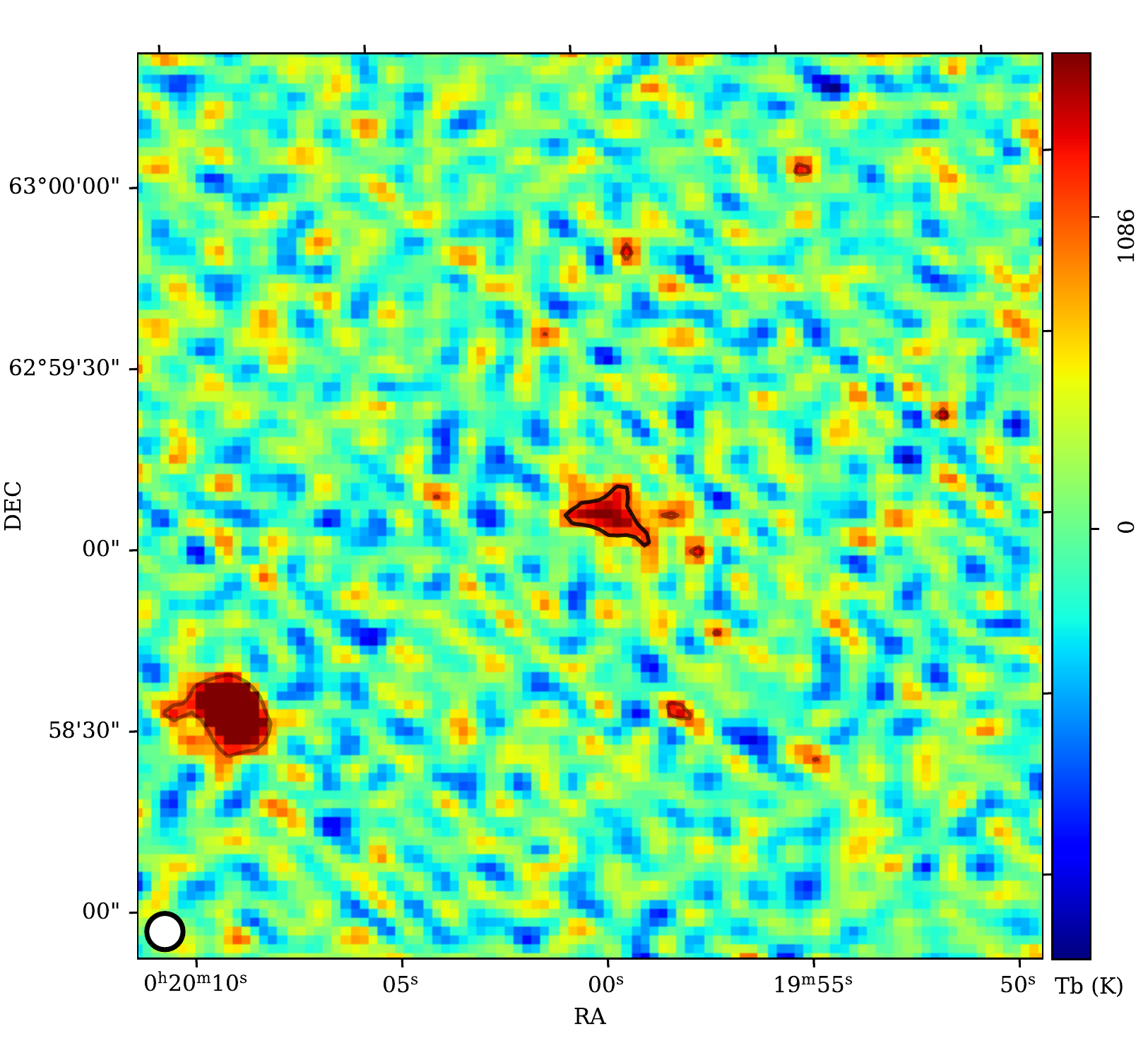}\includegraphics[height=5cm]{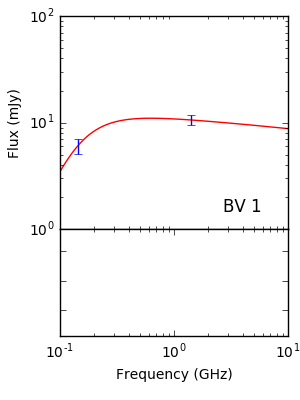}\includegraphics[height=5cm]{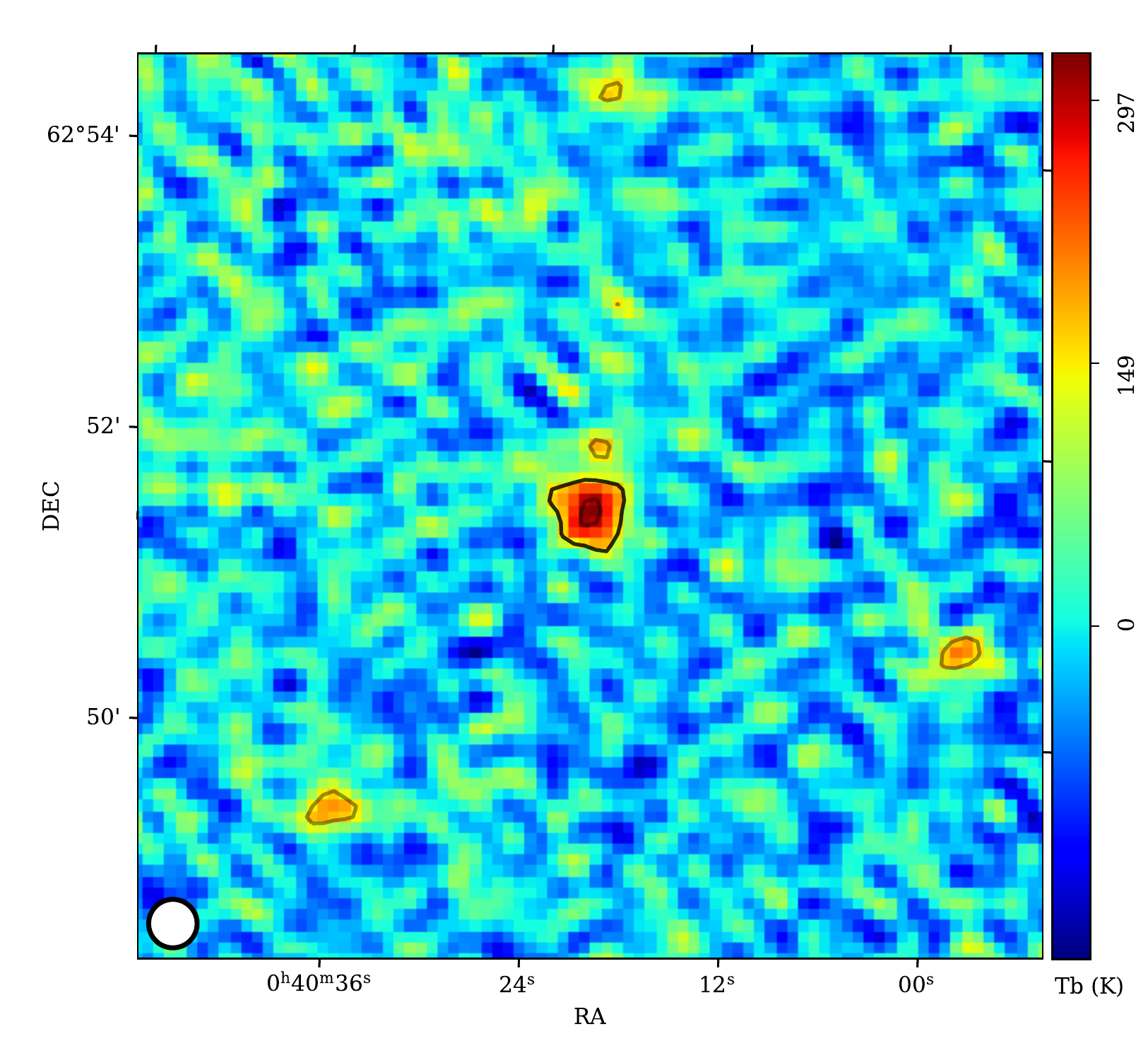}\includegraphics[height=5cm]{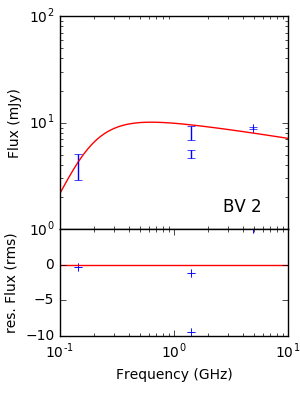}

\includegraphics[height=5cm]{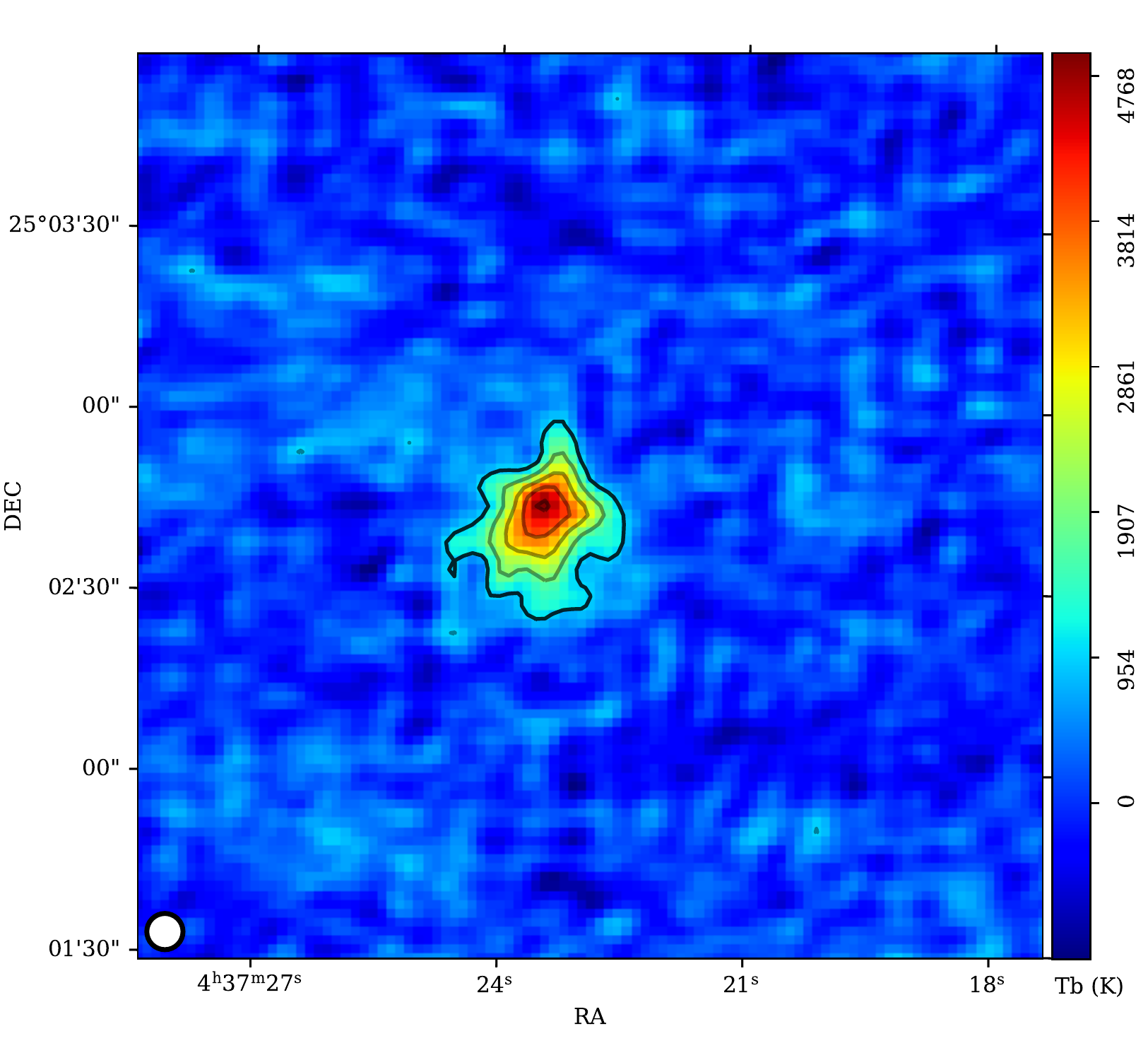}\includegraphics[height=5cm]{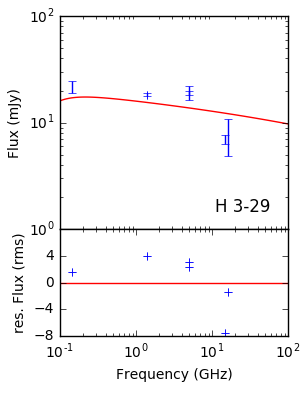}\includegraphics[height=5cm]{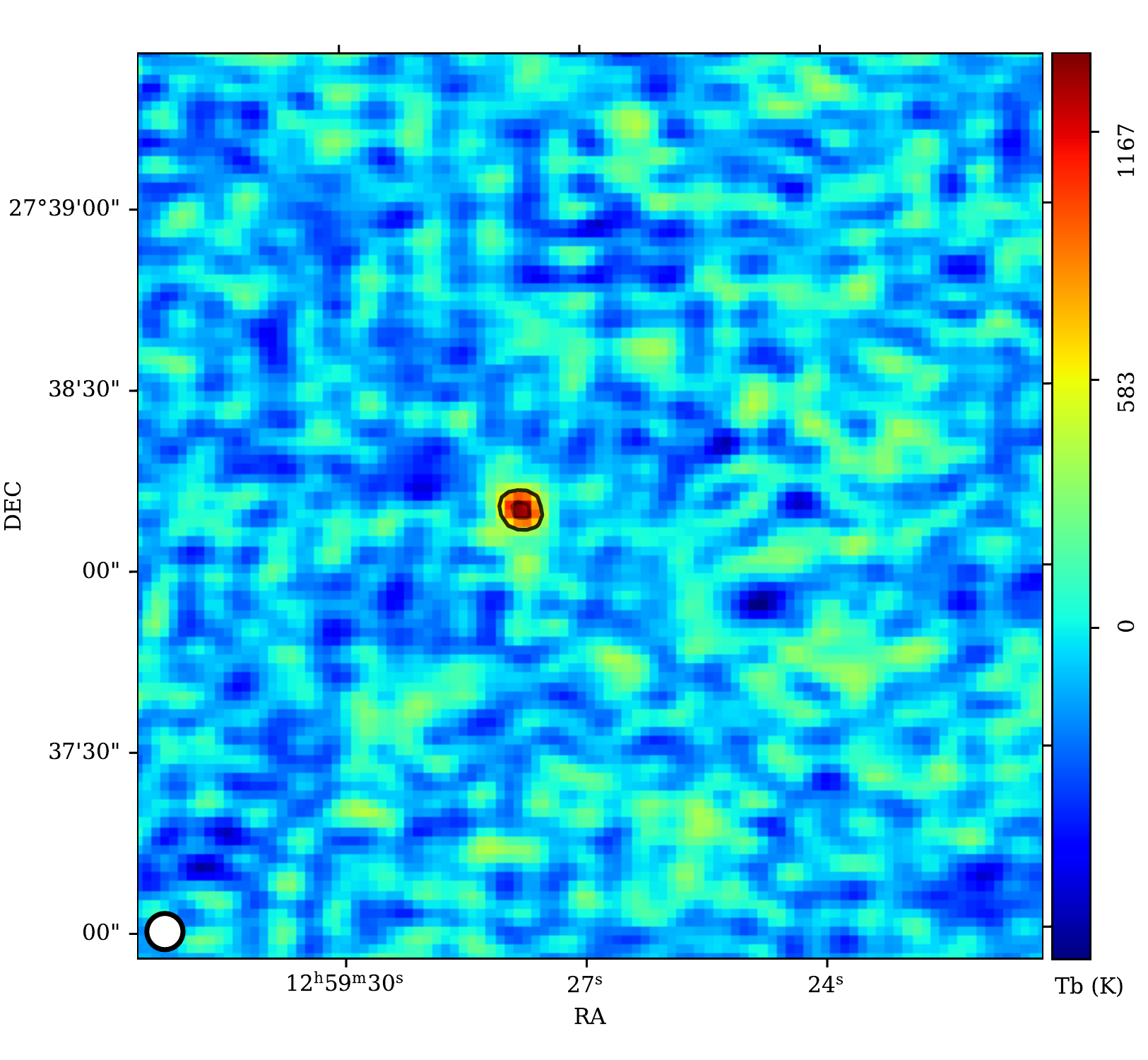}\includegraphics[height=5cm]{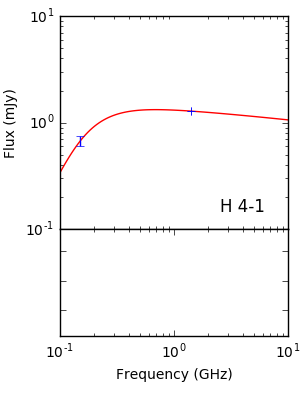}

\includegraphics[height=5cm]{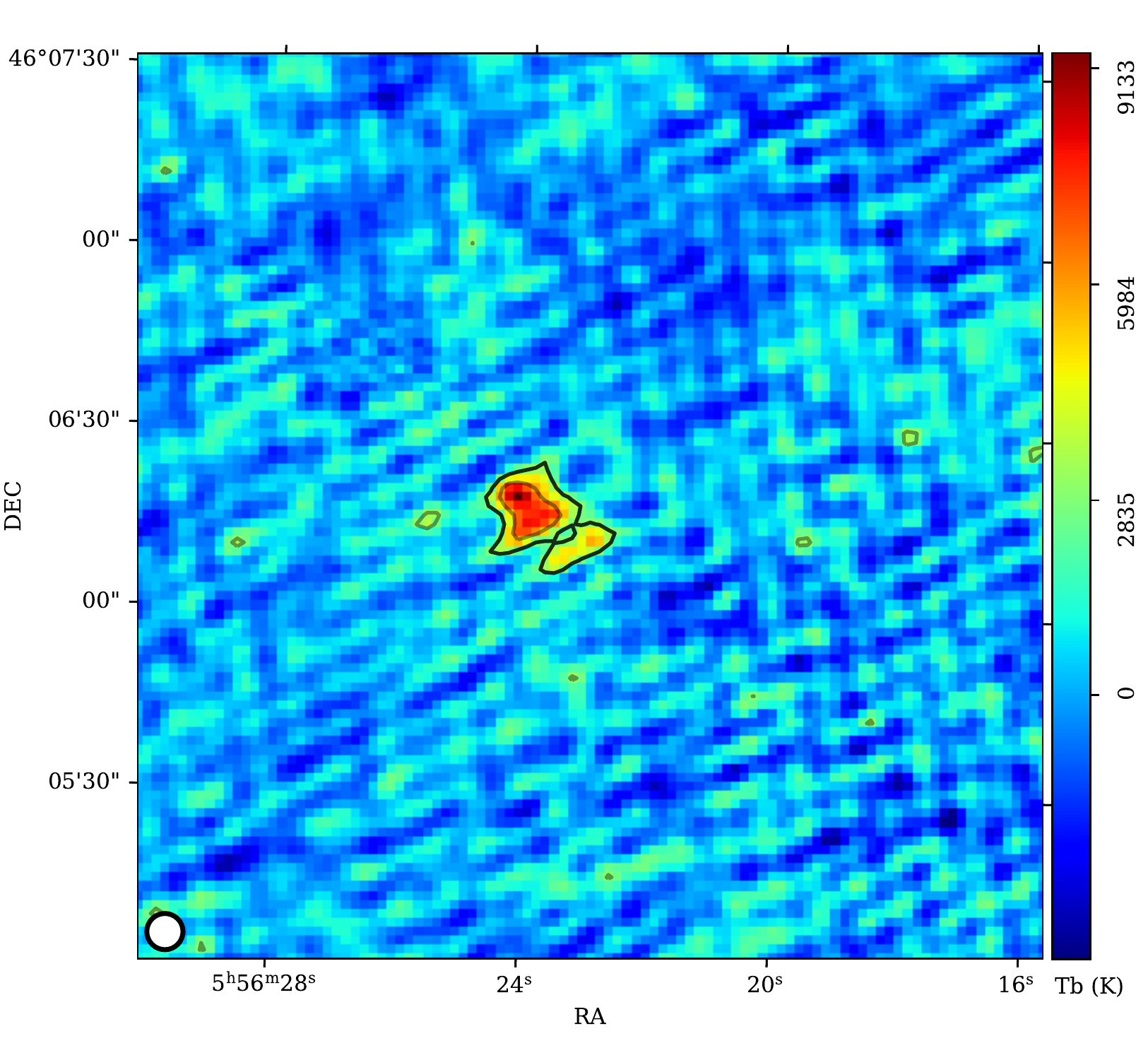}\includegraphics[height=5cm]{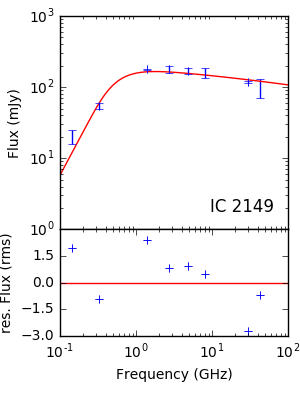}\includegraphics[height=5cm]{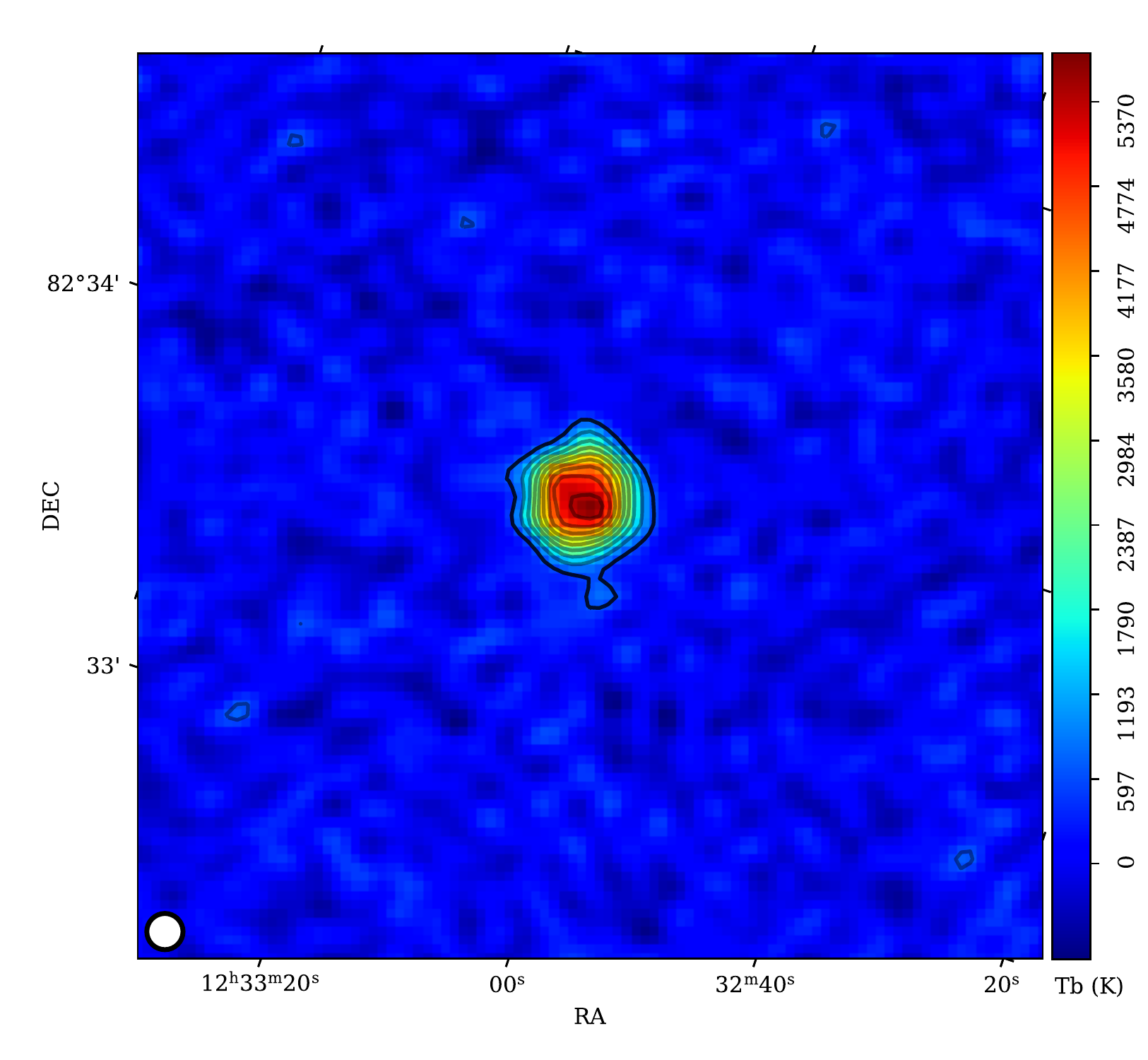}\includegraphics[height=5cm]{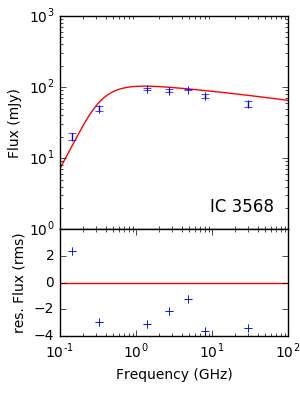}

\includegraphics[height=5cm]{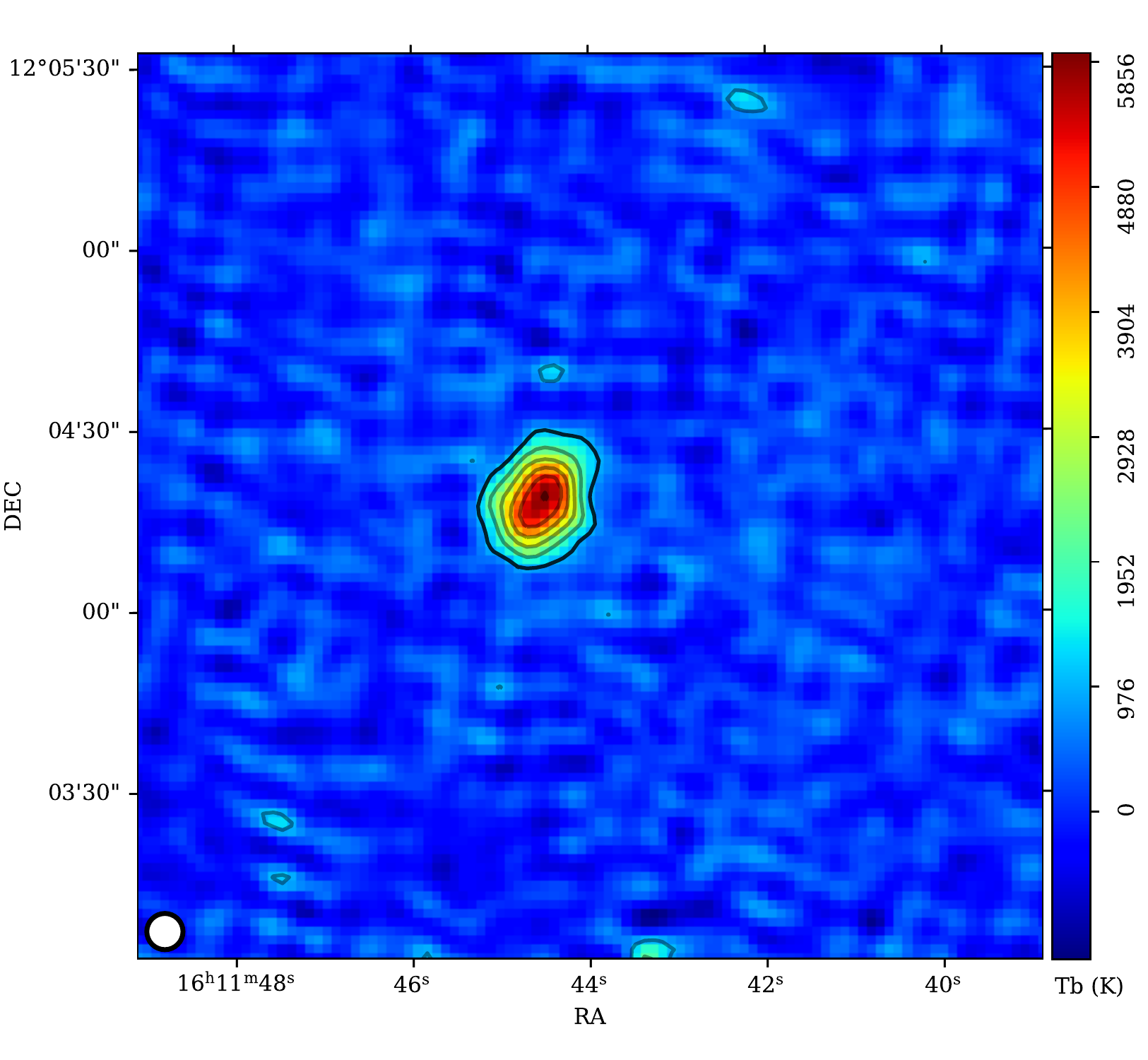}\includegraphics[height=5cm]{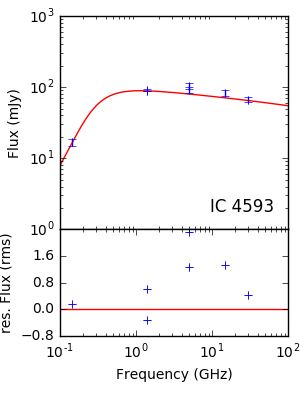}\includegraphics[height=5cm]{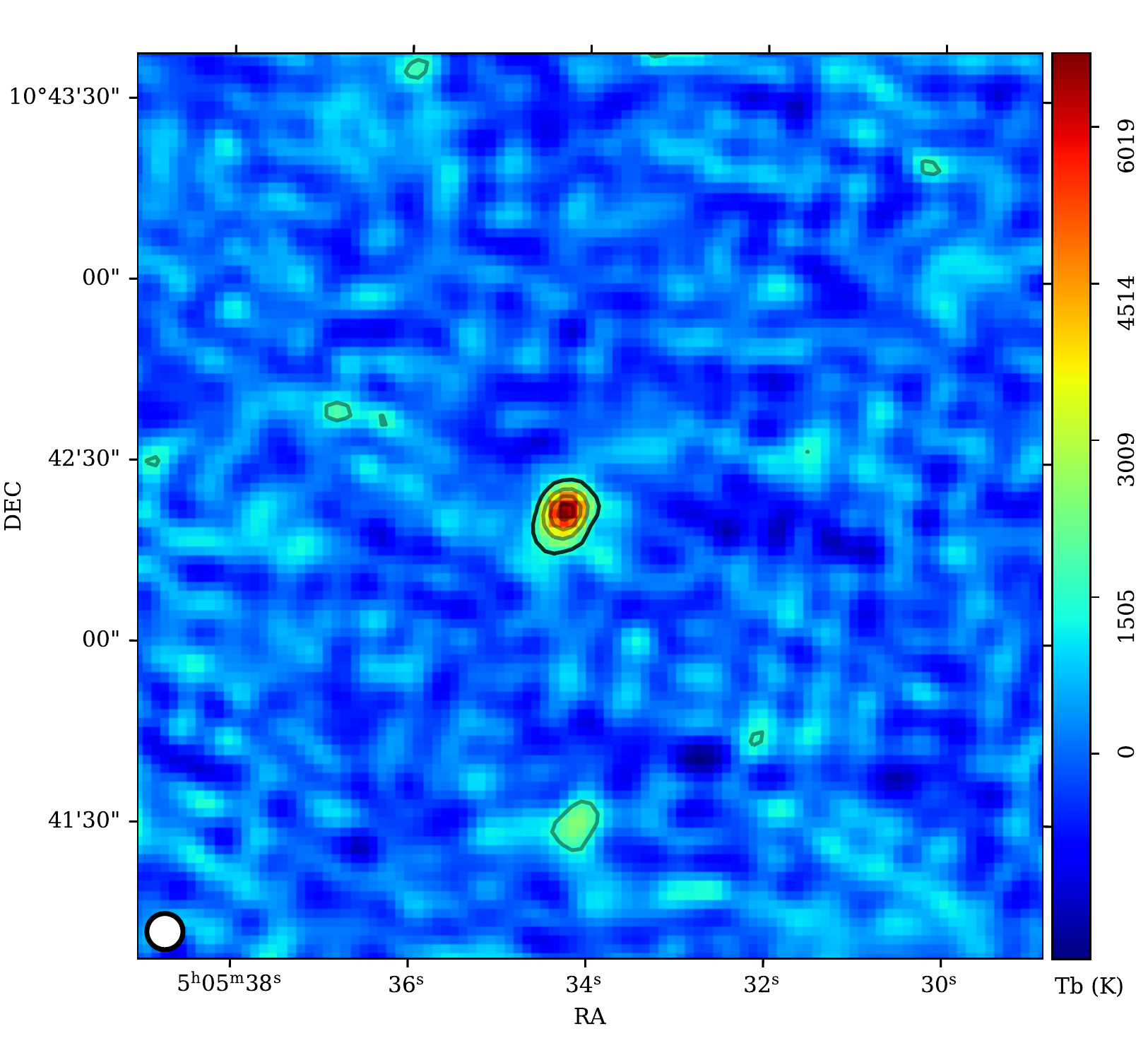}\includegraphics[height=5cm]{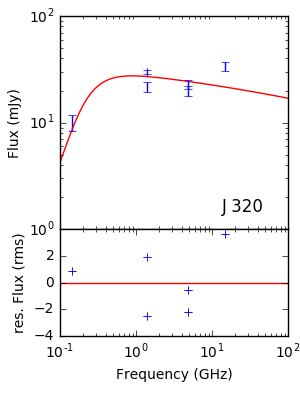}

\label{fig:spectra1}
\end{figure*}

\begin{figure*}
\caption{Images and spectra of PNe - continued.} 
\includegraphics[height=5cm]{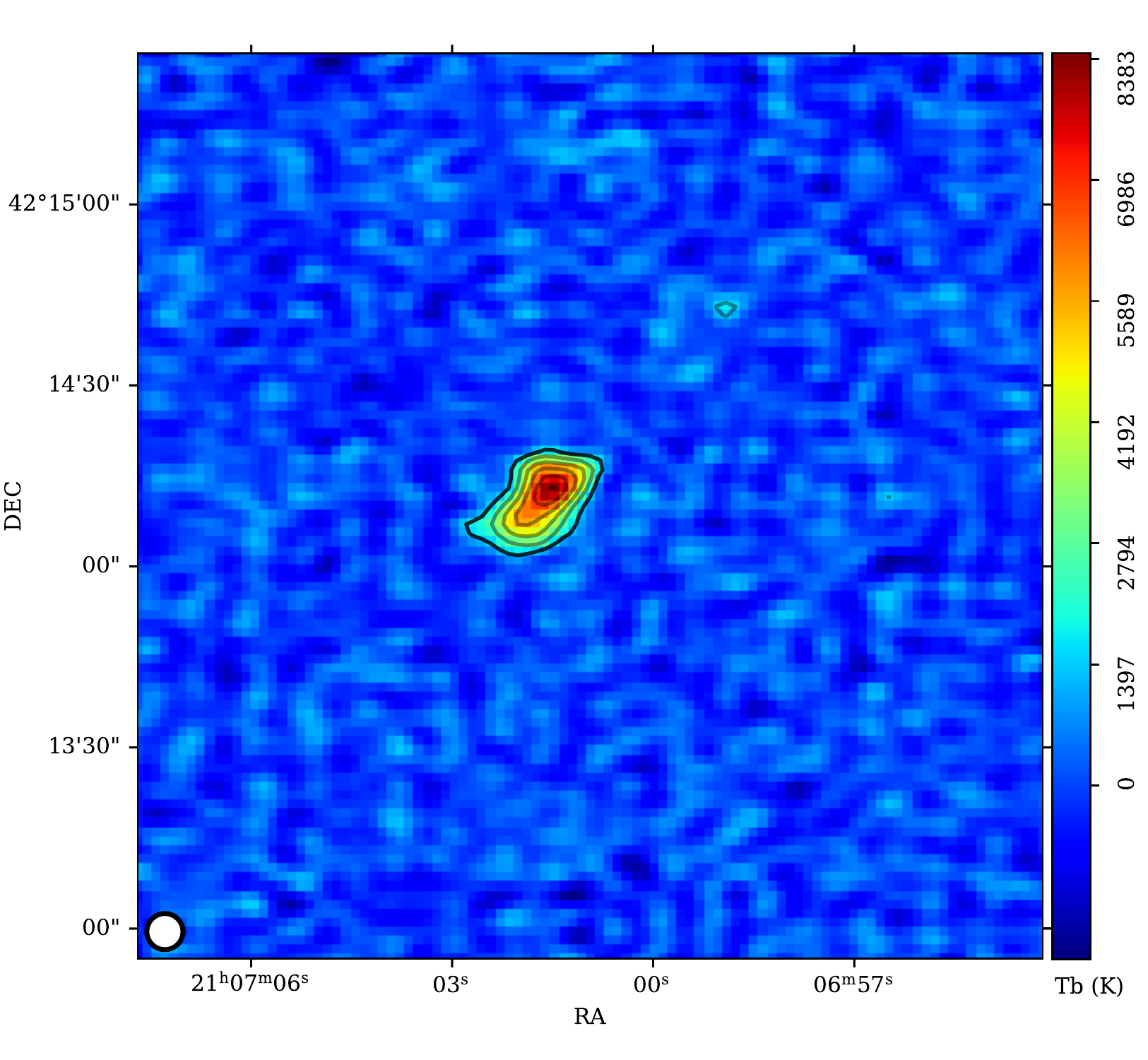}\includegraphics[height=5cm]{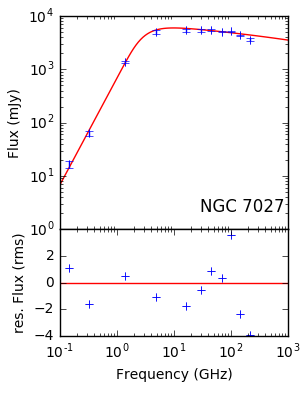}\includegraphics[height=5cm]{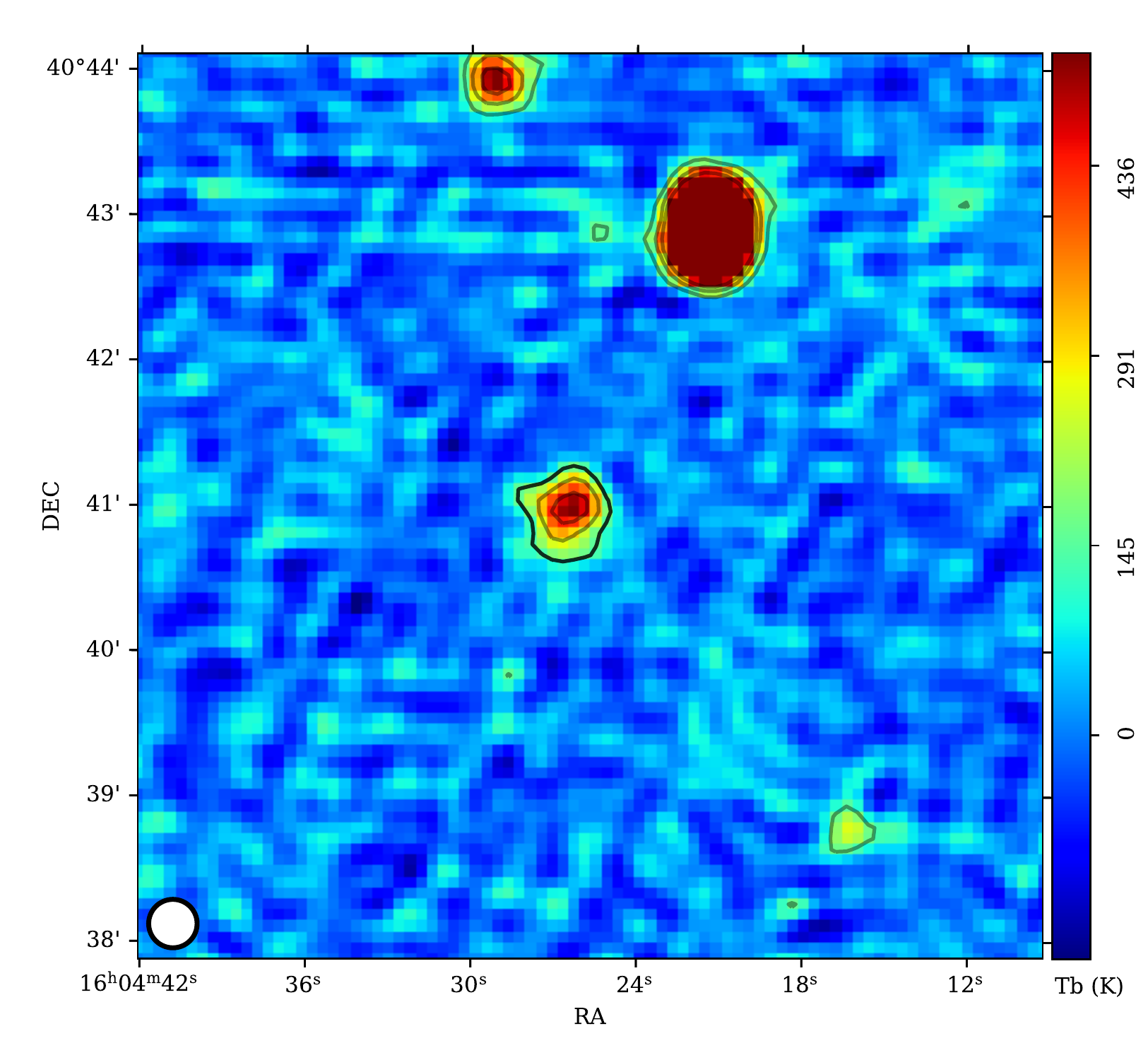}\includegraphics[height=5cm]{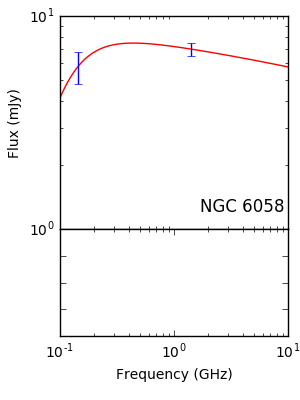}

\includegraphics[height=5cm]{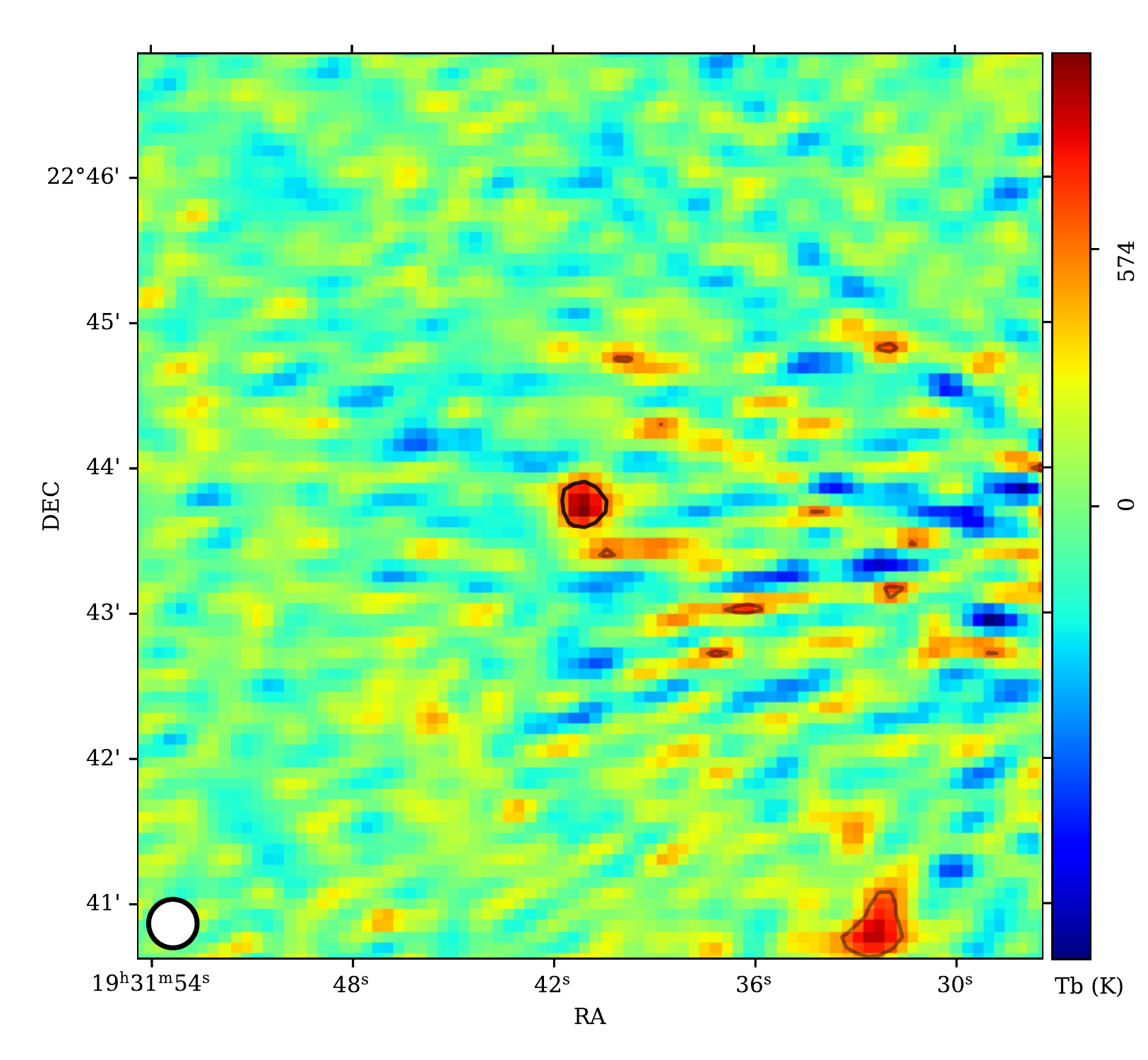}\includegraphics[height=5cm]{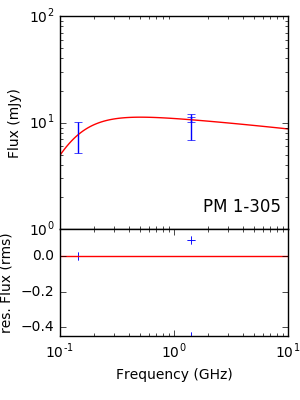}\includegraphics[height=5cm]{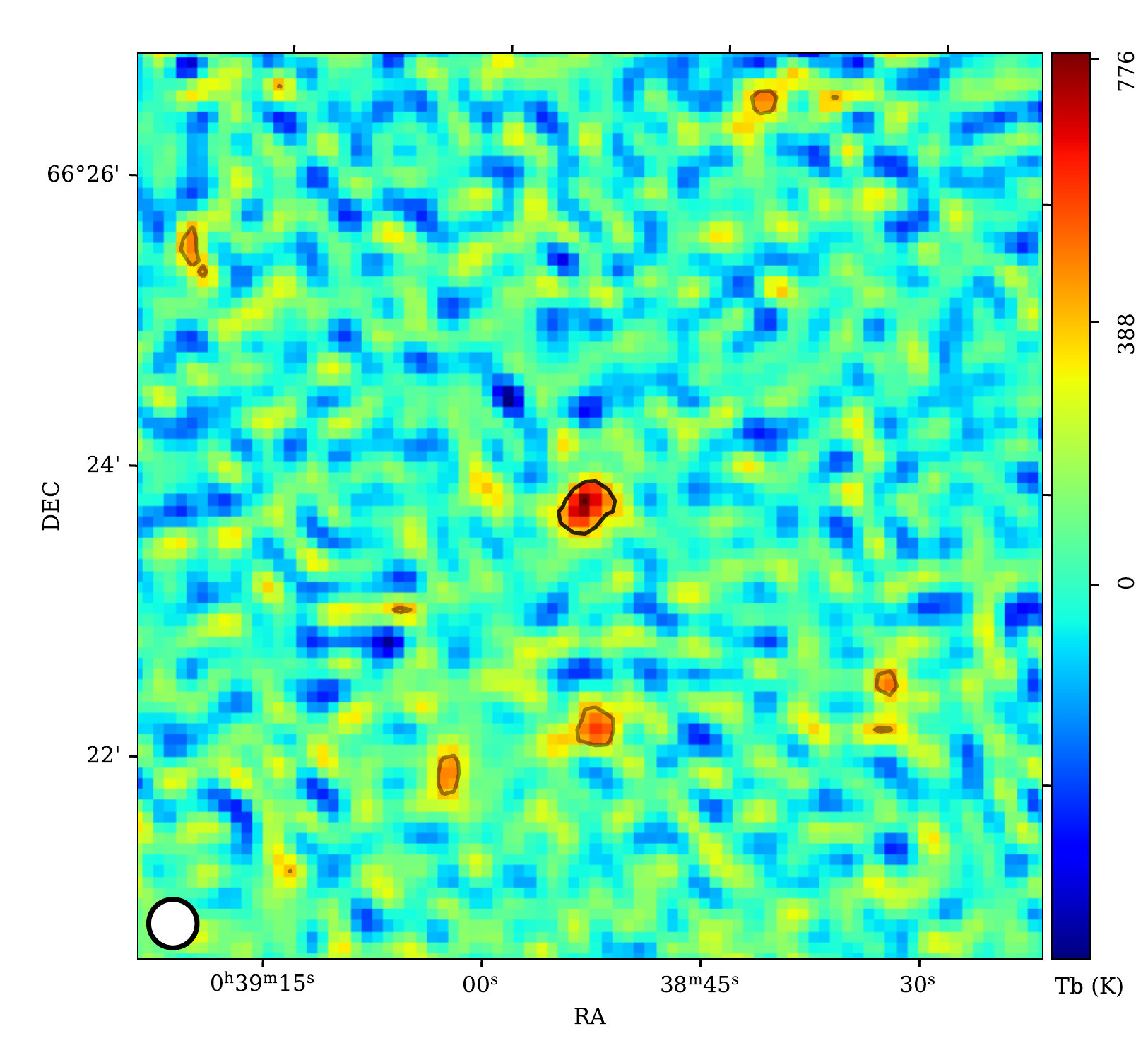}\includegraphics[height=5cm]{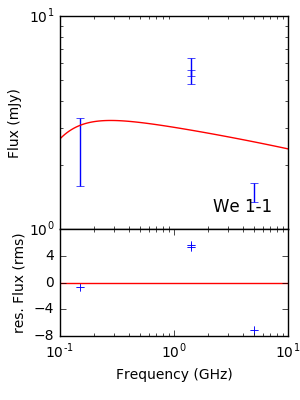}

\includegraphics[height=5cm]{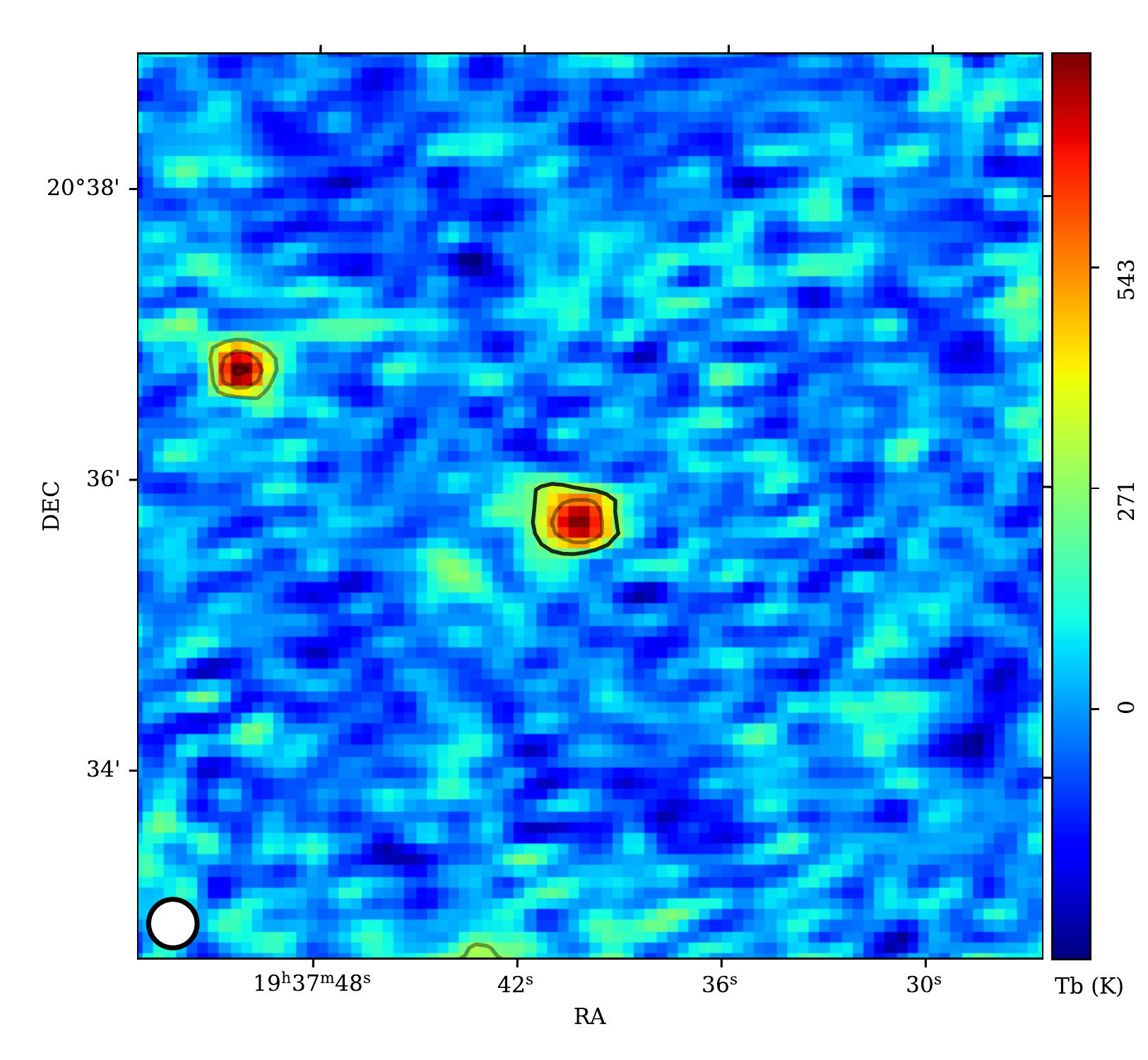}\includegraphics[height=5cm]{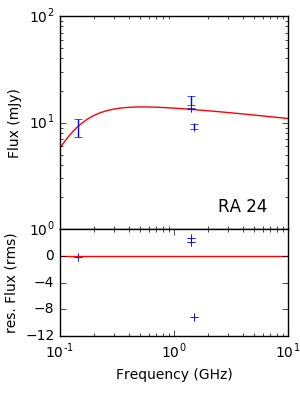}\includegraphics[height=5cm]{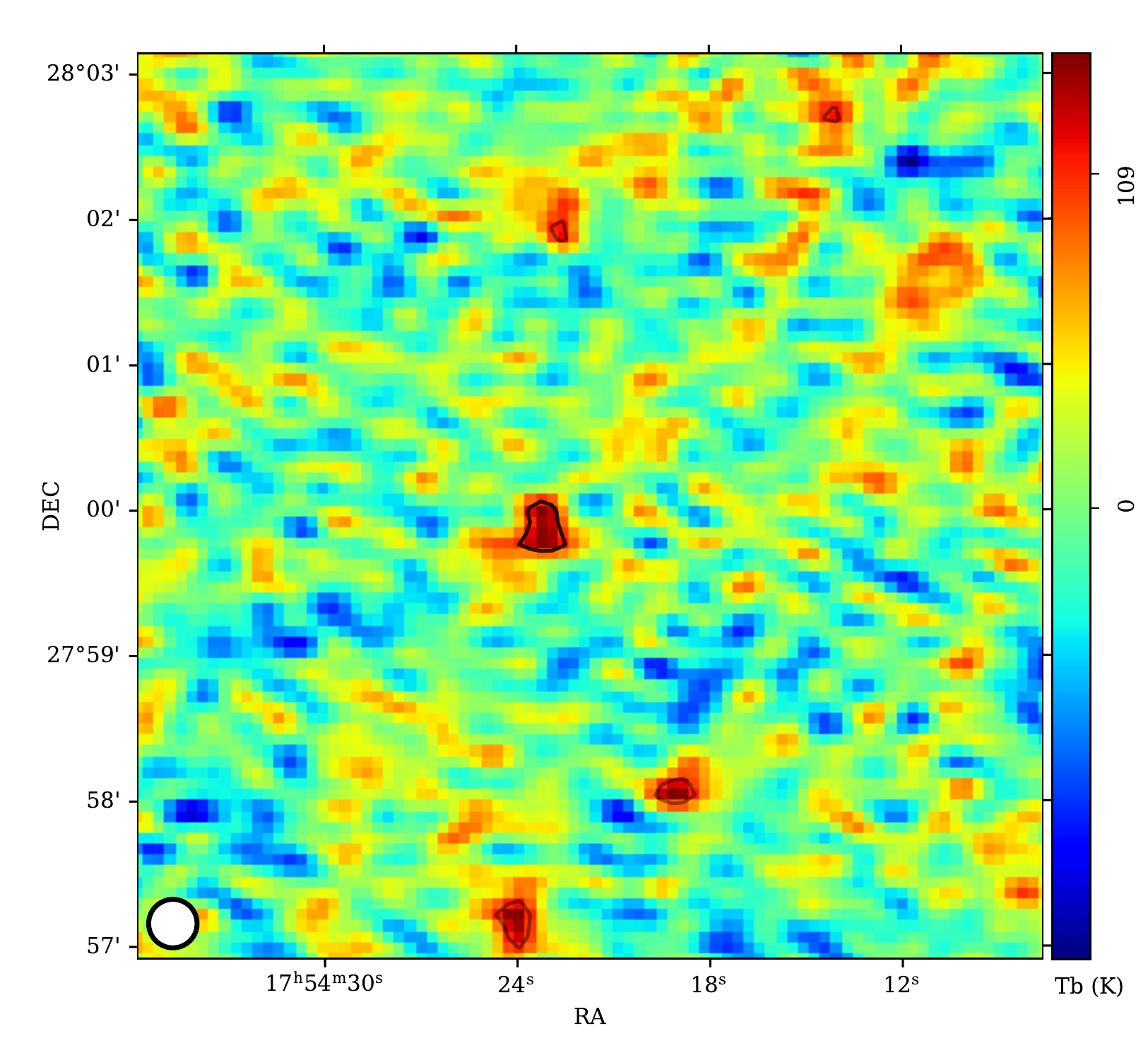}\includegraphics[height=5cm]{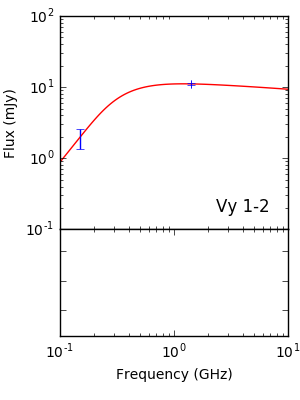}

\label{fig:spectra4}
\end{figure*}

The nebular images and spectra are shown in Figures\,{\ref{fig:spectra1}} through \ref{fig:spectra4}. We converted the intensity scale in the images from $F_\mathrm{144\,MHz}/\mathrm{beam}$ to $T_\mathrm{B}$/beam. The converted images map electron temperature for resolved and optically thick PNe. However, for PNe more compact than the instrument beam, the peak flux and brightness temperature are dilluted by the squared ratio of the source size to the beam size $(\Theta / \Theta _\mathrm{beam}) ^2$.

We determined average electron temperatures for well resolved PNe which are optically thick at 144\,MHz. For this purpose, we substituted the measured size and the integrated 144\,MHz flux to equation\,\ref{eq:tb} with $T_\mathrm{B} = T_\mathrm{e}$. The flux and diameter uncertainties propagate to the calculated $T_\mathrm{e}$ error. The derived temperatures are presented in Table\,\ref{tab:temperatures} along with electron temperatures from the literature obtained using alternative methods. The electron temperatures derived from the 144\,MHz images do not exceed 9.6\,kK. Table \ref{tab:temperatures} shows, that our derived temperatures are 20\% to 60\% lower than the temperatures derived from CELs of [O\,{\sc iii}] and [N\,{\sc ii}]. They are also lower from the temperatures derived from the Balmer jump, although they agree within $1 \, \sigma$ in two cases (NGC\,6543 and NGC\,6826). The mean temperature determined from the 144\,MHz images is about 7.0\,kK, which is about 35\% lower than the 10.7\,kK mean temperature derived for [O\,{\sc iii}]. The low $T_\mathrm{e}$ derived from 144\,MHz optically thick emission results from the presence of the cold plasma component, which is observed in RLs \citep{2000MNRAS.312..585L}. Radio flux is strongly affected by the coldest and most dense regions, even if they contain only a small fraction of the total ionized mass in PNe. The optical thickness of plasma at radio wavelengths is approximately proportional to $T_{\mathrm e}^{-1.35}$. Thus, low electron-temperature plasma has much higher opacity from hot plasma and can become a strong opacity source for low-frequency radio emission.

\begin{figure}
\includegraphics[width=\columnwidth]{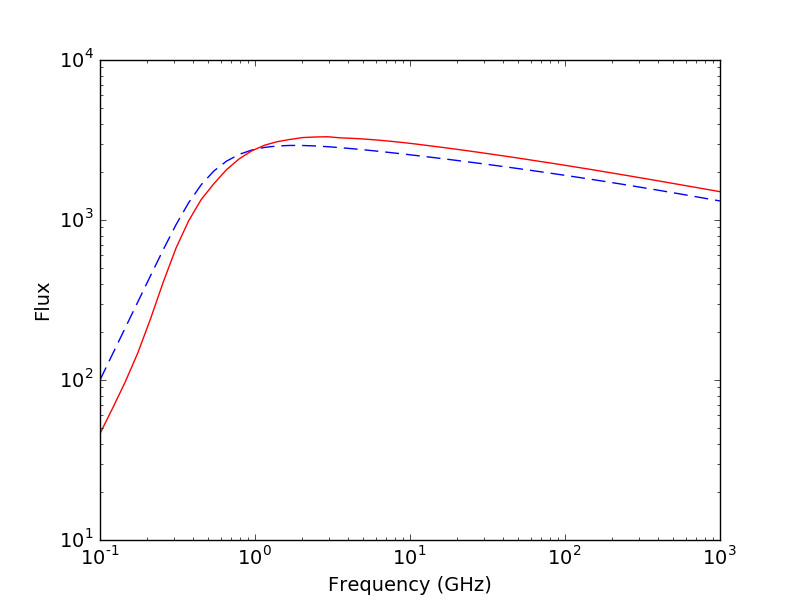}
\includegraphics[width=\columnwidth]{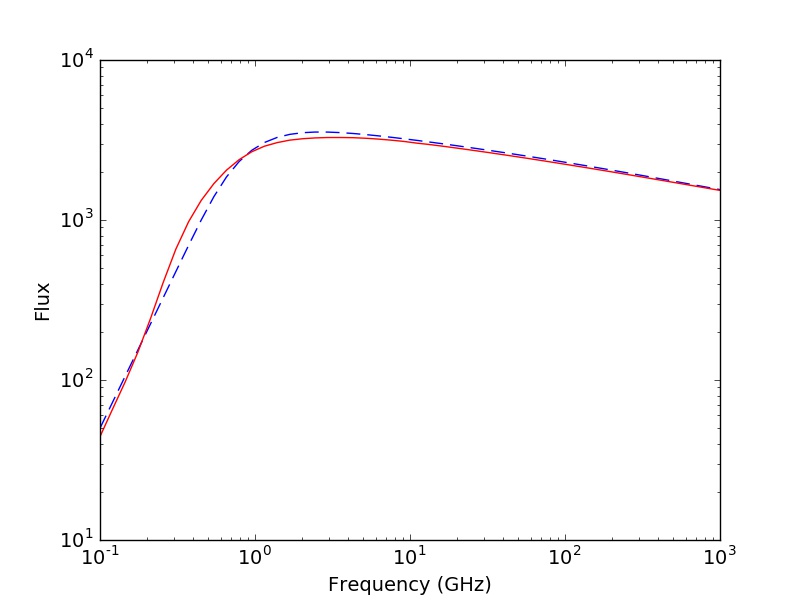}
\caption{Top: Comparison of a computed radio spectrum of a PN with uniform electron temperature of 10\,kK (blue dashed line) and a PN with an inclusion of cold plasma with $T_\mathrm{e} = 2 \, \mathrm{kK}$ randomly distributed filling 10\% of the radius (red solid line). Bottom: the same as above, with the electron temperature of the uniform model is 5\,kK.}
\label{fig:te}
\end{figure}

We modelled a radio spectrum from an ionized nebula filled with the plasma with $T_\mathrm{e}$ of 10\,kK, a typical value for PNe, with 10\% of the volume filled with randomly distributed cool plasma component with $T_\mathrm{e}$ of 2\,kK. The electron temperature averaged over volume is thus 9.2\,kK. Analysis of RLs confirms, that the cold component can indeed have electron temperature as low as ${T}_\mathrm{e} \approx 1 \, \mathrm{kK}$ \citep{2015ApJ...803...99C}. The resulting free-free radio continuum spectrum is compared to the spectrum of the homogeneous model with a temperature of 10\,kK (Equation \ref{eq:snu}) in Figure\,\ref{fig:te}, upper panel.

An inclusion of cold plasma increases flux emitted at optically thin high frequencies with respect to the homogeneous model (Figure\,\ref{fig:te}). This is because optically thin cold plasma emits much more flux than hot plasma. The turnover is shifted to higher frequency. The inclusion of low temperature plasma reduces the brightness temperature in the optically thick part of the spectrum compared to that expected from an homogeneous model by as much as 50\% at 144\,MHz. Hence, in this scenario the electron temperature determined from an optically thick radio image would be around 50\% lower than the temperature of the hot component. This is consistent with our observations, with $T_\mathrm{e}$(144\,MHz) lower by 20-60\% than $T_\mathrm{e}$([O\,{\sc iii}]), which represents the hot component.


The model with two plasma components is compared with the homogeneous model with lower electron temperature of $T_\mathrm{e} = 5 \, \mathrm{kK}$ (Figure\,\ref{fig:te}, lower panel). The spectra appear quite similar. The turnover is approximately at the same frequency. Both models emit similar amounts of optically thick and optically thin flux. However, the optical depth effects become visible already before the turnover frequency in the two component model. This would produce an excess of the 5\,GHz to 1.4\,GHz flux ratio with respect to a homogeneous model.

The distribution, temperature, chemical composition, and density of cold plasma may vary from one PN to another. In every case cold plasma adds more flux to the optically thin part of the spectrum compared to the homogeneous model assuming $T_\mathrm{e}$ derived from CELs, and suppresses optically thick flux, if it extends to the outer boundary of the PN. $T_\mathrm{e}$ should be treated as a free parameter in modeling of radio spectra of PNe, but unfortunately it is not independent from $\xi$ and $EM$.

\subsection{Comparison of radio and optical emission}

\begin{figure*}
\includegraphics[width=0.65\columnwidth]{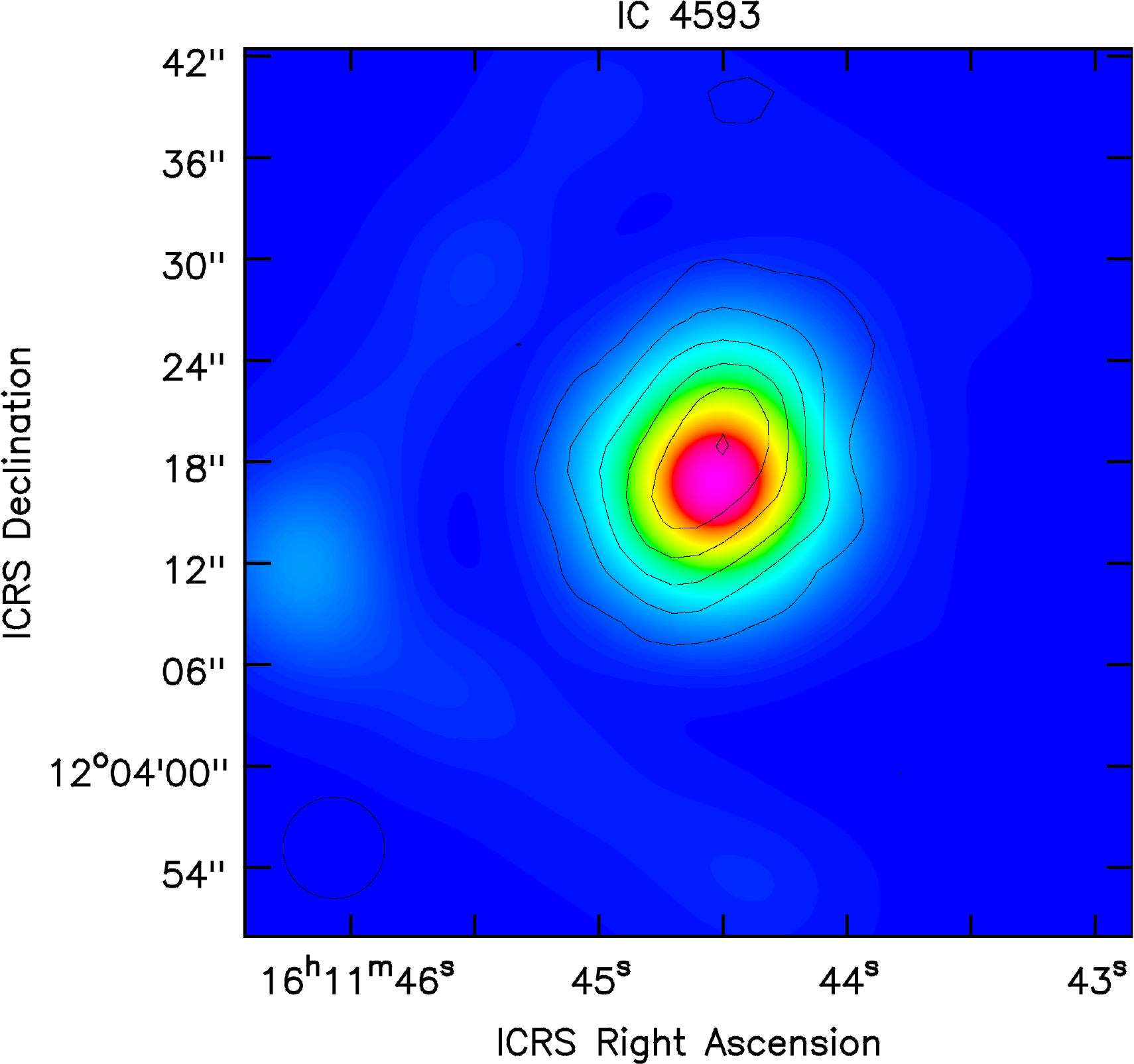}\includegraphics[width=0.65\columnwidth]{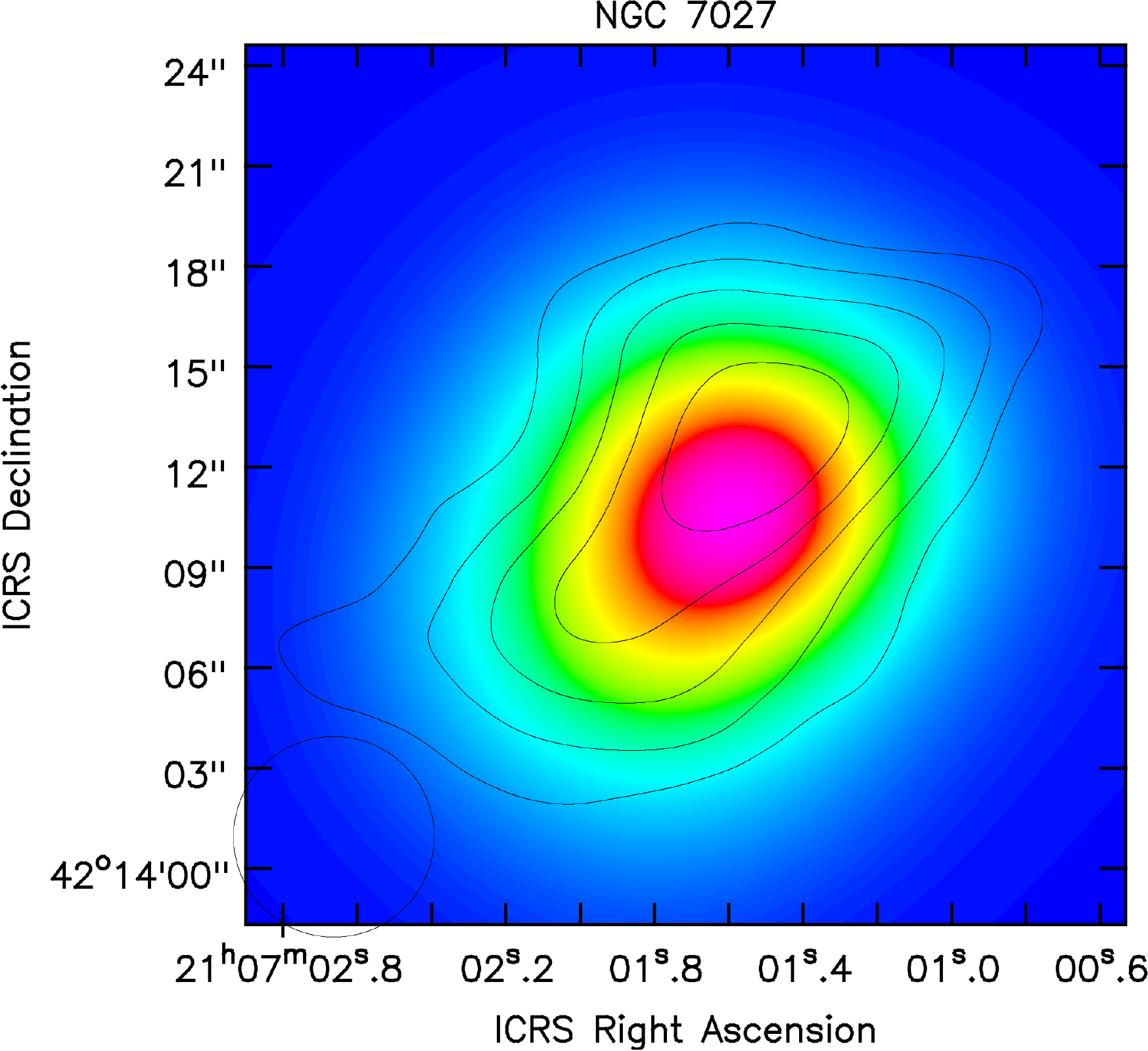}\includegraphics[width=0.65\columnwidth]{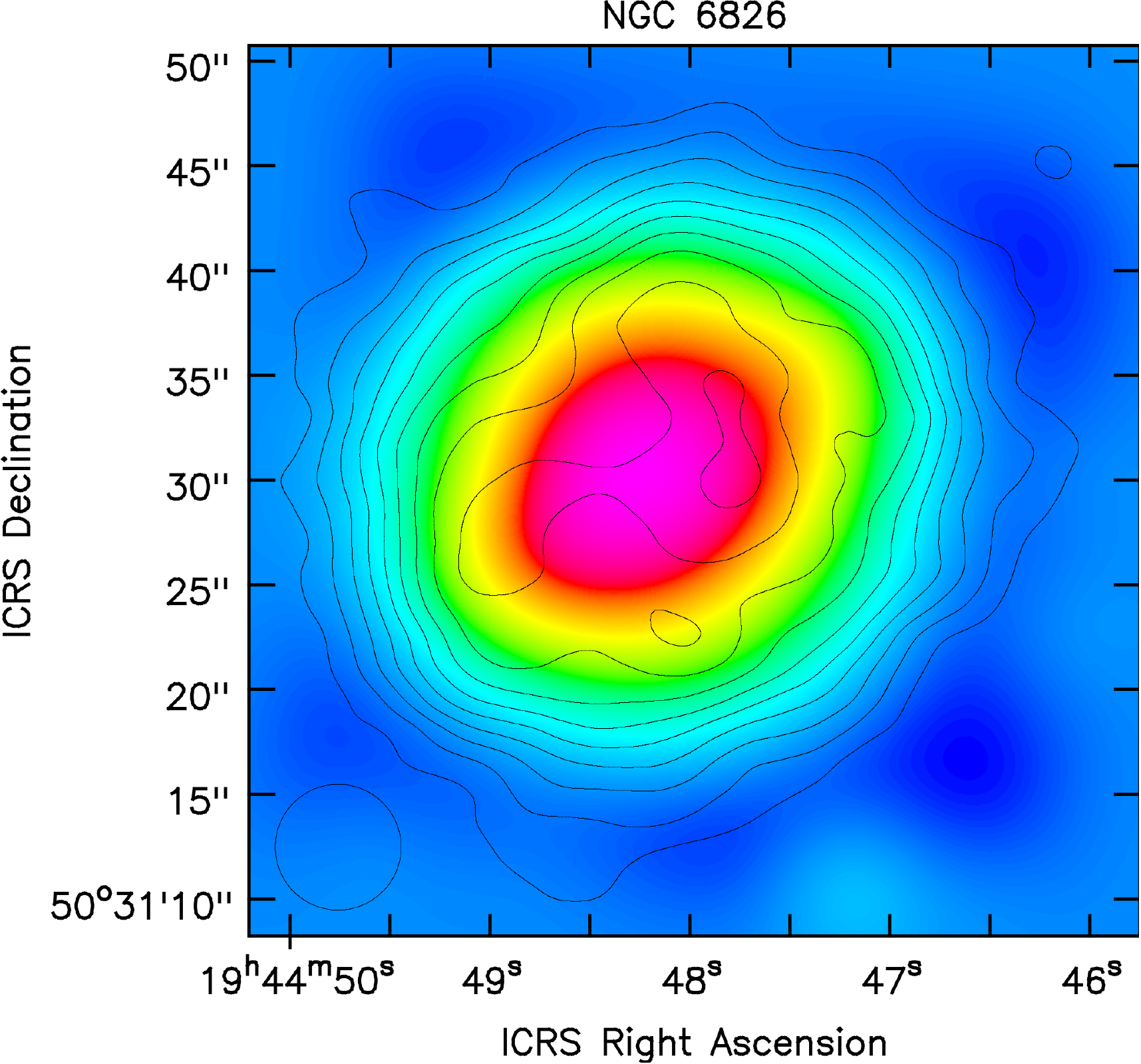}
\includegraphics[width=0.65\columnwidth]{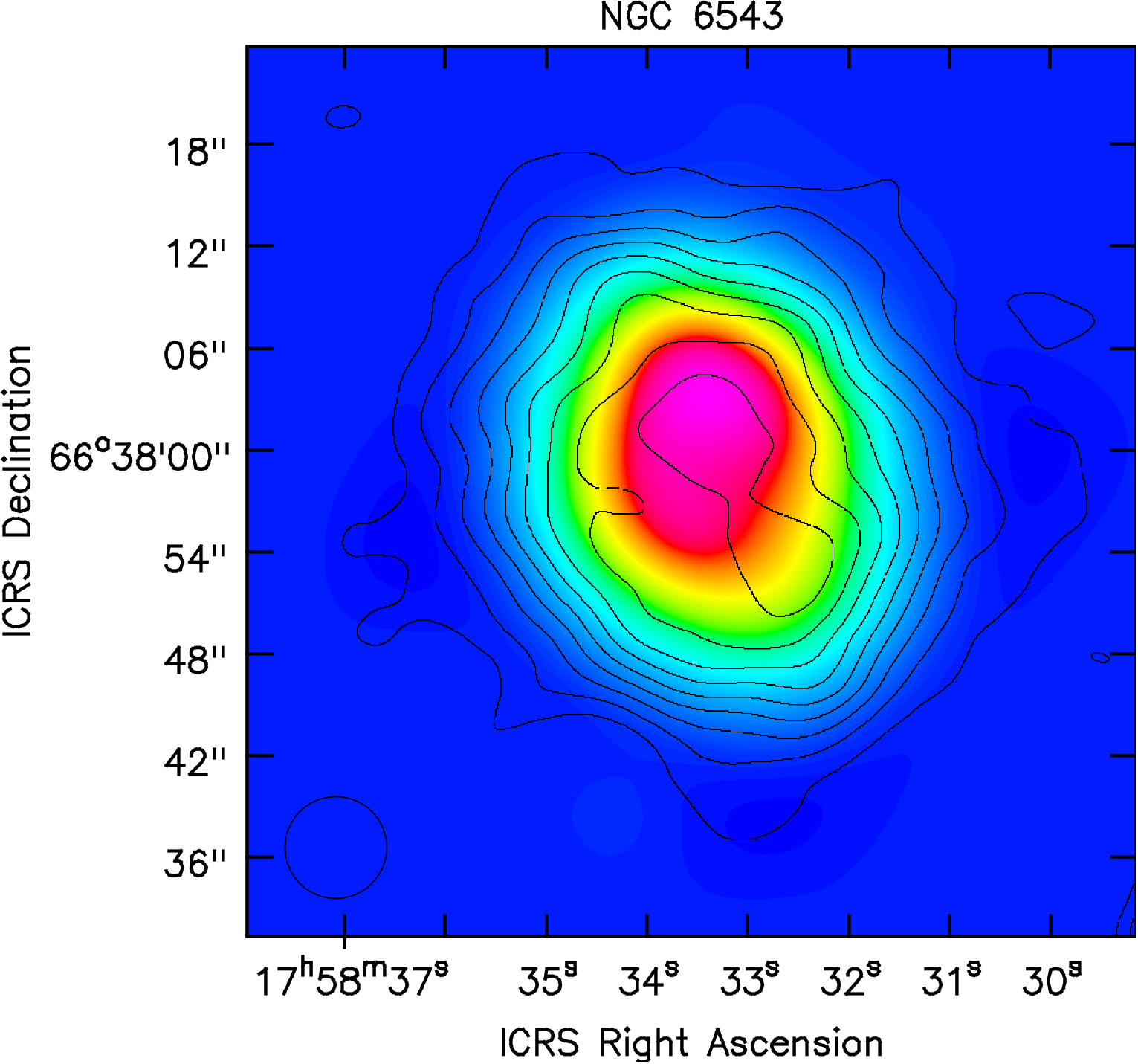}\includegraphics[width=0.65\columnwidth]{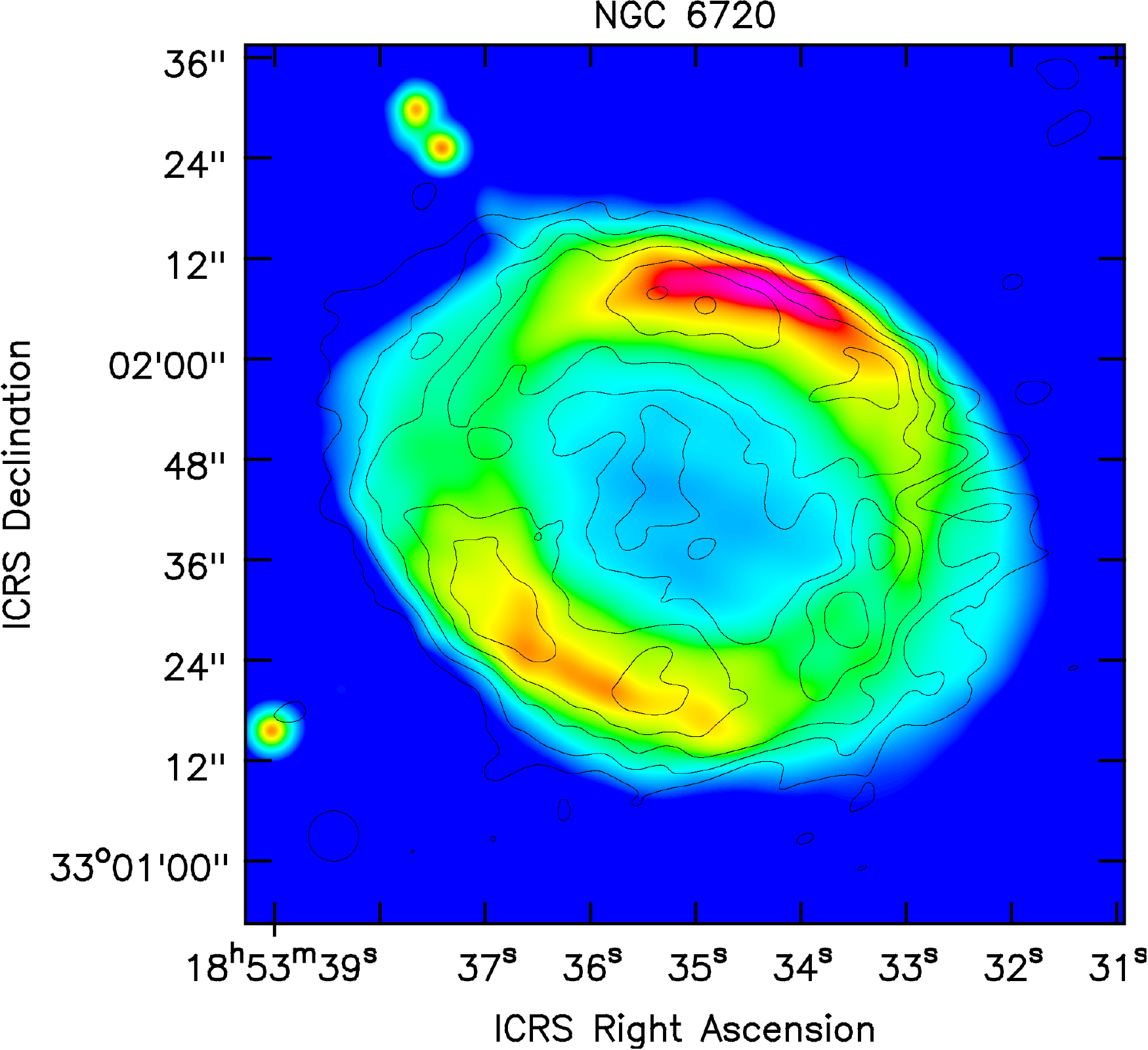}\includegraphics[width=0.65\columnwidth]{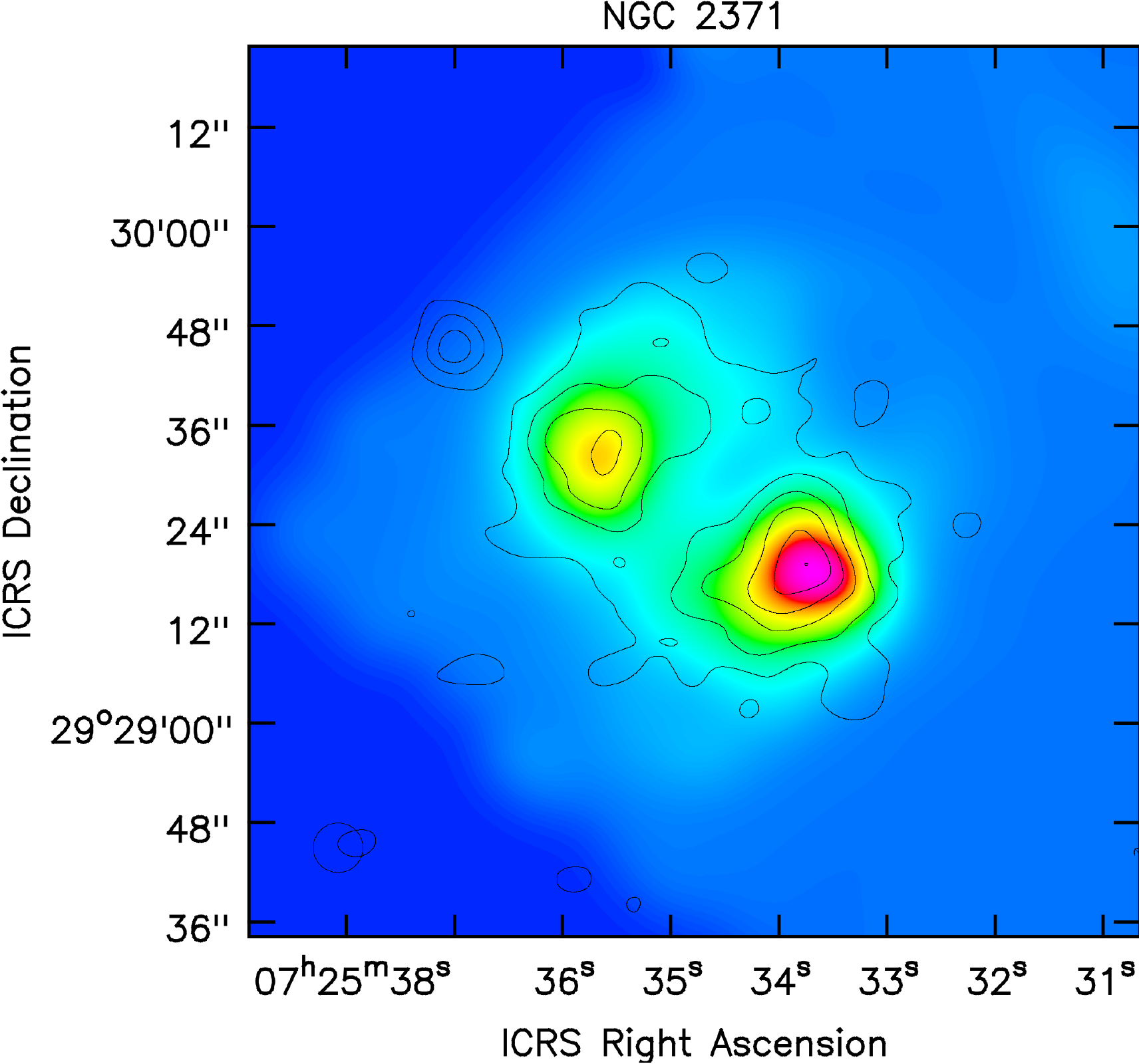}
\caption{Radio contours and optical HST images convolved with a 6 arcsec gaussian} in the F656N filter. The contours are separated by $3\sigma$. North is at the top, east is to the left. The size of the restoring beam of  6\,arcsec is marked in the bottom left corner.
\label{fig:hst}
\end{figure*}

\begin{figure*}
\includegraphics[width=0.65\columnwidth]{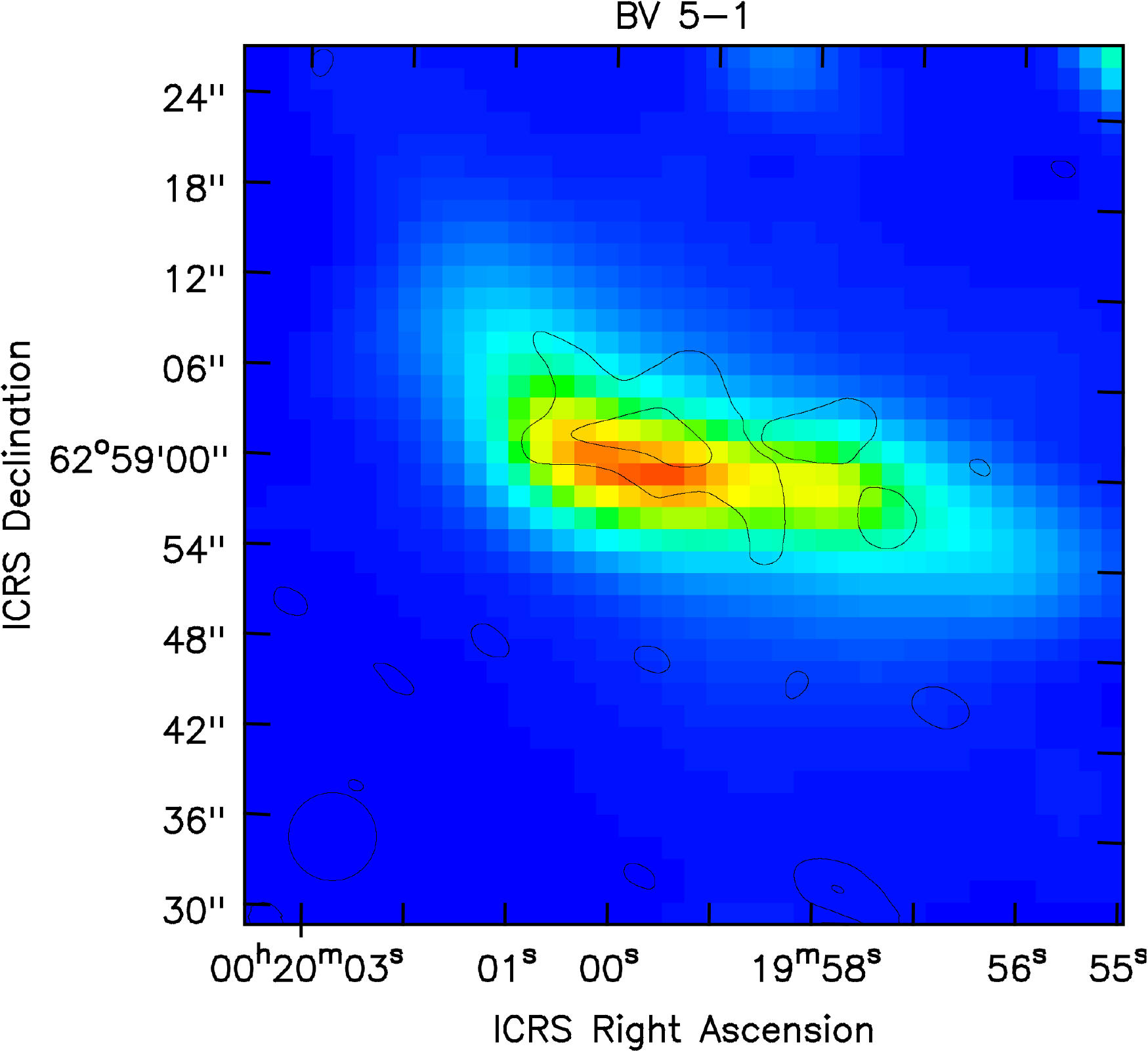}\includegraphics[width=0.65\columnwidth]{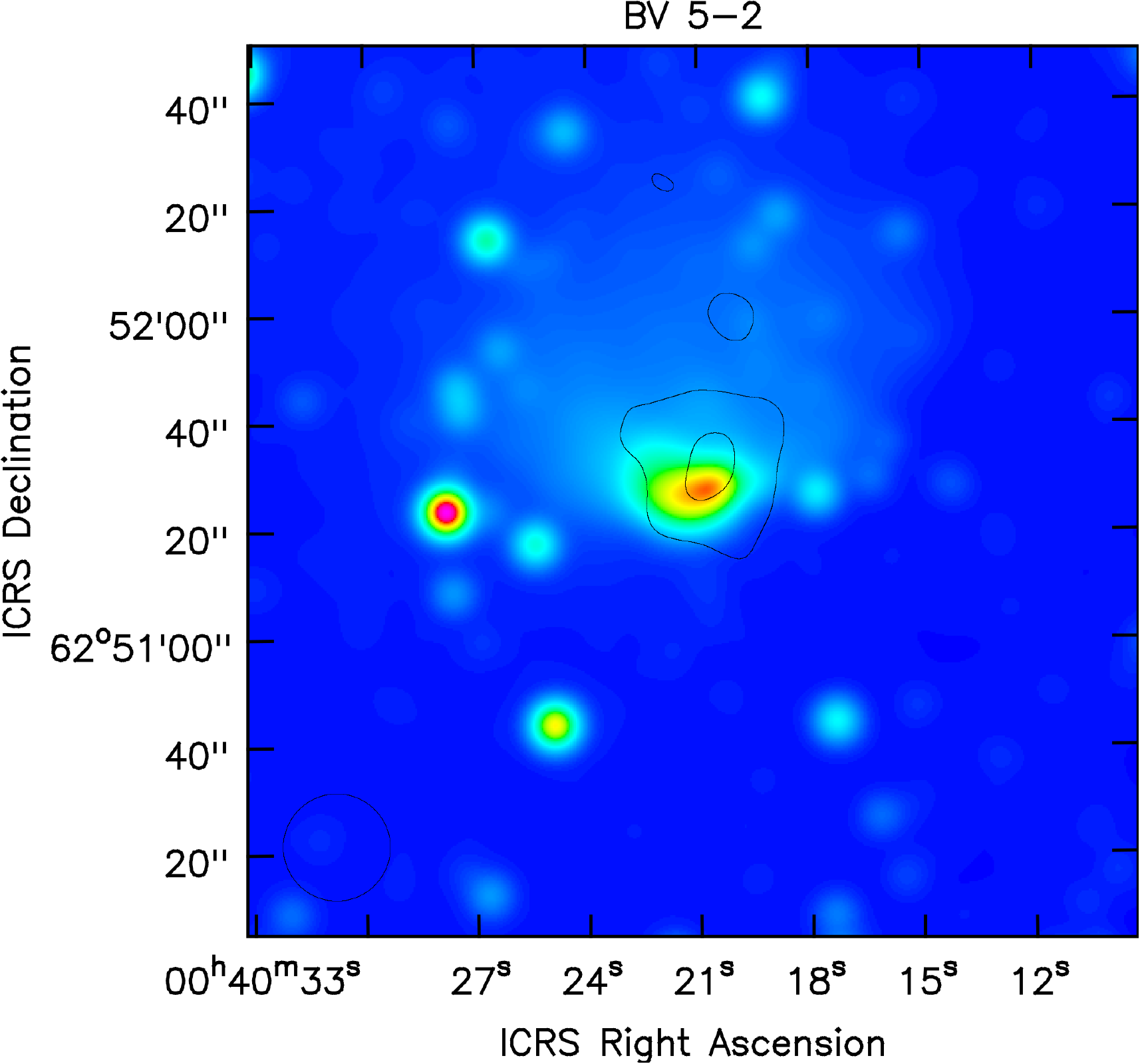}\includegraphics[width=0.65\columnwidth]{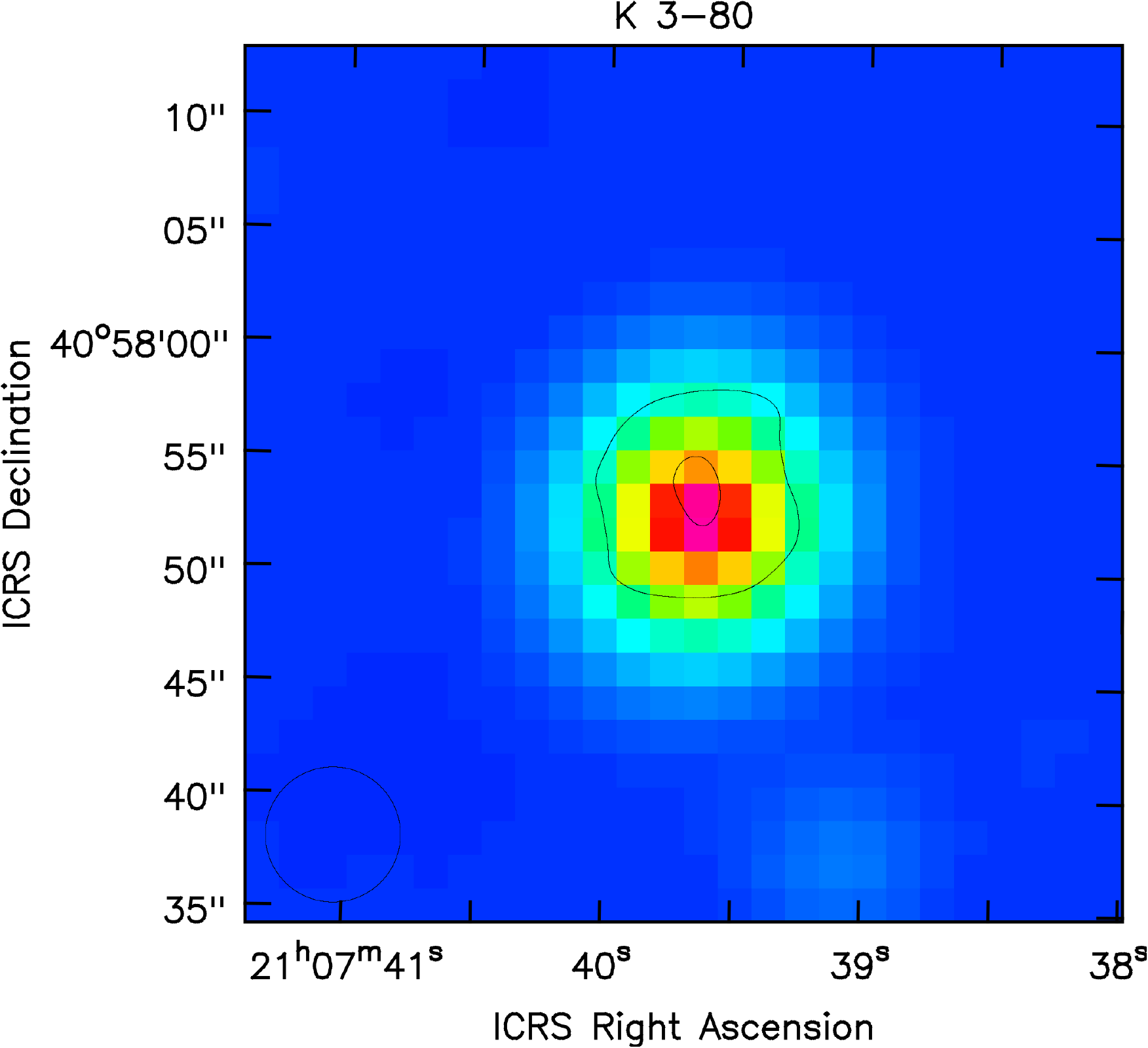}
\includegraphics[width=0.65\columnwidth]{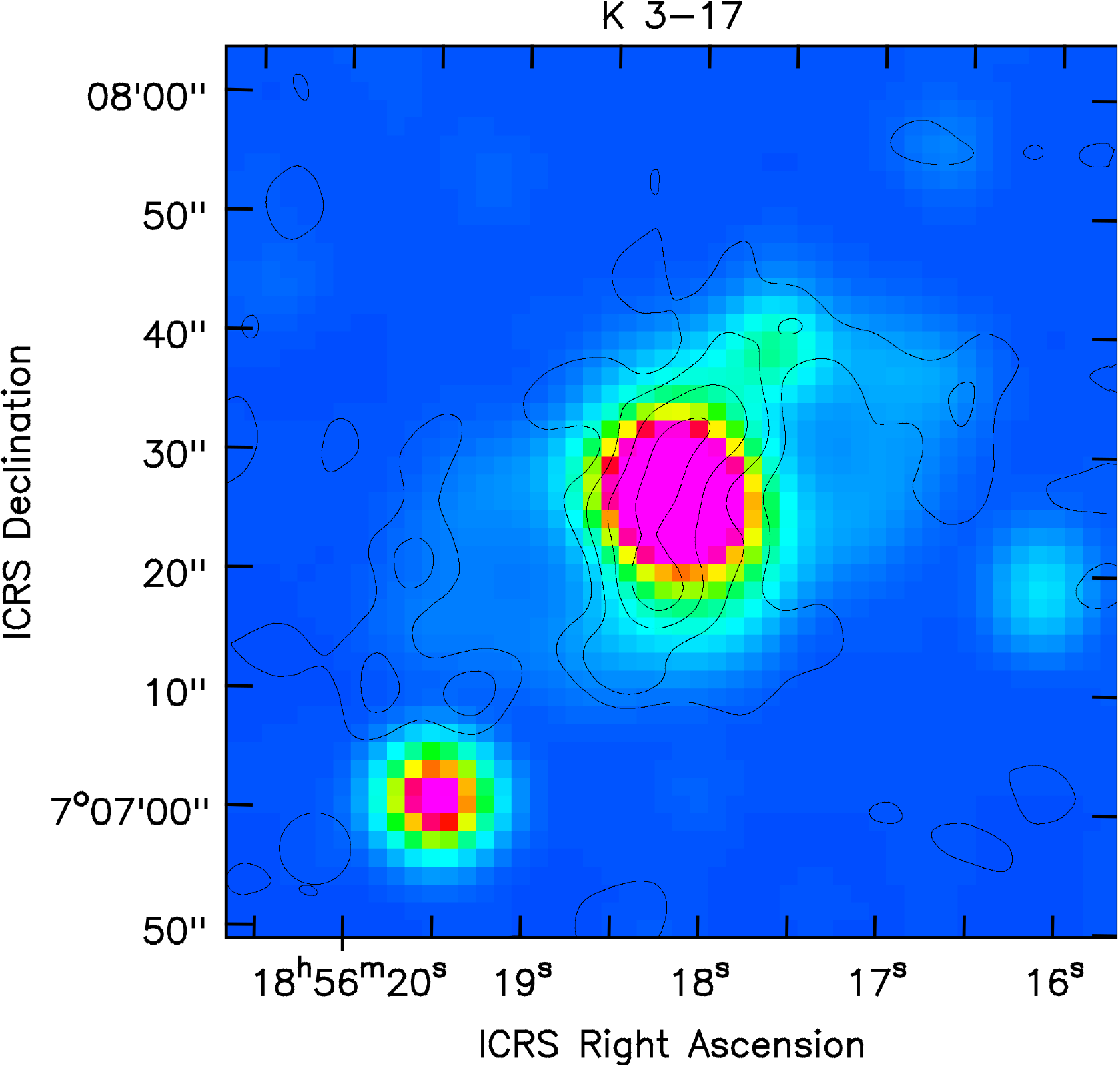}\includegraphics[width=0.65\columnwidth]{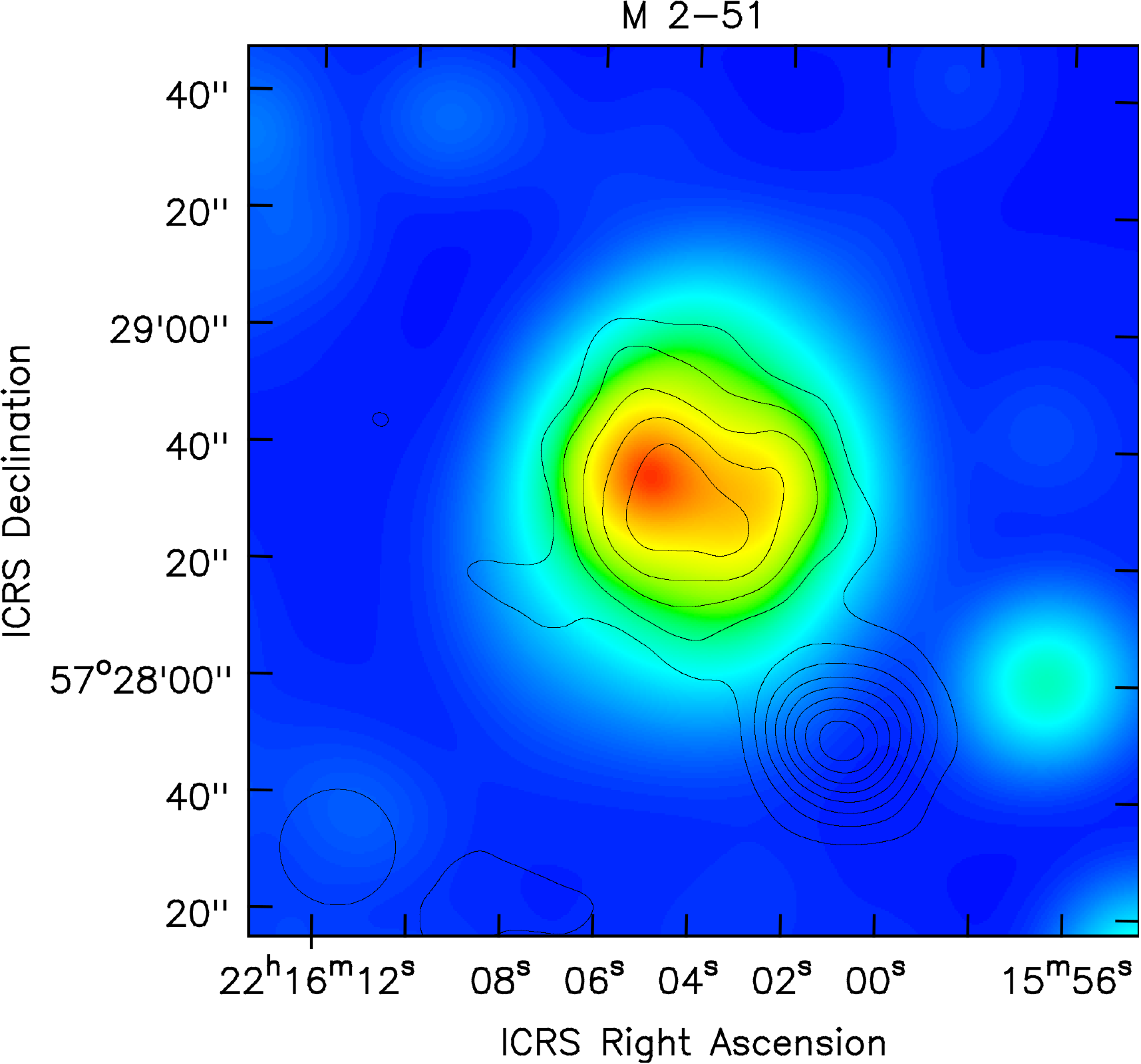}\includegraphics[width=0.65\columnwidth]{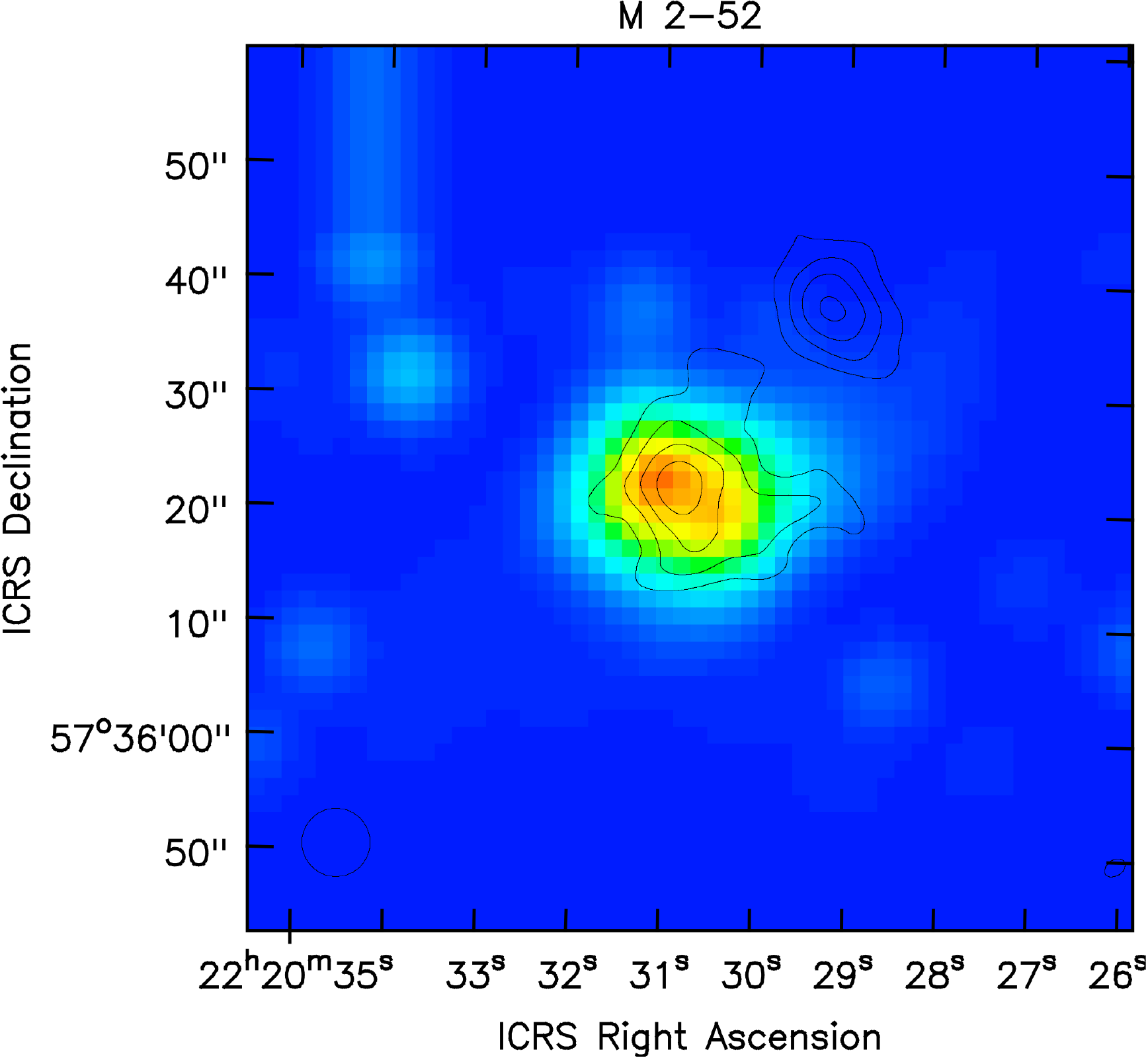}
\includegraphics[width=0.65\columnwidth]{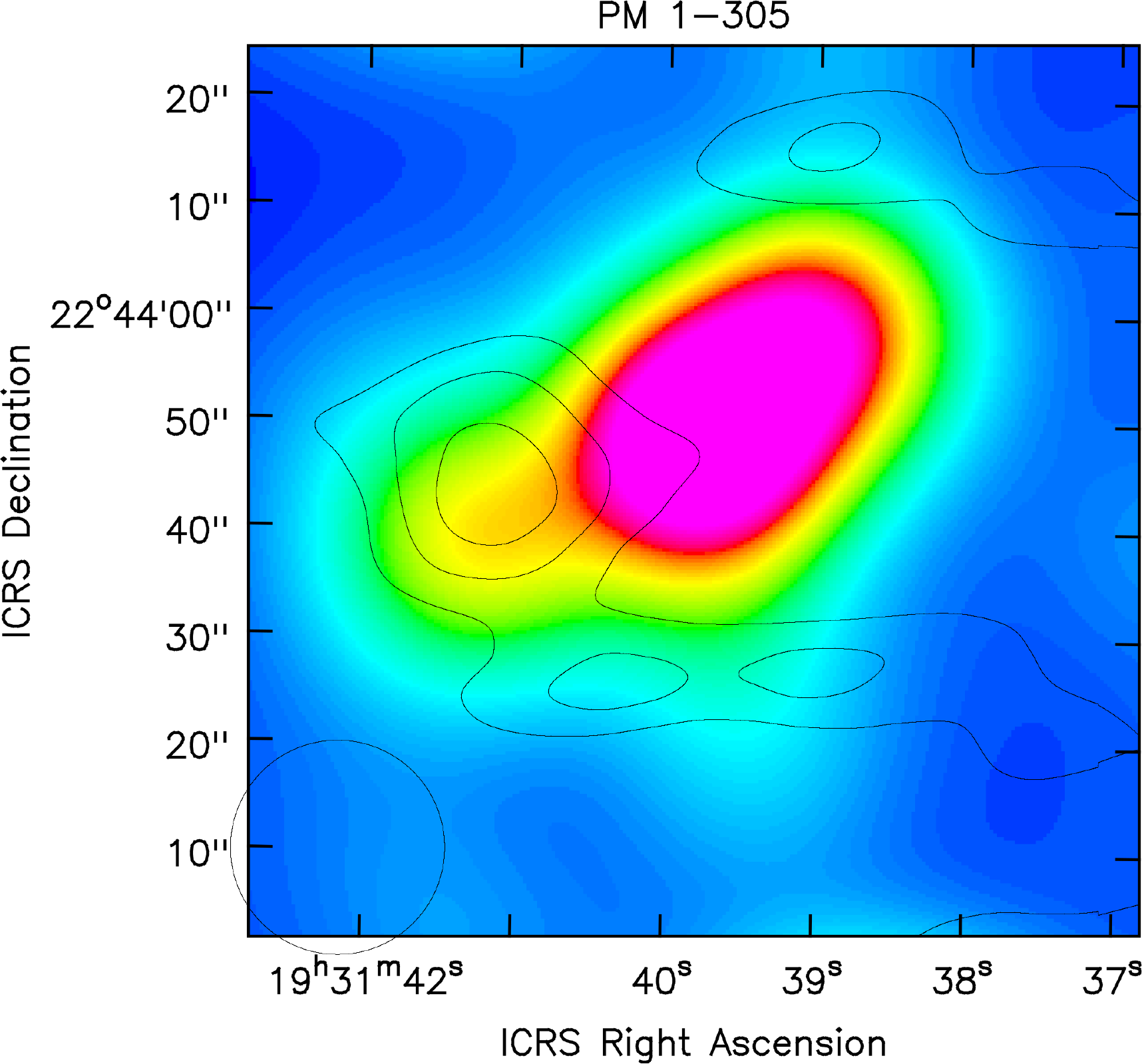}\includegraphics[width=0.65\columnwidth]{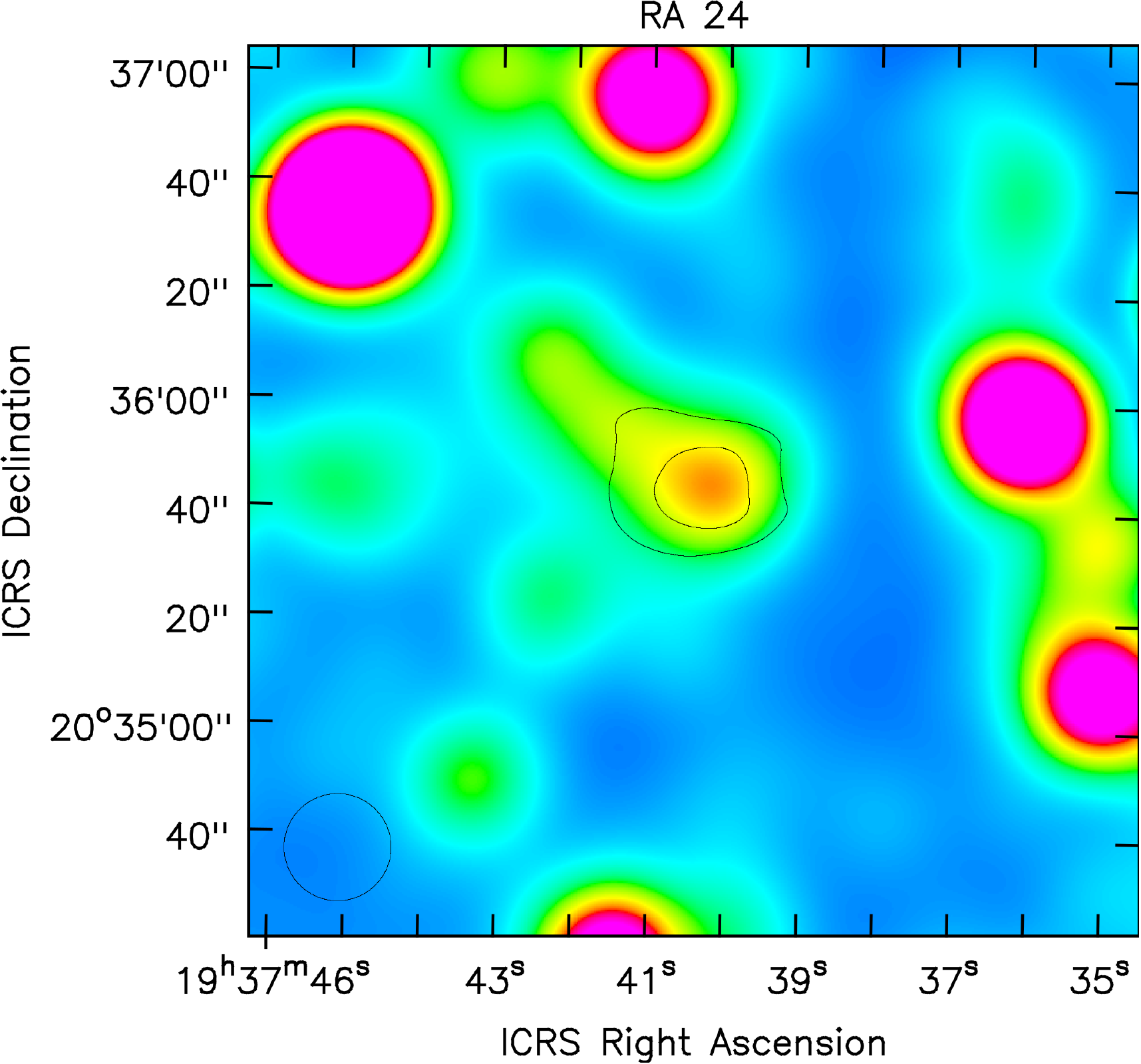}\includegraphics[width=0.65\columnwidth]{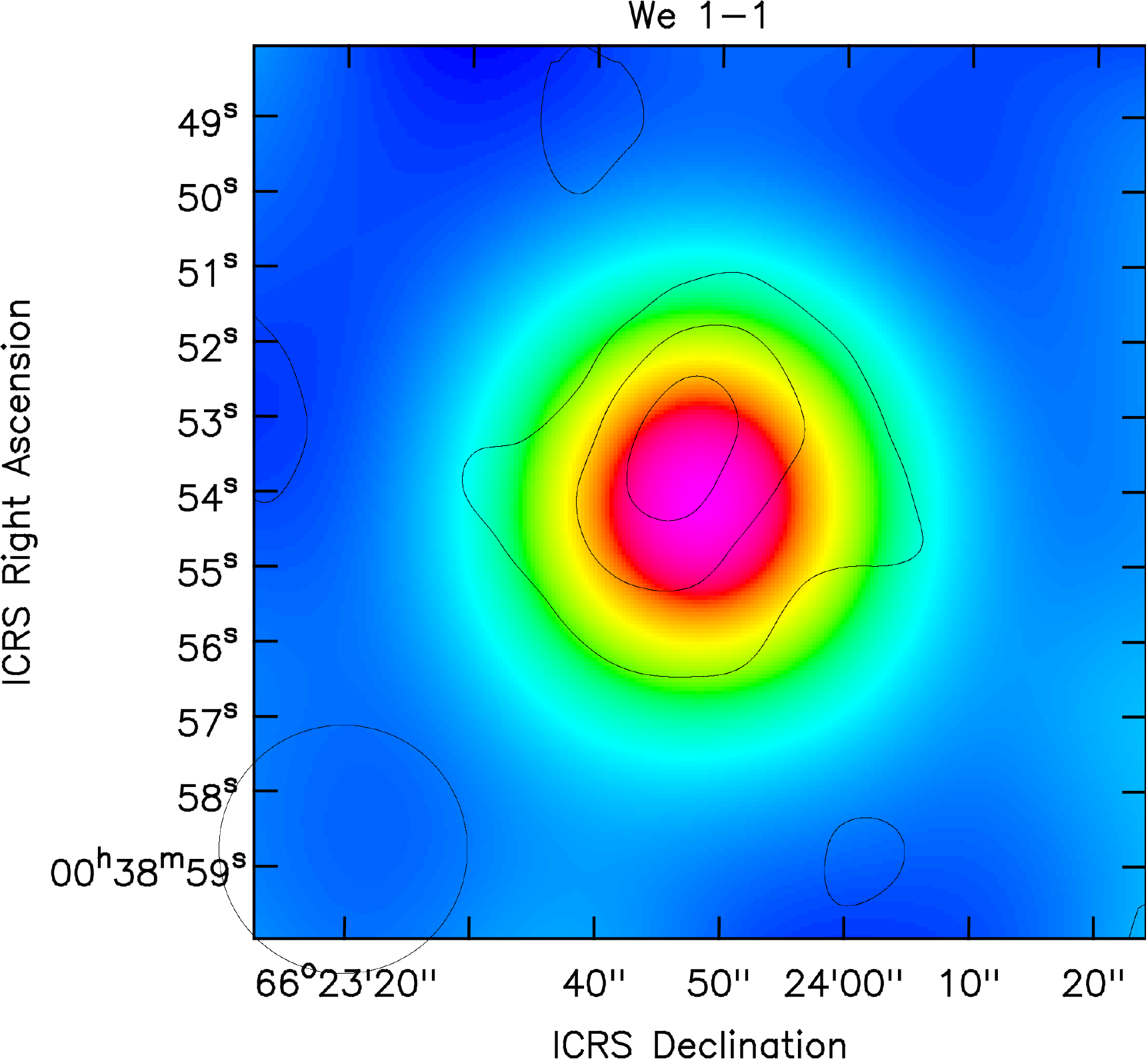}
\caption{Radio contours and optical IPHAS images in the $\mathrm{H}\alpha$ filter. The contours are separated by $3\sigma$ for most of PNe. For K\,3-17 the contours are plotted at the level of 2, 3, 4.5, and $6\,\sigma$ which better show the weak radio emission from the bipolar lobes. PM\,1-305 is not centered in the image and is affected by two background stars in the $\mathrm{H}\alpha$ image. The size of the restoring beam of 6\,arcsec for high-resolution images or 20\,arcsec for low-resolution images is shown in the bottom left corner. North is at the top, east is to the left.}
\label{fig:iphas}
\end{figure*}


\begin{figure}
\includegraphics[width=\columnwidth]{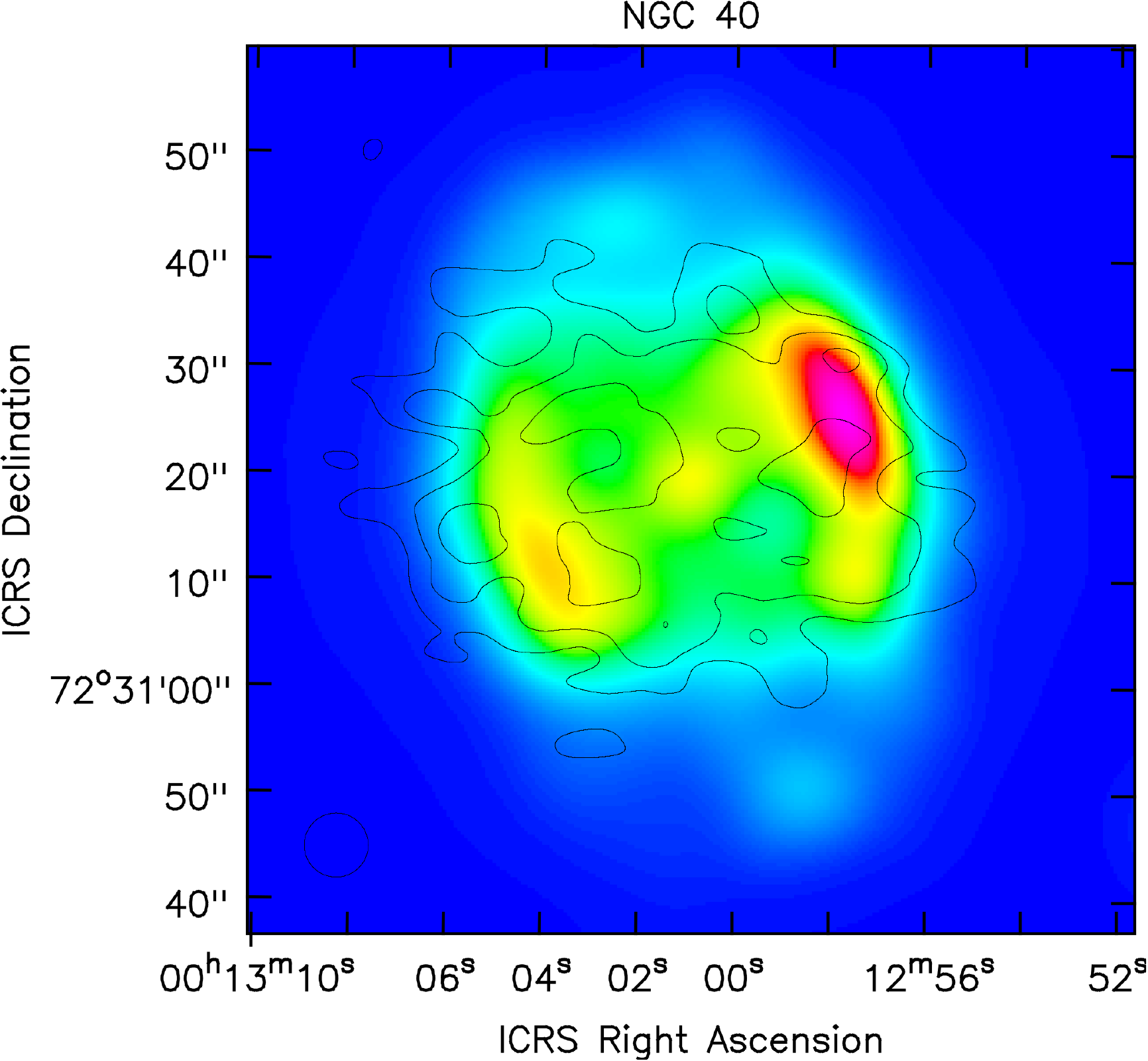}
\caption{Radio contours and optical INT images in the $\mathrm{H}\alpha$ filter. The contours are separated by $3\sigma$. The size of the restoring beam of 6\,arcsec is shown in the bottom left corner.}
\label{fig:int}
\end{figure}

We compared radio continuum images of PNe with the Hubble Space Telescope (HST) and the Isaac Newton Telescope (INT), collected in most cases by the INT Photometric H-Alpha Survey \citep[IPHAS]{2005MNRAS.362..753D} $\mathrm{H}\alpha$ images in Figures\,\ref{fig:hst}, \ref{fig:iphas}, and \ref{fig:int}. We convolved $\mathrm{H}\alpha$ images with 6 or 20 arcsec Gaussian in order to match the resolution of radio and optical images. $\mathrm{H}\alpha$ emission is optically thin in PNe. The brightness distribution in $\mathrm{H}\alpha$ should be similar to optically thin at 144\,MHz emission, as both of them depend primarily on the emission measure. However, cold and hydrogen deficient regions should stand out in the 144\,MHz images. On the other hand, the 144\,MHz surface brightness distribution for optically thick PNe is not expected to correlate with the $\mathrm{H}\alpha$ image since it depends on the local electron temperature close to the outer radius of a PN.

PNe K\,3-17, IC\,4593, NGC\,6543, NGC\,6826, and NGC\,7027 are optically thick in radio. IC\,4593 (Figure\,\ref{fig:hst}) is not very well resolved and does not allow for a detailed comparison of optical and radio surface brightness distribution. NGC\,7027 (Figure\,\ref{fig:hst}) is slightly more resolved. It shows a maximum of the radio emission in the north western part of the nebula. NGC\,6826, NGC\,6543 (Figure\,\ref{fig:hst}), and NGC\,40 (Figure\,\ref{fig:int}) have similar size in radio and $\mathrm{H}\alpha$ (Table\,\ref{tab:observations}). Low brightness temperatures of these three PNe indicate that cold plasma exists near the outer radius of the shell. Otherwise their brightness temperature would reflect the temperature of the hot plasma component, close to the [O\,{\sc iii}] or [N\,{\sc ii}] temperatures (Tab.\,\ref{tab:temperatures}). We were able to measure temperature fluctuations \citep{1967ApJ...150..825P}, since these three PNe are well resolved and optically thick. The fluctuations clearly exceed the $3 \sigma$ noise in the 144\,MHz images (Figure\,\ref{fig:hst}). The point-to-point temperature fluctuations measured in the plane of sky in the central part of the disk are $t^2=0.006 \pm 0.001$ in NGC\,6826, $t^2=0.004 \pm 0.001$ for NGC\,6543, and $t^2=0.013 \pm 0.006$ for NGC\,40. $t^2$ is defined as the standard deviation of the temperature distribution computed close to the nebular center, so that it would not be affected by drop of the flux at the edges. We subtracted the background root mean square error (rms). In order to compute the background rms we first measured the rms in the whole image. Then we flagged all the regions exceeding 3 rms and we computed a new value of the rms in the unflagged image. We repeated the process if there were still regions exceeding 3 rms in the image. The temperature fluctuations in NGC\,6543 agree with the 0.004 estimate by \citet{2004MNRAS.351.1026W} from optical imaging. Temperature fluctuations which could explain the abundance discrepancy would need to be one order of magnitude bigger. However, the temperature fluctuations which could be responsible for the dichotomy of abundance determination \citep{1971BOTT....6...29P} may exist on lower spatial scales.

NGC\,40 shows patchy structure in the 144\,MHz image. The optical and radio images are not correlated. NGC\,40 a born-again candidate \citep{2019MNRAS.485.3360T}. In such a case, a new, hydrogen-free ejecta can be mixed with the previously ejected hydrogen-rich envelope and cause significant inhomogenities of the chemical composition and electron temperature within the nebula.

K\,3-17 shows a weak trace of the bipolar structure in radio (Figure\,\ref{fig:iphas}). The waist of the hourglass nebula is optically thick in radio, whereas the bipolar structure is optically thin.

The radio images of NGC\,6720 (Figure\,\ref{fig:hst}) and NGC\,2371 (Figure\,\ref{fig:hst}) resemble the $\mathrm{H}\alpha$ images. However, their radio spectra indicate considerable optical thickness at 144\,MHz. The 144\,MHz flux is dominated by the brightest regions in these two nebulae. This is confirmed by non-uniform brightness distributions observed in the 144\,MHz images and small values of $\xi$ derived in the fit of the radio spectra. The small value of $\xi$ indicates that most of the emission comes from a small fraction of the nebulae, while the remaining part remains optically thin. Both nebulae have significantly smaller $\tau_{144\,\rm{MHz}}$ than PNe which are fully optically thick.


NGC\,6720 is very well resolved in the LOFAR image. Both 144\,MHz and $\mathrm{H}\alpha$ images show an oval ring. The brightest part, reaching a temperature of 5000\,K at maximum, is optically thick. The center of the nebula and the part of the ring close to the long axis, which contribute less to the radio flux, remain optically thin. \citet{2013AJ....145...92O} modelled the optical image of NGC\,6720 with a triaxial ellipsoid seen nearly pole-on. The projected ring is brighter on its shorter axis. The maximum of the optical emission along the shorter axis is close to the outer edge of the ring. The maximum of radio emission is shifted toward the center of the nebula with respect to $\mathrm{H}\alpha$ emission. The reason for this can be that more cold plasma exists closer to the central star. As a result, the 144\,MHz opacity increases toward the central star, so the inner part of the ring is brighter.

\begin{figure}
\includegraphics[width=\columnwidth]{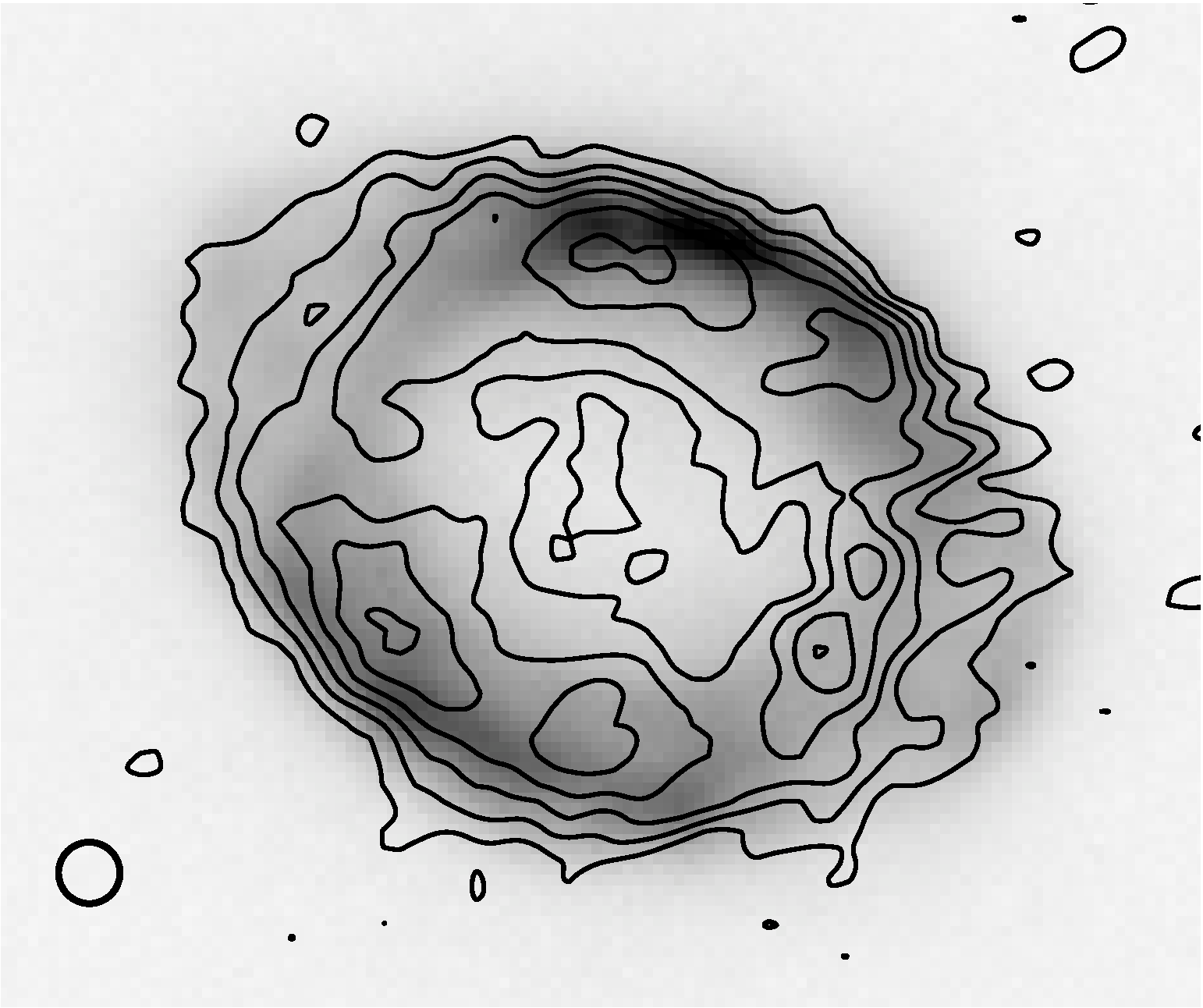}
\caption{Comparison of the 144\,MHz radio continuum (contours) and Herschel 70\,$\mu \mathrm{m}$ image (background) of NGC\,6720. The Herschel image is taken from \citet{2010A&A...518L.137V}. The size of the box is 100\,arcsec height and 120\,arcsec width. The 144\,MHz beam is marked with a circle.} \label{fig:m57herschel}
\end{figure}

Figure \ref{fig:m57herschel} compares the 70\,$\mu$m and 144\,MHz continuum images of NGC\,6720. The maximum of the 144\,MHz emission is closer to the central star than 70\,$\mu$m emission. It appears, that cold plasma is not associated with dust emitting at 70\,$\mu$m.

\begin{figure}
\includegraphics[width=\columnwidth]{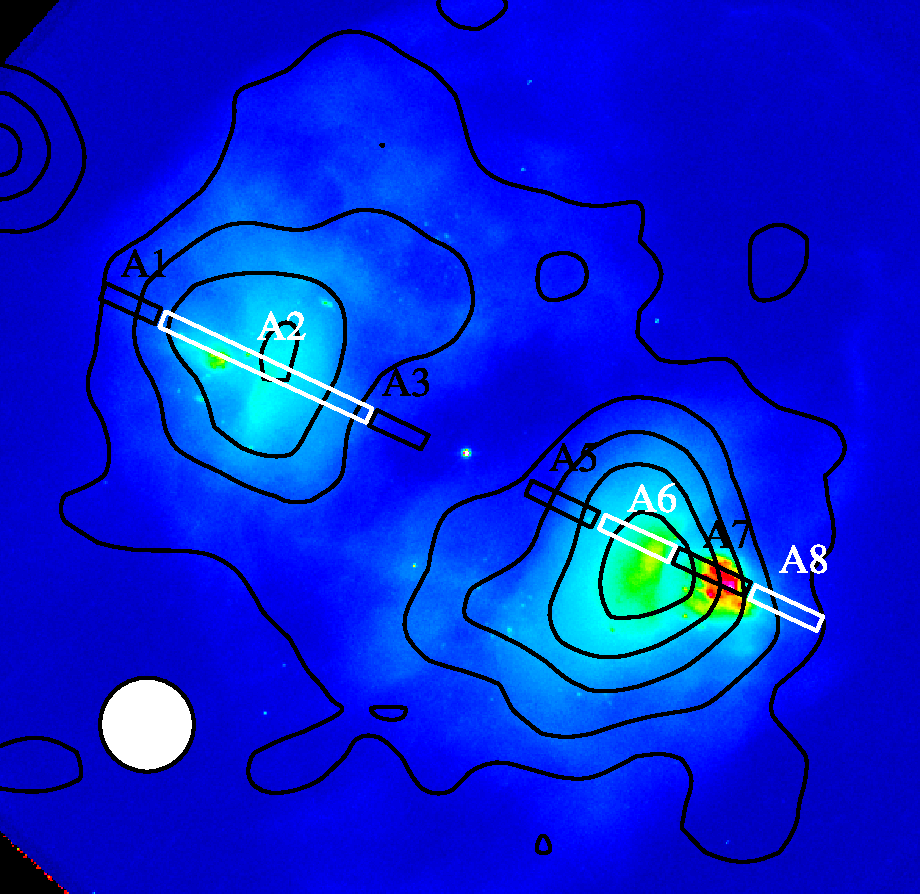}
\caption{Comparison of the 144\,MHz radio continuum (contours) and $\mathrm{H}\alpha$ HST image (background) of NGC\,6371. The regions mark the extraction boxes used in \citet{2020MNRAS.496..959G}.} \label{fig:ngc2371}
\end{figure}

NGC\,2371 is a bipolar nebula. The brightest, barrel-like structure contains a collection of knots. Two pair of brightest knots in the $\mathrm{H}\alpha$ image lie at the position angle of 60 degree in the NW and SE direction from the central star, although they are not perfectly aligned with the central star (Figure\,\ref{fig:hst}). The 144\,MHz image of NGC\,2371 does not trace the $\mathrm{H}\alpha$ emission in detail. In particular, the maxima of 144\,MHz emission are not centered on the brightest knots in the $\mathrm{H}\alpha$ image, but located on fainter knots closer to the central star. This is more clearly seen when comparing a full resolution optical image with the radio emission (Fig.\,\ref{fig:ngc2371}).

\citet{2020MNRAS.496..959G} used spatially resolved spectroscopy to study the electron temperature in NGC\,2371. The brightest clump in the nebula, located in the NE direction from the central star (designed by them as A7, Fig.\,\ref{fig:ngc2371}) has a temperature of 13.8\,kK. It is classified as a low-ionization knot \citep{2001ApJ...547..302G}. For comparison, the neighbouring region A6, which is the brightest region in the 144\,MHz map, has significantly higher temperature of 18\,kK. The A7 clump should stand out at 144\,MHz since it is brighter in optical and cooler. However, it is at least 2 times fainter in radio than the A6 clump. This suggests, that the low-ionization knot A7 does not contain cold plasma or contains less cold plasma than the A6 clump.




\section{Discussion and conclusions}

We observed 144\,MHz free-free radio emission in a sample of PNe using LOFAR. Optically thick emission allows for relatively straightforward measurement of local electron temperature. The observations confirm a presence of significant amount of cold plasma, which was first proposed from the study of RLs and CELs. Cold and hot components remain spatially unresolved in the 144\,MHz images. However, cold plasma has much higher opacity compared to hot plasma. In the result, the determined electron temperatures are significantly weighted toward cold plasma component. Thus, the previous approaches which assumed a homogeneous model of PNe with the temperature derived from CEL or an arbitrary value of 10\,kK are incorrect.

Different studies assumed homogeneous models with the temperature corresponding to the hot plasma component. In particular, ionized mass determination relies on the assumption that PNe are optically thin at 5\,GHz and their plasma temperature is similar to $T_\mathrm{e}$ derived from CELs \citep{1995ApJ...446..279B}. This approximation remains valid as soon as the electron temperatures used in the computation are reduced by 40\% on average. Taking this into account it would scale down the ionized masses of PNe derived from optically thin radio emission. Lower $T_\mathrm{e}$ would also result in lower diameters derived from the radio SED fit compared to observed diameters \citep{2018MNRAS.479.5657H,2021MNRAS.503.2887B}.

\citet{1992A&A...266..486S} used an homogeneous model for extinction determination from the ratio of the optically thin radio to hydrogen flux. They used electron temperatures from \citet{1986ApJ...308..322K} or derived using his formulae. The radio flux was used to determine the dereddened $\mathrm{H}\beta_0$ flux. This dereddened flux was compared with the observed $\mathrm{H}\beta$ flux to derive the extinction $C_\mathrm{rad}$. Another extinction determination $C_\mathrm{opt}$ comes from the observed $\mathrm H \alpha$ to $\mathrm H \beta$ ratio. \citet{1992A&A...266..486S} showed that $C_\mathrm{opt}$ is systematically larger by a factor of about 1.2 than $C_\mathrm{rad}$ for the PNe in the direction to the Galactic center. \citet{2004MNRAS.353..796R} postulated steeper extinction law toward the Galactic center to explain the difference between radio and optical extinction, which was later confirmed by \citet{2012IAUS..283..380H}. Finally, \citet{2013A&A...550A..35P} suggested that the 5\,GHz emission in PNe is not optically thin, which, however, contradicted most of other studies. 

The ratio of the optically thin radio to $\mathrm{H}\beta$ flux depends on electron temperature $\sim T_\mathrm{e}^{0.53}$ \citep{1984ASSL..107.....P}. If PNe were modeled with the temperature lower by 40\%, the dereddened $\mathrm{H}\beta$ fluxes would be higher by a factor of 1.3. This would decrease $C_\mathrm{rad}$ by 0.12 and improve the consistency of both extinction determinations, though not yet fully explain it.

The 144\,MHz images allowed us to constrain the spatial distribution of cold plasma. It extends to the outer radii of the nebulae. However, the partially optically thick image of NGC\,6720 shows, that cool plasma is more abundant toward the center of the nebula. It is noteworthy, that the abundance discrepancy factor also increases toward the center of the PN \citep{2001ApJ...558..145G}. The $70\, \mathrm{\mu m}$ image has also different brightness distribution from radio emission, which suggests that dusty regions observed at $70\, \mathrm{\mu m}$ do not harbour cold plasma. Another example, in which radio emisson is more concentrated in the inner ring of the nebula than dust emission is the Helix nebula \citep{2015A&A...573A...6P}. 

Low-ionization structures are often present in PNe hosting binary central stars \citep{2009A&A...505..249M}, and they are one of the candidates to explain the abundance discrepancy \citep{2015ApJ...803...99C}. If low-ionization structures in NGC\,2371 contained cold plasma, they would stand out in the 144\,MHz radio image. Instead, they are fainter. This could be explained if they did not contain cold plasma, or contained significantly less cold plasma than other regions of the nebula. 
Observations of a larger sample when LoTSS is completed will allow for more deep and statistically important study of the cold plasma component in PNe. Multi-frequency analysis of PNe radio continuum images may allow us to better constrain the spatial distribution of cold plasma in PNe.

We will continue to study low-frequency radio emission of PNe using more complete data from the LoTSS survey. The number of observed PNe will increase rapidly when the survey improves completeness at low Galactic latitudes. The low-frequency survey LoLSS will observe at $42-66\,\mathrm{MHz}$, but with reduced spatial resolution of 15 arcsec \citep{2021A&A...648A.104D} and sensitivity (1 mJy/beam) compared to LoTSS. Further advance will be made with the Square Kilometer Array (SKA) \citep{2015aska.confE.118U}, which will carry out an extremely sensitive radio continuum survey at 1.4\,GHz. Multi-frequency images will allow us to obtain accurate spectral index maps of PNe and possibly model the spatial distribution of cold plasma in PNe.

\acknowledgments

MHaj was supported by the Polish National Agency for Academic Exchange (NAWA) within the Bekker programme under grant No PPN/BEK/2019/1/00431 and by National Science Centre, Poland, under grant No. 2016/23/B/ST9/01653. MO acknowledges the MSHE for granting funds for the Polish contribution to the International LOFAR Telescope (MSHE decision no. DIR/WK/2016/2017/05-1) and for maintenance of the LOFAR PL-612 Baldy (MSHE decision no.~59/E-383/SPUB/SP/2019.1). MHav acknowledges funding from the European Research Council (ERC) under the European Union's Horizon 2020 research and innovation programme (grant agreement No 772663). GJW gratefully acknowledges the receipt of a Leverhulme Emeritus Fellowship. The J\"{u}lich LOFAR Long Term Archive and the German LOFAR network are both coordinated and operated by the J\"{u}lich Supercomputing Centre (JSC), and computing resources on the supercomputer JUWELS at JSC were provided by the Gauss Centre for Supercomputing e.V. (grant CHTB00) through the John von Neumann Institute for Computing (NIC).This research has made use of the SIMBAD database, operated at CDS, Strasbourg, France.

\bibliography{lofar_pne}{}

\begin{thebibliography}{}
\expandafter\ifx\csname natexlab\endcsname\relax\def\natexlab#1{#1}\fi
\providecommand{\url}[1]{\href{#1}{#1}}
\providecommand{\dodoi}[1]{doi:~\href{http://doi.org/#1}{\nolinkurl{#1}}}
\providecommand{\doeprint}[1]{\href{http://ascl.net/#1}{\nolinkurl{http://ascl.net/#1}}}
\providecommand{\doarXiv}[1]{\href{https://arxiv.org/abs/#1}{\nolinkurl{https://arxiv.org/abs/#1}}}

\bibitem[{{Aaquist} \& {Kwok}(1996)}]{1996ApJ...462..813A}
{Aaquist}, O.~B., \& {Kwok}, S. 1996, \apj, 462, 813, \dodoi{10.1086/177196}

\bibitem[{{Boji{\v{c}}i{\'c}} {et~al.}(2021){Boji{\v{c}}i{\'c}},
  {Filipovi{\'c}}, {Uro{\v{s}}evi{\'c}}, {Parker}, \&
  {Galvin}}]{2021MNRAS.503.2887B}
{Boji{\v{c}}i{\'c}}, I.~S., {Filipovi{\'c}}, M.~D., {Uro{\v{s}}evi{\'c}}, D.,
  {Parker}, Q.~A., \& {Galvin}, T.~J. 2021, \mnras, 503, 2887,
  \dodoi{10.1093/mnras/stab687}

\bibitem[{{Buckley} \& {Schneider}(1995)}]{1995ApJ...446..279B}
{Buckley}, D., \& {Schneider}, S.~E. 1995, \apj, 446, 279,
  \dodoi{10.1086/175787}

\bibitem[{{Condon} \& {Kaplan}(1998)}]{1998ApJS..117..361C}
{Condon}, J.~J., \& {Kaplan}, D.~L. 1998, \apjs, 117, 361,
  \dodoi{10.1086/313128}

\bibitem[{{Corradi} {et~al.}(2015){Corradi}, {Garc{\'\i}a-Rojas}, {Jones}, \&
  {Rodr{\'\i}guez-Gil}}]{2015ApJ...803...99C}
{Corradi}, R. L.~M., {Garc{\'\i}a-Rojas}, J., {Jones}, D., \&
  {Rodr{\'\i}guez-Gil}, P. 2015, \apj, 803, 99,
  \dodoi{10.1088/0004-637X/803/2/99}

\bibitem[{{de Gasperin} {et~al.}(2019){de Gasperin}, {Dijkema}, {Drabent},
  {Mevius}, {Rafferty}, {van Weeren}, {Br{\"u}ggen}, {Callingham}, {Emig},
  {Heald}, {Intema}, {Morabito}, {Offringa}, {Oonk}, {Orr{\`u}},
  {R{\"o}ttgering}, {Sabater}, {Shimwell}, {Shulevski}, \&
  {Williams}}]{2019A&A...622A...5D}
{de Gasperin}, F., {Dijkema}, T.~J., {Drabent}, A., {et~al.} 2019, \aap, 622,
  A5, \dodoi{10.1051/0004-6361/201833867}

\bibitem[{{de Gasperin} {et~al.}(2021){de Gasperin}, {Williams}, {Best},
  {Br{\"u}ggen}, {Brunetti}, {Cuciti}, {Dijkema}, {Hardcastle}, {Norden},
  {Offringa}, {Shimwell}, {van Weeren}, {Bomans}, {Bonafede}, {Botteon},
  {Callingham}, {Cassano}, {Chy{\.z}y}, {Emig}, {Edler}, {Haverkorn}, {Heald},
  {Heesen}, {Iacobelli}, {Intema}, {Kadler}, {Ma{\l}ek}, {Mevius}, {Miley},
  {Mingo}, {Morabito}, {Sabater}, {Morganti}, {Orr{\'u}}, {Pizzo}, {Prandoni},
  {Shulevski}, {Tasse}, {Vaccari}, {Zarka}, \&
  {R{\"o}ttgering}}]{2021A&A...648A.104D}
{de Gasperin}, F., {Williams}, W.~L., {Best}, P., {et~al.} 2021, \aap, 648,
  A104, \dodoi{10.1051/0004-6361/202140316}

\bibitem[{{Draine} \& {Kreisch}(2018)}]{2018ApJ...862...30D}
{Draine}, B.~T., \& {Kreisch}, C.~D. 2018, \apj, 862, 30,
  \dodoi{10.3847/1538-4357/aac891}

\bibitem[{{Drew} {et~al.}(2005){Drew}, {Greimel}, {Irwin}, {Aungwerojwit},
  {Barlow}, {Corradi}, {Drake}, {G{\"a}nsicke}, {Groot}, {Hales}, {Hopewell},
  {Irwin}, {Knigge}, {Leisy}, {Lennon}, {Mampaso}, {Masheder}, {Matsuura},
  {Morales-Rueda}, {Morris}, {Parker}, {Phillipps}, {Rodriguez-Gil}, {Roelofs},
  {Skillen}, {Sokoloski}, {Steeghs}, {Unruh}, {Viironen}, {Vink}, {Walton},
  {Witham}, {Wright}, {Zijlstra}, \& {Zurita}}]{2005MNRAS.362..753D}
{Drew}, J.~E., {Greimel}, R., {Irwin}, M.~J., {et~al.} 2005, \mnras, 362, 753,
  \dodoi{10.1111/j.1365-2966.2005.09330.x}

\bibitem[{{Frew} {et~al.}(2016){Frew}, {Parker}, \&
  {Boji{\v{c}}i{\'c}}}]{2016MNRAS.455.1459F}
{Frew}, D.~J., {Parker}, Q.~A., \& {Boji{\v{c}}i{\'c}}, I.~S. 2016, \mnras,
  455, 1459, \dodoi{10.1093/mnras/stv1516}

\bibitem[{{Garc{\'\i}a-Rojas} {et~al.}(2019){Garc{\'\i}a-Rojas}, {Wesson},
  {Boffin}, {Jones}, {Corradi}, {Esteban}, \&
  {Rodr{\'\i}guez-Gil}}]{2019arXiv190406763G}
{Garc{\'\i}a-Rojas}, J., {Wesson}, R., {Boffin}, H.~M.~J., {et~al.} 2019, arXiv
  e-prints, arXiv:1904.06763.
\newblock \doarXiv{1904.06763}

\bibitem[{{Garnett} \& {Dinerstein}(2001)}]{2001ApJ...558..145G}
{Garnett}, D.~R., \& {Dinerstein}, H.~L. 2001, \apj, 558, 145,
  \dodoi{10.1086/322452}

\bibitem[{{G{\'o}mez-Gonz{\'a}lez} {et~al.}(2020){G{\'o}mez-Gonz{\'a}lez},
  {Toal{\'a}}, {Guerrero}, {Todt}, {Sabin}, {Ramos-Larios}, \&
  {Mayya}}]{2020MNRAS.496..959G}
{G{\'o}mez-Gonz{\'a}lez}, V.~M.~A., {Toal{\'a}}, J.~A., {Guerrero}, M.~A.,
  {et~al.} 2020, \mnras, 496, 959, \dodoi{10.1093/mnras/staa1542}

\bibitem[{{Gon{\c{c}}alves} {et~al.}(2001){Gon{\c{c}}alves}, {Corradi}, \&
  {Mampaso}}]{2001ApJ...547..302G}
{Gon{\c{c}}alves}, D.~R., {Corradi}, R. L.~M., \& {Mampaso}, A. 2001, \apj,
  547, 302, \dodoi{10.1086/318364}

\bibitem[{{Hajduk} {et~al.}(2018){Hajduk}, {van Hoof}, {{\'S}niadkowska},
  {Krankowski}, {B{\l}aszkiewicz}, {D{\k{a}}browski}, \&
  {Zijlstra}}]{2018MNRAS.479.5657H}
{Hajduk}, M., {van Hoof}, P.~A.~M., {{\'S}niadkowska}, K., {et~al.} 2018,
  \mnras, 479, 5657, \dodoi{10.1093/mnras/sty1673}

\bibitem[{{Hajduk} \& {Zijlstra}(2012)}]{2012IAUS..283..380H}
{Hajduk}, M., \& {Zijlstra}, A.~A. 2012, IAU Symposium, 283, 380,
  \dodoi{10.1017/S1743921312011520}

\bibitem[{{Kaler}(1986)}]{1986ApJ...308..322K}
{Kaler}, J.~B. 1986, \apj, 308, 322, \dodoi{10.1086/164503}

\bibitem[{{Kaler} {et~al.}(1996){Kaler}, {Kwitter}, {Shaw}, \&
  {Browning}}]{1996PASP..108..980K}
{Kaler}, J.~B., {Kwitter}, K.~B., {Shaw}, R.~A., \& {Browning}, L. 1996, \pasp,
  108, 980, \dodoi{10.1086/133823}

\bibitem[{{Kingdon} \& {Ferland}(1995)}]{1995ApJ...450..691K}
{Kingdon}, J.~B., \& {Ferland}, G.~J. 1995, \apj, 450, 691,
  \dodoi{10.1086/176175}

\bibitem[{{Liu} {et~al.}(2000){Liu}, {Storey}, {Barlow}, {Danziger}, {Cohen},
  \& {Bryce}}]{2000MNRAS.312..585L}
{Liu}, X.~W., {Storey}, P.~J., {Barlow}, M.~J., {et~al.} 2000, \mnras, 312,
  585, \dodoi{10.1046/j.1365-8711.2000.03167.x}

\bibitem[{{Masson}(1990)}]{1990ApJ...348..580M}
{Masson}, C.~R. 1990, \apj, 348, 580, \dodoi{10.1086/168264}

\bibitem[{{McMullin} {et~al.}(2007){McMullin}, {Waters}, {Schiebel}, {Young},
  \& {Golap}}]{2007ASPC..376..127M}
{McMullin}, J.~P., {Waters}, B., {Schiebel}, D., {Young}, W., \& {Golap}, K.
  2007, in Astronomical Society of the Pacific Conference Series, Vol. 376,
  Astronomical Data Analysis Software and Systems XVI, ed. R.~A. {Shaw},
  F.~{Hill}, \& D.~J. {Bell}, 127

\bibitem[{{McNabb} {et~al.}(2013){McNabb}, {Fang}, {Liu}, {Bastin}, \&
  {Storey}}]{2013MNRAS.428.3443M}
{McNabb}, I.~A., {Fang}, X., {Liu}, X.~W., {Bastin}, R.~J., \& {Storey}, P.~J.
  2013, \mnras, 428, 3443, \dodoi{10.1093/mnras/sts283}

\bibitem[{{Miszalski} {et~al.}(2009){Miszalski}, {Acker}, {Parker}, \&
  {Moffat}}]{2009A&A...505..249M}
{Miszalski}, B., {Acker}, A., {Parker}, Q.~A., \& {Moffat}, A.~F.~J. 2009,
  \aap, 505, 249, \dodoi{10.1051/0004-6361/200912176}

\bibitem[{{O'Dell} {et~al.}(2013){O'Dell}, {Ferland}, {Henney}, \&
  {Peimbert}}]{2013AJ....145...92O}
{O'Dell}, C.~R., {Ferland}, G.~J., {Henney}, W.~J., \& {Peimbert}, M. 2013,
  \aj, 145, 92, \dodoi{10.1088/0004-6256/145/4/92}

\bibitem[{{Olnon}(1975)}]{1975A&A....39..217O}
{Olnon}, F.~M. 1975, \aap, 39, 217

\bibitem[{{Parker} {et~al.}(2016){Parker}, {Boji{\v{c}}i{\'c}}, \&
  {Frew}}]{2016JPhCS.728c2008P}
{Parker}, Q.~A., {Boji{\v{c}}i{\'c}}, I.~S., \& {Frew}, D.~J. 2016, in Journal
  of Physics Conference Series, Vol. 728, Journal of Physics Conference Series,
  032008, \dodoi{10.1088/1742-6596/728/3/032008}

\bibitem[{{Pazderska} {et~al.}(2009){Pazderska}, {Gawro{\'n}ski}, {Feiler},
  {Birkinshaw}, {Browne}, {Davis}, {Kus}, {Lancaster}, {Lowe}, {Pazderski},
  {Peel}, \& {Wilkinson}}]{2009A&A...498..463P}
{Pazderska}, B.~M., {Gawro{\'n}ski}, M.~P., {Feiler}, R., {et~al.} 2009, \aap,
  498, 463, \dodoi{10.1051/0004-6361/200811369}

\bibitem[{{Peimbert}(1967)}]{1967ApJ...150..825P}
{Peimbert}, M. 1967, \apj, 150, 825, \dodoi{10.1086/149385}

\bibitem[{{Peimbert}(1971)}]{1971BOTT....6...29P}
---. 1971, Boletin de los Observatorios Tonantzintla y Tacubaya, 6, 29

\bibitem[{{Phillips}(2007)}]{2007MNRAS.378..231P}
{Phillips}, J.~P. 2007, \mnras, 378, 231,
  \dodoi{10.1111/j.1365-2966.2007.11764.x}

\bibitem[{{Planck Collaboration} {et~al.}(2015){Planck Collaboration},
  {Arnaud}, {Atrio-Barandela}, {Aumont}, {Baccigalupi}, {Banday}, {Barreiro},
  {Battaner}, {Benabed}, {Benoit-L{\'e}vy}, {Bernard}, {Bersanelli},
  {Bielewicz}, {Bonaldi}, {Bond}, {Borrill}, {Bouchet}, {Buemi}, {Burigana},
  {Cardoso}, {Casassus}, {Catalano}, {Cerrigone}, {Chamballu}, {Chiang},
  {Colombi}, {Colombo}, {Couchot}, {Crill}, {Curto}, {Cuttaia}, {Davies},
  {Davis}, {de Bernardis}, {de Rosa}, {de Zotti}, {Delabrouille}, {Dickinson},
  {Diego}, {Donzelli}, {Dor{\'e}}, {Dupac}, {En{\ss}lin}, {Eriksen}, {Finelli},
  {Frailis}, {Franceschi}, {Galeotta}, {Ganga}, {Giard}, {Gonz{\'a}lez-Nuevo},
  {G{\'o}rski}, {Gregorio}, {Gruppuso}, {Hansen}, {Harrison}, {Hildebrandt},
  {Hivon}, {Holmes}, {Hora}, {Hornstrup}, {Hovest}, {Huffenberger}, {Jaffe},
  {Jones}, {Juvela}, {Keih{\"a}nen}, {Keskitalo}, {Kisner}, {Knoche}, {Kunz},
  {Kurki-Suonio}, {L{\"a}hteenm{\"a}ki}, {Lamarre}, {Lasenby}, {Lawrence},
  {Leonardi}, {Leto}, {Lilje}, {Linden-V{\o}rnle}, {L{\'o}pez-Caniego},
  {Mac{\'\i}as-P{\'e}rez}, {Maffei}, {Maino}, {Mandolesi}, {Martin}, {Masi},
  {Massardi}, {Matarrese}, {Mazzotta}, {Mendes}, {Mennella}, {Migliaccio},
  {Miville-Desch{\^e}nes}, {Moneti}, {Montier}, {Morgante}, {Mortlock},
  {Munshi}, {Murphy}, {Naselsky}, {Nati}, {Natoli}, {Noviello}, {Novikov},
  {Novikov}, {Pagano}, {Pajot}, {Paladini}, {Paoletti}, {Peel}, {Perdereau},
  {Perrotta}, {Piacentini}, {Piat}, {Pietrobon}, {Plaszczynski},
  {Pointecouteau}, {Polenta}, {Popa}, {Pratt}, {Procopio}, {Prunet}, {Puget},
  {Rachen}, {Reinecke}, {Remazeilles}, {Ricciardi}, {Riller}, {Ristorcelli},
  {Rocha}, {Rosset}, {Roudier}, {Rubi{\~n}o-Mart{\'\i}n}, {Rusholme}, {Sandri},
  {Savini}, {Scott}, {Spencer}, {Stolyarov}, {Sutton}, {Suur-Uski}, {Sygnet},
  {Tauber}, {Terenzi}, {Toffolatti}, {Tomasi}, {Trigilio}, {Tristram},
  {Trombetti}, {Tucci}, {Umana}, {Valiviita}, {Van Tent}, {Vielva}, {Villa},
  {Wade}, {Wandelt}, {Zacchei}, {Zijlstra}, \& {Zonca}}]{2015A&A...573A...6P}
{Planck Collaboration}, {Arnaud}, M., {Atrio-Barandela}, F., {et~al.} 2015,
  \aap, 573, A6, \dodoi{10.1051/0004-6361/201423836}

\bibitem[{{Pottasch}(1984)}]{1984ASSL..107.....P}
{Pottasch}, S.~R. 1984, {Planetary nebulae. A study of late stages of stellar
  evolution}, Vol. 107, \dodoi{10.1007/978-94-009-7233-9}

\bibitem[{{Pottasch} \& {Bernard-Salas}(2013)}]{2013A&A...550A..35P}
{Pottasch}, S.~R., \& {Bernard-Salas}, J. 2013, \aap, 550, A35,
  \dodoi{10.1051/0004-6361/201219647}

\bibitem[{{Ruffle} {et~al.}(2004){Ruffle}, {Zijlstra}, {Walsh}, {Gray},
  {Gesicki}, {Minniti}, \& {Comeron}}]{2004MNRAS.353..796R}
{Ruffle}, P.~M.~E., {Zijlstra}, A.~A., {Walsh}, J.~R., {et~al.} 2004, \mnras,
  353, 796, \dodoi{10.1111/j.1365-2966.2004.08113.x}

\bibitem[{{Savini} {et~al.}(2018){Savini}, {Bonafede}, {Br{\"u}ggen}, {Wilber},
  {Harwood}, {Murgia}, {Shimwell}, {Rafferty}, {Shulevski}, {Brienza},
  {Hardcastle}, {Morganti}, {R{\"o}ttgering}, {Clarke}, {de Gasperin}, {van
  Weeren}, {Best}, {Botteon}, {Brunetti}, \& {Cassano}}]{2018MNRAS.474.5023S}
{Savini}, F., {Bonafede}, A., {Br{\"u}ggen}, M., {et~al.} 2018, \mnras, 474,
  5023, \dodoi{10.1093/mnras/stx2876}

\bibitem[{{Shimwell} {et~al.}(2017){Shimwell}, {R{\"o}ttgering}, {Best},
  {Williams}, {Dijkema}, {de Gasperin}, {Hardcastle}, {Heald}, {Hoang},
  {Horneffer}, {Intema}, {Mahony}, {Mandal}, {Mechev}, {Morabito}, {Oonk},
  {Rafferty}, {Retana-Montenegro}, {Sabater}, {Tasse}, {van Weeren},
  {Br{\"u}ggen}, {Brunetti}, {Chy{\.z}y}, {Conway}, {Haverkorn}, {Jackson},
  {Jarvis}, {McKean}, {Miley}, {Morganti}, {White}, {Wise}, {van Bemmel},
  {Beck}, {Brienza}, {Bonafede}, {Calistro Rivera}, {Cassano}, {Clarke},
  {Cseh}, {Deller}, {Drabent}, {van Driel}, {Engels}, {Falcke}, {Ferrari},
  {Fr{\"o}hlich}, {Garrett}, {Harwood}, {Heesen}, {Hoeft}, {Horellou},
  {Israel}, {Kapi{\'n}ska}, {Kunert-Bajraszewska}, {McKay}, {Mohan},
  {Orr{\'u}}, {Pizzo}, {Prandoni}, {Schwarz}, {Shulevski}, {Sipior}, {Smith},
  {Sridhar}, {Steinmetz}, {Stroe}, {Varenius}, {van der Werf}, {Zensus}, \&
  {Zwart}}]{2017A&A...598A.104S}
{Shimwell}, T.~W., {R{\"o}ttgering}, H.~J.~A., {Best}, P.~N., {et~al.} 2017,
  \aap, 598, A104, \dodoi{10.1051/0004-6361/201629313}

\bibitem[{{Shimwell} {et~al.}(2019){Shimwell}, {Tasse}, {Hardcastle}, {Mechev},
  {Williams}, {Best}, {R{\"o}ttgering}, {Callingham}, {Dijkema}, {de Gasperin},
  {Hoang}, {Hugo}, {Mirmont}, {Oonk}, {Prandoni}, {Rafferty}, {Sabater},
  {Smirnov}, {van Weeren}, {White}, {Atemkeng}, {Bester}, {Bonnassieux},
  {Br{\"u}ggen}, {Brunetti}, {Chy{\.z}y}, {Cochrane}, {Conway}, {Croston},
  {Danezi}, {Duncan}, {Haverkorn}, {Heald}, {Iacobelli}, {Intema}, {Jackson},
  {Jamrozy}, {Jarvis}, {Lakhoo}, {Mevius}, {Miley}, {Morabito}, {Morganti},
  {Nisbet}, {Orr{\'u}}, {Perkins}, {Pizzo}, {Schrijvers}, {Smith}, {Vermeulen},
  {Wise}, {Alegre}, {Bacon}, {van Bemmel}, {Beswick}, {Bonafede}, {Botteon},
  {Bourke}, {Brienza}, {Calistro Rivera}, {Cassano}, {Clarke}, {Conselice},
  {Dettmar}, {Drabent}, {Dumba}, {Emig}, {En{\ss}lin}, {Ferrari}, {Garrett},
  {G{\'e}nova-Santos}, {Goyal}, {G{\"u}rkan}, {Hale}, {Harwood}, {Heesen},
  {Hoeft}, {Horellou}, {Jackson}, {Kokotanekov}, {Kondapally},
  {Kunert-Bajraszewska}, {Mahatma}, {Mahony}, {Mandal}, {McKean}, {Merloni},
  {Mingo}, {Miskolczi}, {Mooney}, {Nikiel-Wroczy{\'n}ski}, {O'Sullivan},
  {Quinn}, {Reich}, {Roskowi{\'n}ski}, {Rowlinson}, {Savini}, {Saxena},
  {Schwarz}, {Shulevski}, {Sridhar}, {Stacey}, {Urquhart}, {van der Wiel},
  {Varenius}, {Webster}, \& {Wilber}}]{2019A&A...622A...1S}
{Shimwell}, T.~W., {Tasse}, C., {Hardcastle}, M.~J., {et~al.} 2019, \aap, 622,
  A1, \dodoi{10.1051/0004-6361/201833559}

\bibitem[{{Si{\'o}dmiak} \& {Tylenda}(2001)}]{2001A&A...373.1032S}
{Si{\'o}dmiak}, N., \& {Tylenda}, R. 2001, \aap, 373, 1032,
  \dodoi{10.1051/0004-6361:20010664}

\bibitem[{{Stasi{\'n}ska}(2002)}]{2002RMxAC..12...62S}
{Stasi{\'n}ska}, G. 2002, in Revista Mexicana de Astronomia y Astrofisica
  Conference Series, Vol.~12, Revista Mexicana de Astronomia y Astrofisica
  Conference Series, ed. W.~J. {Henney}, J.~{Franco}, \& M.~{Martos}, 62--69.
\newblock \doarXiv{astro-ph/0102403}

\bibitem[{{Stasi{\'n}ska} \& {Szczerba}(2001)}]{2001A&A...379.1024S}
{Stasi{\'n}ska}, G., \& {Szczerba}, R. 2001, \aap, 379, 1024,
  \dodoi{10.1051/0004-6361:20011403}

\bibitem[{{Stasi{\'n}ska} {et~al.}(1992){Stasi{\'n}ska}, {Tylenda}, {Acker}, \&
  {Stenholm}}]{1992A&A...266..486S}
{Stasi{\'n}ska}, G., {Tylenda}, R., {Acker}, A., \& {Stenholm}, B. 1992, \aap,
  266, 486

\bibitem[{{Tasse} {et~al.}(2018){Tasse}, {Hugo}, {Mirmont}, {Smirnov},
  {Atemkeng}, {Bester}, {Hardcastle}, {Lakhoo}, {Perkins}, \&
  {Shimwell}}]{2018A&A...611A..87T}
{Tasse}, C., {Hugo}, B., {Mirmont}, M., {et~al.} 2018, \aap, 611, A87,
  \dodoi{10.1051/0004-6361/201731474}

\bibitem[{{Toal{\'a}} {et~al.}(2019){Toal{\'a}}, {Ramos-Larios}, {Guerrero}, \&
  {Todt}}]{2019MNRAS.485.3360T}
{Toal{\'a}}, J.~A., {Ramos-Larios}, G., {Guerrero}, M.~A., \& {Todt}, H. 2019,
  \mnras, 485, 3360, \dodoi{10.1093/mnras/stz624}

\bibitem[{{Tsamis} {et~al.}(2004){Tsamis}, {Barlow}, {Liu}, {Storey}, \&
  {Danziger}}]{2004MNRAS.353..953T}
{Tsamis}, Y.~G., {Barlow}, M.~J., {Liu}, X.~W., {Storey}, P.~J., \& {Danziger},
  I.~J. 2004, \mnras, 353, 953, \dodoi{10.1111/j.1365-2966.2004.08140.x}

\bibitem[{{Umana} {et~al.}(2015){Umana}, {Trigilio}, {Cerrigone}, {Cesaroni},
  {Zijlstra}, {Hoare}, {Weis}, {Beasley}, {Bomans}, {Hallinan}, {Molinari},
  {Taylor}, {Testi}, \& {Thompson}}]{2015aska.confE.118U}
{Umana}, G., {Trigilio}, C., {Cerrigone}, L., {et~al.} 2015, in Advancing
  Astrophysics with the Square Kilometre Array (AASKA14), 118.
\newblock \doarXiv{1412.5833}

\bibitem[{{van Haarlem} {et~al.}(2013){van Haarlem}, {Wise}, {Gunst}, {Heald},
  {McKean}, {Hessels}, {de Bruyn}, {Nijboer}, {Swinbank}, {Fallows},
  {Brentjens}, {Nelles}, {Beck}, {Falcke}, {Fender}, {H{\"o}randel},
  {Koopmans}, {Mann}, {Miley}, {R{\"o}ttgering}, {Stappers}, {Wijers},
  {Zaroubi}, {van den Akker}, {Alexov}, {Anderson}, {Anderson}, {van Ardenne},
  {Arts}, {Asgekar}, {Avruch}, {Batejat}, {B{\"a}hren}, {Bell}, {Bell}, {van
  Bemmel}, {Bennema}, {Bentum}, {Bernardi}, {Best}, {B{\^\i}rzan}, {Bonafede},
  {Boonstra}, {Braun}, {Bregman}, {Breitling}, {van de Brink}, {Broderick},
  {Broekema}, {Brouw}, {Br{\"u}ggen}, {Butcher}, {van Cappellen}, {Ciardi},
  {Coenen}, {Conway}, {Coolen}, {Corstanje}, {Damstra}, {Davies}, {Deller},
  {Dettmar}, {van Diepen}, {Dijkstra}, {Donker}, {Doorduin}, {Dromer}, {Drost},
  {van Duin}, {Eisl{\"o}ffel}, {van Enst}, {Ferrari}, {Frieswijk}, {Gankema},
  {Garrett}, {de Gasperin}, {Gerbers}, {de Geus}, {Grie{\ss}meier}, {Grit},
  {Gruppen}, {Hamaker}, {Hassall}, {Hoeft}, {Holties}, {Horneffer}, {van der
  Horst}, {van Houwelingen}, {Huijgen}, {Iacobelli}, {Intema}, {Jackson},
  {Jelic}, {de Jong}, {Juette}, {Kant}, {Karastergiou}, {Koers}, {Kollen},
  {Kondratiev}, {Kooistra}, {Koopman}, {Koster}, {Kuniyoshi}, {Kramer},
  {Kuper}, {Lambropoulos}, {Law}, {van Leeuwen}, {Lemaitre}, {Loose}, {Maat},
  {Macario}, {Markoff}, {Masters}, {McFadden}, {McKay-Bukowski}, {Meijering},
  {Meulman}, {Mevius}, {Middelberg}, {Millenaar}, {Miller-Jones}, {Mohan},
  {Mol}, {Morawietz}, {Morganti}, {Mulcahy}, {Mulder}, {Munk}, {Nieuwenhuis},
  {van Nieuwpoort}, {Noordam}, {Norden}, {Noutsos}, {Offringa}, {Olofsson},
  {Omar}, {Orr{\'u}}, {Overeem}, {Paas}, {Pand ey-Pommier}, {Pandey}, {Pizzo},
  {Polatidis}, {Rafferty}, {Rawlings}, {Reich}, {de Reijer}, {Reitsma},
  {Renting}, {Riemers}, {Rol}, {Romein}, {Roosjen}, {Ruiter}, {Scaife}, {van
  der Schaaf}, {Scheers}, {Schellart}, {Schoenmakers}, {Schoonderbeek},
  {Serylak}, {Shulevski}, {Sluman}, {Smirnov}, {Sobey}, {Spreeuw}, {Steinmetz},
  {Sterks}, {Stiepel}, {Stuurwold}, {Tagger}, {Tang}, {Tasse}, {Thomas},
  {Thoudam}, {Toribio}, {van der Tol}, {Usov}, {van Veelen}, {van der Veen},
  {ter Veen}, {Verbiest}, {Vermeulen}, {Vermaas}, {Vocks}, {Vogt}, {de Vos},
  {van der Wal}, {van Weeren}, {Weggemans}, {Weltevrede}, {White}, {Wijnholds},
  {Wilhelmsson}, {Wucknitz}, {Yatawatta}, {Zarka}, {Zensus}, \& {van
  Zwieten}}]{2013A&A...556A...2V}
{van Haarlem}, M.~P., {Wise}, M.~W., {Gunst}, A.~W., {et~al.} 2013, \aap, 556,
  A2, \dodoi{10.1051/0004-6361/201220873}

\bibitem[{{van Hoof}(2000)}]{2000MNRAS.314...99V}
{van Hoof}, P.~A.~M. 2000, \mnras, 314, 99,
  \dodoi{10.1046/j.1365-8711.2000.03281.x}

\bibitem[{{van Hoof} {et~al.}(2014){van Hoof}, {Williams}, {Volk}, {Chatzikos},
  {Ferland}, {Lykins}, {Porter}, \& {Wang}}]{2014MNRAS.444..420V}
{van Hoof}, P.~A.~M., {Williams}, R.~J.~R., {Volk}, K., {et~al.} 2014, \mnras,
  444, 420, \dodoi{10.1093/mnras/stu1438}

\bibitem[{{van Hoof} {et~al.}(2010){van Hoof}, {van de Steene}, {Barlow},
  {Exter}, {Sibthorpe}, {Ueta}, {Peris}, {Groenewegen}, {Blommaert}, {Cohen},
  {De Meester}, {Ferland}, {Gear}, {Gomez}, {Hargrave}, {Huygen}, {Ivison},
  {Jean}, {Leeks}, {Lim}, {Olofsson}, {Polehampton}, {Regibo}, {Royer},
  {Swinyard}, {Vandenbussche}, {van Winckel}, {Waelkens}, {Walker}, \&
  {Wesson}}]{2010A&A...518L.137V}
{van Hoof}, P.~A.~M., {van de Steene}, G.~C., {Barlow}, M.~J., {et~al.} 2010,
  \aap, 518, L137, \dodoi{10.1051/0004-6361/201014590}

\bibitem[{{van Weeren} {et~al.}(2016){van Weeren}, {Williams}, {Hardcastle},
  {Shimwell}, {Rafferty}, {Sabater}, {Heald}, {Sridhar}, {Dijkema}, {Brunetti},
  {Br{\"u}ggen}, {Andrade-Santos}, {Ogrean}, {R{\"o}ttgering}, {Dawson},
  {Forman}, {de Gasperin}, {Jones}, {Miley}, {Rudnick}, {Sarazin}, {Bonafede},
  {Best}, {B{\^\i}rzan}, {Cassano}, {Chy{\.z}y}, {Croston}, {Ensslin},
  {Ferrari}, {Hoeft}, {Horellou}, {Jarvis}, {Kraft}, {Mevius}, {Intema},
  {Murray}, {Orr{\'u}}, {Pizzo}, {Simionescu}, {Stroe}, {van der Tol}, \&
  {White}}]{2016ApJS..223....2V}
{van Weeren}, R.~J., {Williams}, W.~L., {Hardcastle}, M.~J., {et~al.} 2016,
  \apjs, 223, 2, \dodoi{10.3847/0067-0049/223/1/2}

\bibitem[{{Wenger} {et~al.}(2000){Wenger}, {Ochsenbein}, {Egret}, {Dubois},
  {Bonnarel}, {Borde}, {Genova}, {Jasniewicz}, {Lalo{\"e}}, {Lesteven}, \&
  {Monier}}]{2000A&AS..143....9W}
{Wenger}, M., {Ochsenbein}, F., {Egret}, D., {et~al.} 2000, \aaps, 143, 9,
  \dodoi{10.1051/aas:2000332}

\bibitem[{{Wesson} \& {Liu}(2004)}]{2004MNRAS.351.1026W}
{Wesson}, R., \& {Liu}, X.~W. 2004, \mnras, 351, 1026,
  \dodoi{10.1111/j.1365-2966.2004.07856.x}

\bibitem[{{Wesson} {et~al.}(2005){Wesson}, {Liu}, \&
  {Barlow}}]{2005MNRAS.362..424W}
{Wesson}, R., {Liu}, X.~W., \& {Barlow}, M.~J. 2005, \mnras, 362, 424,
  \dodoi{10.1111/j.1365-2966.2005.09325.x}

\bibitem[{{Wright} \& {Barlow}(1975)}]{1975MNRAS.170...41W}
{Wright}, A.~E., \& {Barlow}, M.~J. 1975, \mnras, 170, 41,
  \dodoi{10.1093/mnras/170.1.41}

\bibitem[{{Zhang} {et~al.}(2005){Zhang}, {Liu}, {Liu}, \&
  {Rubin}}]{2005MNRAS.358..457Z}
{Zhang}, Y., {Liu}, X.~W., {Liu}, Y., \& {Rubin}, R.~H. 2005, \mnras, 358, 457,
  \dodoi{10.1111/j.1365-2966.2005.08810.x}

\bibitem[{{Zhang} {et~al.}(2004){Zhang}, {Liu}, {Wesson}, {Storey}, {Liu}, \&
  {Danziger}}]{2004MNRAS.351..935Z}
{Zhang}, Y., {Liu}, X.~W., {Wesson}, R., {et~al.} 2004, \mnras, 351, 935,
  \dodoi{10.1111/j.1365-2966.2004.07838.x}

\end{thebibliography}
\bibliographystyle{aasjournal}

\appendix

\begin{longtable}{ccc}
\caption{Upper limits for nebular flux densities at 144\,MHz and corresponding brightness temperatures.}
\label{tab:upper} \\
\hline
\hline
\\[-1.5ex]
{Name} & {3 $\times$ RMS} & {$T_\mathrm{B}$}  \\
 & {[mJy/beam]} & K  \\
\\[-1.5ex]
\hline
\\[-1.5ex]
A\,16	&	0.791	&	1853	\\
A\,28	&	0.722	&	1692	\\
A\,30	&	0.570	&	1335	\\
A\,39	&	3.330	&	7806	\\
A\,43	&	3.180	&	7452	\\
A\,43	&	0.762	&	1786	\\
A\,46	&	0.669	&	1569	\\
A\,53	&	4.980	&	11660	\\
A\,59	&	5.090	&	11932	\\
A\,61	&	0.927	&	2172	\\
A\,63	&	3.890	&	9109	\\
A\,73	&	3.810	&	8935	\\
A\,74	&	0.726	&	1700	\\
A\,79	&	5.810	&	13606	\\
A\,84	&	5.340	&	12520	\\
Bl\,2-1	&	0.639	&	1496	\\
Cn\,3-1	&	5.790	&	13563	\\
DDDM\,1	&	0.391	&	916	\\
EGB\,1	&	0.411	&	962	\\
ETHOS\,4	&	1.060	&	2486	\\
FBP\,8	&	1.560	&	3644	\\
FSMV\,1	&	1.380	&	3226	\\
GLMP\,879	&	10.200	&	23976	\\
HaTr\,12	&	3.280	&	7674	\\
HaTr\,14	&	7.210	&	16897	\\
Hen\,1-1	&	1.730	&	4061	\\
Hen\,1-2	&	1.350	&	3173	\\
Hen\,2-447	&	5.360	&	12561	\\
IC\,2003	&	0.955	&	2237	\\
IC\,351	&	2.150	&	5040	\\
IPHAS\,J185321.76+055641.9	&	1.850	&	4333	\\
IPHAS\,J185322.1+083018	&	1.750	&	4090	\\
IPHAS\,J185744.4+105053	&	1.690	&	3949	\\
IPHAS\,J185815.8+073753	&	1.820	&	4271	\\
IPHAS\,J185957.0+073544	&	1.940	&	4537	\\
IPHAS\,J190718.1+044056	&	7.120	&	16681	\\
IPHAS\,J192553.53+165331.4	&	5.060	&	11854	\\
IPHAS\,J193517.8+223120 	&	8.250	&	19320	\\
IPHAS\,J193652.96+171940.7	&	2.350	&	5495	\\
IPHAS\,J193718.6+202102	&	1.150	&	2701	\\
IPHAS\,J221118.0+552841	&	5.810	&	13606	\\
IPHASX\,J185225.0+080843	&	1.950	&	4565	\\
IPHASX\,J185309.4+075241	&	1.740	&	4069	\\
IPHASX\,J190340.7+094639	&	4.280	&	10017	\\
IPHASX\,J190417.9+084916	&	3.220	&	7555	\\
IPHASX\,J190432.9+091656	&	4.400	&	10305	\\
IPHASX\,J190454.0+101801	&	6.500	&	15224	\\
IPHASX\,J192146.7+172055	&	8.250	&	19320	\\
IPHASX\,J193009.3+192129	&	1.040	&	2437	\\
IPHASX\,J194301.3+215424	&	3.610	&	8456	\\
IPHASX\,J194648.2+193608	&	5.630	&	13187	\\
IRAS\,19086+0603	&	1.940	&	4537	\\
IRAS\,19297+1954	&	0.929	&	2176	\\
Jn\,1	&	0.289	&	678	\\
JnEr\,1	&	2.060	&	4837	\\
K\,1-15	&	1.240	&	2906	\\
K\,1-16	&	0.422	&	988	\\
K\,1-20	&	4.070	&	9527	\\
K\,3-14	&	9.270	&	21720	\\
K\,3-15	&	3.240	&	7603	\\
K\,3-31	&	4.880	&	11441	\\
K\,3-32	&	5.320	&	12471	\\
K\,3-35	&	1.280	&	3004	\\
K\,3-38	&	2.380	&	5566	\\
K\,3-40	&	5.260	&	12325	\\
K\,3-42	&	1.400	&	3284	\\
K\,3-43	&	1.720	&	4039	\\
K\,3-58	&	2.330	&	5448	\\
K\,3-73	&	2.860	&	6702	\\
K\,3-76	&	6.470	&	15150	\\
K4\,-30	&	2.080	&	4881	\\
KLSS\,1-1	&	2.850	&	6672	\\
KLSS\,1-2	&	1.170	&	2737	\\
KLSS\,2-1	&	2.990	&	7008	\\
KLSS\,2-6	&	2.130	&	4990	\\
Kn\,132	&	0.865	&	2026	\\
Kn\,20	&	2.600	&	6087	\\
Kn\,21	&	2.360	&	5524	\\
Kn\,23	&	0.675	&	1580	\\
Kn\,43	&	1.700	&	3988	\\
Kn\,49	&	0.708	&	1659	\\
Kn\,58	&	1.700	&	3979	\\
Kn\,59	&	1.230	&	2873	\\
Kn\,68	&	0.694	&	1626	\\
Kn\,7	&	2.760	&	6461	\\
Kn\,9	&	8.750	&	20495	\\
KnFe\,1	&	5.270	&	12347	\\
LoTr\,5	&	0.401	&	940	\\
M\,1-64	&	2.340	&	5486	\\
M\,1-71	&	1.040	&	2429	\\
M\,1-72	&	2.630	&	6172	\\
M\,2-53	&	2.580	&	6044	\\
M\,3-35	&	6.070	&	14215	\\
MSX\,6c	&	2.580	&	6048	\\
NGC\,6742	&	5.190	&	12166	\\
NGC\,6765	&	3.670	&	8589	\\
NGC\,6833	&	0.902	&	2114	\\
Ou\,2	&	1.090	&	2558	\\
Ou\,3	&	6.870	&	16095	\\
Ou\,5	&	1.560	&	3665	\\
Pa\,157	&	1.540	&	3619	\\
Pa\,18	&	1.440	&	3373	\\
Pa\,27	&	2.390	&	5595	\\
Pa\,4	&	5.500	&	12883	\\
Pa\,5	&	6.340	&	14853	\\
PK\,020-02 1	&	1.680	&	3930	\\
PM\,1-262	&	1.460	&	3424	\\
PM\,1-264	&	1.770	&	4146	\\
PM\,1-267	&	1.780	&	4163	\\
PM\,1-273	&	2.040	&	4785	\\
PM\,1-276	&	4.410	&	10329	\\
PM\,1-279	&	3.940	&	9233	\\
PM\,1-335	&	3.420	&	8019	\\
Pre\,8	&	1.270	&	2983	\\
Ra\,1	&	1.120	&	2625	\\
Rai\,1	&	0.980	&	2296	\\
Sh\,1-118	&	1.810	&	4248	\\
SkAc\,1	&	0.546	&	1279	\\
StDr\,11	&	3.490	&	8185	\\
StDr\,12	&	7.070	&	16569	\\
StDr\,25	&	1.450	&	3395	\\
StDr\,28	&	7.010	&	16413	\\
StDr\,30	&	1.480	&	3473	\\
Te\,8	&	9.540	&	22359	\\
Tk\,1	&	0.273	&	639	\\
Tk\,2	&	2.160	&	5050	\\
TS\,1	&	1.310	&	3067	\\
UWISH\,2	&	0.970	&	2272	\\
We\,1-2	&	0.489	&	1146	\\
We\,1-3	&	0.649	&	1521	\\
We\,2-245	&	2.540	&	5947	\\
We\,92	&	0.643	&	1505	\\
WOW\,1	&	3.040	&	7124	\\
WSLS\,1	&	1.040	&	2435	\\
YM\,16	&	1.720	&	4023	\\
\hline
\end{longtable}

\newpage
\begin{longtable}{cccc}
\caption{Lower limits for 144\,MHz-1.4\,GHz spectral indices.}
\label{tab:si} \\
\hline
\hline
\\[-1.5ex]
{Name} & {3 $\times$ RMS} & $F_\mathrm{1.4\,GHz}$ & $SI_{\mathrm{1.4-0.144\,GHz}}$ \\
 &  [mJy/beam] & [mJy] & \\
\\[-1.5ex]
\hline
\\[-1.5ex]
A66 53	&	4.98	&	33.6	$\pm$	1.1	&	0.84	\\
A66 63	&	3.89	&	4.5	$\pm$	5	&	0.06	\\
A66 73	&	3.81	&	11	$\pm$	1.3	&	0.47	\\
A66 79	&	5.81	&	16.6	$\pm$	2.1	&	0.46	\\
Cn 3-1	&	5.79	&	59.5	$\pm$	1.8	&	1.02	\\
EGB 1	&	0.41	&	8.6	$\pm$	1.7	&	1.34	\\
Hen 1-1	&	1.73	&	16	$\pm$	0.7	&	0.98	\\
Hen 1-2	&	1.35	&	14.6	$\pm$	0.6	&	1.05	\\
Hen 2-447	&	5.36	&	22.9	$\pm$	1.9	&	0.64	\\
IC 2003	&	0.96	&	54.8	$\pm$	1.7	&	1.78	\\
IC 351	&	2.15	&	31.9	$\pm$	1.3	&	1.19	\\
IPHASX J185309.4+075241	&	1.74	&	9	$\pm$	0.5	&	0.72	\\
IPHASX J185815.8+073753	&	1.82	&	16.6	$\pm$	1.1	&	0.97	\\
IPHASX J190340.7+094639	&	4.28	&	6.1	$\pm$	0.6	&	0.16	\\
IPHASX J190432.9+091656	&	4.40	&	17.5	$\pm$	0.7	&	0.61	\\
IPHASX J192553.5+165331	&	5.06	&	44.6	$\pm$	1.8	&	0.96	\\
IPHASX J193718.6+202102	&	1.15	&	6.6	$\pm$	0.5	&	0.77	\\
IPHASX J221118.0+552841	&	5.81	&	3	$\pm$	1.5	&	-0.29	\\
IRAS 19086+0603	&	1.94	&	8.5	$\pm$	0.7	&	0.65	\\
IRAS 19297+1954	&	0.93	&	2.1	$\pm$	0.5	&	0.36	\\
K 3-31	&	4.88	&	16.8	$\pm$	0.7	&	0.54	\\
K 3-35	&	1.28	&	14.5	$\pm$	0.6	&	1.07	\\
K 3-38	&	2.38	&	28.7	$\pm$	1	&	1.09	\\
K 3-40	&	5.26	&	17.1	$\pm$	0.7	&	0.52	\\
K 3-42	&	1.40	&	11.2	$\pm$	1	&	0.91	\\
K 3-43	&	1.72	&	4.8	$\pm$	0.5	&	0.45	\\
K 4-30	&	2.08	&	20	$\pm$	1	&	1.00	\\
NGC 6742	&	5.19	&	3.7	$\pm$	0.5	&	-0.15	\\
NGC 6765	&	3.67	&	10.1	$\pm$	0.5	&	0.45	\\
NGC 6833	&	0.90	&	4.2	$\pm$	0.5	&	0.68	\\
Sh 1-118	&	1.81	&	39.8	$\pm$	3.1	&	1.36	\\
Tk 1	&	0.27	&	14	$\pm$	0.6	&	1.73	\\
\hline
\end{longtable}

\end{document}